\documentclass[12pt,hyperfootnotes=false, hidelinks=true]{article}
\usepackage{mathptmx}
\usepackage[latin9]{inputenc}
\usepackage{color}
\usepackage{float}
\usepackage{dsfont}
\usepackage{amsmath}
\usepackage{amsthm}
\usepackage{amssymb}
\usepackage{graphicx}
\usepackage{geometry}
\usepackage{setspace}
\usepackage[authoryear]{natbib}
\usepackage[]
 {hyperref}

\makeatletter
\usepackage{amsthm}
\usepackage{bm}
\usepackage{setspace}
\usepackage{sectsty}
\usepackage{booktabs}
\usepackage{longtable}

\usepackage{titling}
\usepackage{datetime}
\newdateformat{ourdate}{\monthname[\THEMONTH] \THEDAY , \THEYEAR}
\usepackage{pdflscape}
\usepackage[english]{babel}
\usepackage{times}
\usepackage[small]{caption}
\usepackage[bottom,hang,flushmargin]{footmisc}
\usepackage[position=top]{subfig}
\usepackage{titlesec}
\titlespacing*{\section}{0pt}{0.2\baselineskip}{0.2\baselineskip}
\titlespacing*{\subsection}{0pt}{0.2\baselineskip}{0.2\baselineskip}
\titlespacing*{\subsubsection}{0pt}{0.2\baselineskip}{0.2\baselineskip}

\titleformat{\section}
  {\normalfont\large\bfseries}{\thesection}{1em}{}
\titleformat{\subsection}
  {\normalfont\normalsize\bfseries}{\thesubsection}{1em}{}
\titleformat{\subsubsection}
  {\normalfont\it}{\thesubsubsection}{1em}{}

\newcommand{\taumeanaligned}{4.1}
\newcommand{\taumeanopposed}{5.5}
\newcommand{\taumeanalignedlessopposedpval}{0.15}
\newcommand{\attentionshposmeanaligned}{20.7}
\newcommand{\attentionshposmeanopposed}{20.8}

\newcommand{\womexposureyzero}{12.4}
\newcommand{\adawarenessyzero}{5.8}
\newcommand{\otheraffiliationpct}{11}
\newcommand{\articlepremonthavg}{161}
\newcommand{\articlediffmean}{87}
\newcommand{\articlediffmedian}{15}
\newcommand{\articlediffpsevenfive}{66}
\newcommand{\articlediffmeanpct}{54}
\newcommand{\countyagreementpct}{70}
\newcommand{\intupct}{11}
\newcommand{\doubledonorpct}{0.7}
\newcommand{\donorcountfloor}{30}
\newcommand{\xgboostntrees}{11288}
\newcommand{\donorpredictionsbalancedaccuracyoos}{0.82}

\newcommand{\donorpredictionsbalancedaccuracyoosthreedigits}{0.824}
\newcommand{\donorpredictionsrocaucoos}{0.91}

\newcommand{\donorpredictionsrocaucoosthreedigits}{0.910}

\newcommand{\donorpredictionsbalancedaccuracyooscountygeothreedigits}{0.660}

\newcommand{\donorpredictionsbalancedaccuracyoostudemothreedigits}{0.525}

\newcommand{\donorpredictionsbalancedaccuracyoostransactionsthreedigits}{0.817}

\newcommand{\donorpredictionsrocaucooscountygeothreedigits}{0.722}

\newcommand{\donorpredictionsrocaucoostudemothreedigits}{0.546}

\newcommand{\donorpredictionsrocaucoostransactionsthreedigits}{0.904}
\newcommand{\coefaligneddonormatchtotal}{19}
\newcommand{\negcoefopposeddonormatchtotal}{12}
\newcommand{\donordiff}{31}
\newcommand{\seconddiff}{7}
\newcommand{\coefaligneddonorlevels}{18}
\newcommand{\negcoefopposeddonorlevels}{13}
\newcommand{\coefaligneddecilelevels}{9}
\newcommand{\negcoefopposeddecilelevels}{6}
\newcommand{\coefoverall}{3}
\newcommand{\seconddiffalignmentshare}{0.59}
\newcommand{\coefalignedsharemonthzero}{19}
\newcommand{\coefalignedsharemonthnine}{9}
\newcommand{\bootstrapiterations}{ten thousand}

\newcommand{\llmStanceOrgOrProductHighPreMean}{9}

\newcommand{\llmStanceOrgHighEventSizeMean}{74}

\newcommand{\sharetauimputed}{14}
\newcommand{\taumean}{5}
\newcommand{\taumedian}{3}
\newcommand{\taupsevenfive}{5}
\newcommand{\taupnineninerounded}{40}
\newcommand{\nsocialevents}{116}
\newcommand{\nsocialfirms}{95}

\newcommand{\sharesocialeventsgtonly}{32.8}
\newcommand{\sharesocialeventsnewsonly}{26.7}
\newcommand{\sharesocialeventsgptonly}{3.4}
\newcommand{\sharesocialeventsbionly}{4.3}
\newcommand{\sharesocialeventsmultiplesources}{32.8}
\newcommand{\sharesocialeventsmultiplesourcestauweighted}{61.5}
\newcommand{\sharesocialeventsgptbiadded}{9.5}
\newcommand{\eventsizecomparisonstr}{fourth}

\newcommand{\awarenessimputationRMSE}{0.04}
\newcommand{\llmStanceOrgHighPctDiff}{822}

\ifdefined\showcaptionsetup
 \PassOptionsToPackage{caption=false}{subfig}
\fi
\usepackage{subfig}
\makeatother

\begin{document}
\title{\noindent\textbf{Consuming Values\thanks{{\footnotesize Email: jacob.conway@chicagobooth.edu, leviboxell@gmail.com.
We particularly thank Matthew Gentzkow, Liran Einav, and Amit Seru
for their invaluable mentorship and advice. We thank Elliott Ash,
Efraim Benmelech, Marianne Bertrand, Julia Cag\'e, Emanuele Colonnelli,
Jos\'e Ignacio Cuesta, Daniel Diermeier, Kwabena Donkor, James Druckman,
Wesley Hartmann, Caroline Hoxby, Guido Imbens, Emir Kamenica, John
List, Neale Mahoney, Timothy McQuade, Muriel Niederle, Paola Sapienza,
Jann Spiess, Francesco Trebbi, Ali Yurukoglu, Guangli Zhang, Luigi
Zingales, and seminar participants at Stanford, the University of
Chicago, and other universities and conferences for valuable suggestions.
We thank Peng Zhang for excellent research assistance. We gratefully
acknowledge financial support from the National Science Foundation
(grant DGE-1656518), the Institute for Research in the Social Sciences
at Stanford University, the Institute for Humane Studies, the Stanford
Center on Philanthropy and Civil Society, the B.F. Haley and E.S.
Shaw Fellowship for Economics (through a grant at the Stanford Institute
for Economic Policy Research), and the George P. Shultz Dissertation
Support Fund at SIEPR. This research was supported in part through
the computational resources and staff contributions provided for the
Mercury high-performance computing cluster at The University of Chicago
Booth School of Business, which is supported by the Office of the
Dean. We thank the data science team at a payment card company for
generously providing data used in this paper. Researchers' own secondary
analyses (of advertising and prices) were conducted in part using
data from The Nielsen Company (US), LLC, from Market Track, LLC dba
Numerator, and from marketing databases provided through the Nielsen
and Numerator Datasets at the Kilts Center for Marketing Data at The
University of Chicago Booth School of Business. All conclusions drawn
are those of the researchers and do not reflect data providers' views.
Data providers are not responsible for and had no role in analyzing
and preparing the results reported herein. Levi Boxell worked on this
paper while a Ph.D.\ student at Stanford University. Our research
was approved by the Stanford IRB (eProtocol 53506).}}}\vspace*{-0.4cm}}
\author{\noindent Jacob Conway\vspace*{-0.2cm}\\
\small \emph{University of Chicago}
\and
Levi Boxell\vspace*{-0.2cm}\\
\small \emph{Independent}}
\date{April 2026\vspace*{-1cm}}
\maketitle
\begin{abstract}
\begin{onehalfspace}
\noindent We study the extent to which individuals' consumption decisions
are influenced by firms' stances on controversial social issues and
the implied incentives for firms to take such stances. We use transactions
from a major payment card company to predict cardholders' likely social
alignment with firm stances and to quantify effects on consumption.
The social stances taken by firms increase revenue on average, with
significant heterogeneity across consumers and firm stances. Consumers
most aligned with a firm's social stance increase their consumption
at the firm by $\coefaligneddonormatchtotal$ percent in the month
following widely known social stance events, and consumers most opposed
to the firm's stance decrease their consumption by $\negcoefopposeddonormatchtotal$
percent. These diverging consumption responses attenuate over time
but persist even a year later. Firms tend to take stances that align
with their consumers' and employees' social preferences and that correlate
with the firm's ownership structure. Together, our results show that
consumers meaningfully respond to their social alignment with firms,
and that this consumer response can incentivize profit-maximizing
firms to engage with social issues.
\end{onehalfspace}
\end{abstract}
\pagebreak{}

\section{Introduction}

In recent years, both consumers and firms have frequently engaged
with social issues and espoused social motives in their decision-making
\citep{roundtable2019statement,lin2022capitalist,rajan2022purpose}.
This has led to active debates about the extent to which socially
conscious consumerism can meaningfully incentivize firm behavior,
and about whether firms' social engagement is consistent with traditional
profit-maximization motives \citep{friedman1970social,graff2005modigliani,ramaswamy2021woke}.

Most consumers state that they have made or avoided a purchase due
to the social values of a company and that they are attracted by
firm values that align with their own \citep{anes2016,barton_ishikawa_theofilou_quiring_2018},
but some researchers hypothesize that consumers' self-reported demand
may be ``cheap talk\textquotedblright{} in that stated preferences
may not reflect their actual purchase decisions \citep{auger2007consumers}.
The magnitude of this response matters for society as it determines
the extent to which consumers can incentivize or discipline firm behavior.
Some have argued that socially conscious consumerism can effectively
cause firms to internalize their social externalities \citep{barboza2021shared,economistecoawakening},
while others have argued that it is too weak to do so and may crowd
out other forms of regulation \citep{csutora2012one,wicker2017conscious,sheffi2021climate}.

One increasingly relevant setting for potential engagement occurs
when firms take controversial social stances, such as advertising
campaigns featuring divisive racial justice protesters, corporate
policy on contraceptives and reproductive rights, or comments regarding
sexual orientation and gender legislation \citep{lin2022capitalist}.
These controversial stances have become more frequent in recent years
\citep{klostermann2021effect,economist2021}, while consumers have
also become more socially divided and increasingly report caring about
and seeking out information on firm stances \citep{iyengar2015fear,gsg2018}.

In this paper, we study the extent to which individuals' consumption
decisions respond to their social alignment with firms around events
in which firms took salient and controversial social stances, and
we analyze the ensuing incentives for firms to take such stances.
We use consumer transactions from a major payment card company (covering
approximately 20 percent of all U.S.\ consumption) to predict cardholders'
likely social alignment with firm stances and to quantify effects
on consumption. We estimate that observed firm stances increase revenue
on average, with considerable heterogeneity across consumers and events.
Consumers whose social views are likely aligned with a firm's stance
increase their consumption at the firm in the months following the
social stance event, while consumers likely to be opposed to the firm's
stance decrease their consumption. Social stances thus typically have
more positive revenue impacts when the stance better aligns with the
views of the firm's customers, and revenue-maximizing social stance
decisions vary across firms. We also show that firms tend to take
stances that align with their consumers' and employees' social preferences
and that correlate with the firm's ownership structure.

We start by building a dataset of $\nsocialevents$ events in which
controversial social stances were taken by firms within our transaction
data. We identify these events in part by searching systematically
for unusual spikes combining the firm name with keywords indicative
of social stances in either Google Trends searches or in news articles.
We extend this list of events based on contemporaneous brand perception
surveys and queries to a large language model (OpenAI's GPT-4). We
measure consumer awareness of each event in contemporaneous surveys
that asked respondents whether they had recently heard any good or
bad news about the firm.

We then use the transaction data to measure the effect of each stance
on consumption at the firm, quantifying overall impacts as well as
heterogeneity across consumer groups that are likely more aligned
or less aligned with the firm's stance. We account for changes in
consumption at the firm unrelated to its stance by predicting the
counterfactual consumption that would have occurred absent the firm's
stance. Following the synthetic difference-in-differences approach
of \citet{arkhangelsky2021synthetic}, this prediction draws from
contemporaneous consumption at each of the thousands of other firms
in the economy and from past consumption at the firm taking a social
stance. This synthetic series closely tracks consumption at the firm
prior to its social stance.

Observed firm social stances increase revenue on average relative
to the synthetic counterfactual. To illustrate how the magnitude of
this response varies with consumer awareness, consider a stance that
25 percent of consumers report hearing about in contemporaneous surveys,
which would be the \eventsizecomparisonstr\ most salient event
in our data. On average, such a stance significantly increases overall
revenues by $\coefoverall$ percent in the month following the firm's
stance. Our estimated impacts in subsequent months are weakly positive
on average but not statistically significant at the 95 percent level.

We analyze heterogeneous treatment effects across consumers by first
inferring cardholders' likely alignment with firm stances from their
transactions and demographics. To do so, we identify more than 30
million consumers who have clearly expressed their likely alignment
on social issues through their donations to PACs, charitable organizations,
and other non-profits. We use these donors to train a machine learning
model that predicts an individual's likely alignment based on a wider
set of their other transactions and demographics, which we use to
predict likely social alignment among all other consumers (``non-donors'').
We then quantify the distribution of revenues across consumers by
alignment for each firm in the year preceding its social stance event.

Disaggregating the overall consumption response by alignment reveals
starkly diverging responses, thus providing clear evidence of consumer
demand for social alignment with firms. Donors aligned with highly
salient firm stances increase their consumption at the firm in the
following month by $\coefaligneddonormatchtotal$ percent, and donors
opposed to the firm's stance decrease their consumption at the firm
by $\negcoefopposeddonormatchtotal$ percent. These consumption responses
attenuate over time but persist even a year later. There is also a
strongly positive correlation between how likely non-donors are to
be aligned with a firm's stance and their consumption response.

Turning to the supply-side implications of our consumer response estimates,
we analyze the revenue-maximizing stance for hypothetical firms facing
different baseline consumption shares across consumer social alignment
groups. For example, we show how revenue-maximizing social stance
decisions vary for typical firms depending on the state or industry
in which they operate.\emph{}

Decision-makers taking firm social stances may care not only about
impacts on revenues, but may also seek to align with their own preferences
or the preferences of other stakeholders \citep{benabou2010individual}.
For example, companies often face internal pressure from employees
\citep{maks2021corporations} as well as external pressure from their
shareholders and owners \citep{baron2009positive} on social issues.\footnote{Relatedly, there is an active normative debate regarding what the
purpose and goals of a corporation \emph{should} be. Normative theory
in business ethics has long been dominated by the Friedman doctrine
\citep{friedman1970social}, which argues that firms are beholden
to their shareholders (i.e., \textquotedblleft shareholder primacy\textquotedblright ).
More recently, however, the main business association for CEOs has
argued that companies should also commit to benefiting their customers,
employees, suppliers, and communities (\citealp{roundtable2019statement},
i.e., \textquotedblleft stakeholder theory\textquotedblright ).} Combining our stances with measures of the preferences of a firm's
different stakeholders, we analyze which stakeholders' preferences
best predict the direction of a firm's stance and how this interacts
with the firm's corporate governance structure. The direction of a
firm's stance is best predicted by the preferences of its employees
and consumers, as well as by its public vs.\ private ownership
status. In contrast, the social alignment of its corporate board
is not a strong predictor.\emph{}

Our paper contributes to several existing literatures. The first analyzes
socially conscious consumerism, quantifying the extent to which consumers\textquoteright{}
preferences on social or environmental issues impact their purchase
decisions. Closest to our own paper are studies that examine consumer
responses to controversial firm social stances and the net impacts
of these stances. For example, \citet{liaukonyte2023frontiers} analyze
a controversial social stance by Goya, finding evidence of increased
consumption at store locations in counties that are home to many consumers
who are likely aligned with the firm\textquoteright s stance (\textquotedblleft aligned
buycotts\textquotedblright ). They do not find similar evidence of
``opposed boycotts'' in more opposed areas, and thus estimate that
Goya\textquoteright s stance had a positive net impact on its sales
in the following weeks. Similarly, \citet{painter2021consumer} uses
smartphone-location data to quantify foot traffic responses to a Walmart
stance favoring increased gun control. Painter finds increases among
locations in socially aligned counties, but (in contrast to \citealp{liaukonyte2023frontiers})
also estimates decreases in generally socially opposed counties that
result in a negative overall impact on foot traffic to Walmart. \citet{klostermann2021effect}
analyze the impact of controversial firm social stances on self-reported
favorability toward the firm in YouGov BrandIndex data, finding negative
overall impacts on favorability on average. \citet{hydock2020should}
find evidence of aligned buycotts, opposed boycotts, and negative
overall impacts on average when providing information about firm stances
in unincentivized survey experiments. \citet{schoenmueller2023polarized}
find evidence that after some types of firm stances, the firm's Twitter
following shifts toward users who are likely aligned with the firm's
stance.\footnote{\citet{schoenmueller2023polarized} focus primarily on documenting
increased polarization in consumer behavior following the 2016 U.S.\ presidential
election.} These papers provide mixed evidence on the existence and relative
magnitudes of aligned buycott and opposed boycott responses to controversial
firm social stances, and on the net revenue impacts caused by firm
stances.\footnote{Related research on socially conscious consumerism analyzes consumption
responses to other forms of firm social engagement or social impact.
For example, \citet{panagopoulos2020partisan} experimentally manipulate
consumer beliefs about socially impactful behaviors and then observe
subsequent choices between firm gift cards. They find evidence that
consumers are more likely to choose gift cards from firms that align
with personal social preferences. There is also a related literature
examining consumer responses to foreign policy, with some finding
evidence of a significant consumer response (e.g., \citealt{chavis2009consumer,fuchs2013paying,heilmann2016does,pandya2016french,fouka2022collective,chen2023history})
and others finding no such evidence (e.g., \citealt{ashenfelter2007french,davis2011business})
in different contexts. Another group of papers, such as \citet{10.1093/restud/rds012},
\citet{bartling2014markets}, \citet{barrage2020advertising}, \citet{hart2022private},
and \citet{leonelli2024consumers}, investigates consumer responses
to firm activities on less controversial social issues on which most
consumers hold similar views. Theoretical work emphasizes that socially
responsible consumption may reflect altruism, reputation, or self-image
motives \citep{benabou2006incentives}.}

We make several contributions to the existing literature on socially
conscious consumerism. First, we focus on actual consumption choices
made by consumers representing a large and representative portion
of firm revenues, rather than relying on self-reported survey responses
or other proxies that might not reflect true consumer behavior. Second,
we measure individuals' social alignment and heterogeneous consumption
responses at granular levels, thereby strengthening our identification
relative to papers analyzing data at larger temporal or spatial aggregations,
such as quarters or counties.\footnote{\citet{chen2025consumers} study how the distance between firms' and
consumers' social values affects sales, analyzing plausibly exogenous
shocks to county-level social values. They similarly find that sales
decrease with the distance between consumers' and firms' values. This
subsequent analysis complements our own research, as we use a distinct
identification strategy combined with individual-level data that allows
us to more precisely identify individuals' social values and consumption
choices, relative and absolute impacts on firm sales, and firm incentives.} Third, we systematically identify and analyze a larger number of
social stance events. This allows us to provide robust evidence of
aligned buycotts and opposed boycotts, to quantify the ensuing revenue
tradeoff between these two countervailing effects, to explain heterogeneous
impacts across different events, and to better reconcile the mixed
evidence in the existing literature. Lastly, we analyze the supply-side
implications of consumer demand, examining the incentives of profit-maximizing
firms to engage with controversial social issues.

We also build on a literature analyzing the impacts and drivers of
firms\textquoteright{} ESG (Environmental, Social, and Governance)
or CSR (Corporate Social Responsibility) behavior. This includes work
in a variety of research fields including economics (reviewed in \citealp{kitzmueller2012economic}),
marketing (e.g., \citealp{hydock2019consumer}), and management science
(e.g., \citealp{mcwilliams2001corporate}). This literature has analyzed
such firm behavior in relation to other (non-consumer) stakeholders,
including work on employees (e.g., \citealp{hedblom2019toward,burbano2021demotivating,colonnelli2023polarizing,makridis2023discrimination,colonnelli2021selfish,ferreira2025polarization,adrjan2023we}),
on financial performance (e.g., \citealp{dimson2015active,bhagwat2020corporate,bhagat2023corporate,mkrtchyan2023ceo,gangopadhyay2023strategic}),
on investors (e.g., \citealp{larcker2020s,broccardo2022exit,kahn2023divestment,bonnefon2022moral}),
and on local governments (e.g., \citealp{bertrand2020tax}). This
literature provides the strongest support for impacts on employees,
with mixed evidence of impacts on financial performance and investors.
Contemporaneous work in \citet{barari2023political} looks at the
preferences of different stakeholders as potential predictors of firms'
controversial speech,\footnote{Firms have increasingly used controversial language in their corporate
communication online \citep{cassidy2022rise} and have experienced
increased homophily in their executive teams \citep{fos2021political}.
See also \citet{colonnelli2022politics} for evidence of homophily
between owners and their employees.} finding moderate correlations between the firms' choice of language
and proxies for the preferences of potential consumers, employees,
and elected officials, without quantifying connections to firm profits.
An important hypothesis is that firms pursue social goals only to
the extent that doing so increases their profits, whereas others have
argued that firms are pushed to pursue social efforts by other stakeholders
at a cost to shareholders and profits \citep{friedman1970social,graff2005modigliani,ramaswamy2021woke}.
We contribute to this literature by analyzing firms' social stances
on particularly controversial and salient issues, providing evidence
of consumers' and employees' social preferences as drivers of firm
behavior and showing that firms' social stances have typically been
consistent with traditional profit-maximization motives.

By considering the relative importance of different stakeholders\textquoteright{}
preferences as drivers of firm behavior, our paper also contributes
to a literature on corporate governance and agency problems within
the firm. For summaries of this literature, see \citet{shleifer1997survey},
\citet{becht2003corporate}, and \citet{stein2003agency}. Our analysis
also relates to debates around stakeholder theory vs.\ shareholder
primacy as firm objective functions (e.g., \citealp{friedman1970social,hart2017companies}).
We contribute to this literature by analyzing realized firm behavior
in a social stance context in which we can precisely quantify the
revenue impacts of firm actions and in which we can measure the (potentially
competing) personal preferences of different stakeholders.

The remainder of the paper is structured as follows. Section \ref{sec:conceptual-framework}
provides our conceptual framework modeling a revenue-maximizing firm's
decision to take a social stance and highlights key empirical targets.
Section \ref{sec:data} describes the data sources we use to estimate
these empirical targets. Section \ref{sec:event-selection-size} describes
our event selection procedure and quantification of event size. In
Section \ref{sec:social-preferences-baseline-shares}, we turn to
analysis of our transaction dataset and describe our measurement of
individual consumers' social alignment. Section \ref{sec:consumption-reponses}
presents our synthetic difference-in-differences procedure for imputing
no-event counterfactual consumption and our resulting event-study
estimates of overall and disaggregated consumption responses. With
these empirical targets estimated, in Section \ref{sec:supply-side}
we return to our conceptual framework and discuss the supply-side
implications of these consumer response estimates along with other
potential drivers of firm behavior. Section \ref{sec:conclusion}
concludes.

\section{Conceptual Framework\label{sec:conceptual-framework}}

In this section, we provide a conceptual framework to illustrate the
consumer demand elasticity we wish to estimate, the tradeoffs firms
face when deciding whether and how to take social stances, and the
key parameters that determine optimal firm behavior.

\subsection{Firm Problem}

We consider a single firm choosing to either take a stance on a binary
social issue or not to take a stance, denoting its stance decision
as $s\in\{\mathbf{F}or,\ \mathbf{A}gainst,\ \mathbf{N}one\}$. There
exists a continuum of consumers partitioned into groups ($g$). The
net present value of revenues from each group may depend on the firm's
stance and adds up to total revenue at the firm: $\sum_{g\in G}y_{g}(s)=y(s)$.
Consumers are independently aware of the firm's stance with probability
$\tau$, and otherwise believe that the firm has not taken a stance.\footnote{Although consumer awareness could in principle vary across groups,
we show in Section \ref{sec:event-selection-size} that consumer awareness
is empirically similar across groups. Our conceptual framework and
subsequent empirical analysis therefore assumes that consumer awareness
of a given event does not vary across groups.} The firm seeks to maximize its revenue, $y(s)$.\footnote{Revenue maximization may differ from profit maximization in that the
former omits the direct cost of taking a stance as well as any impact
this stance may have on the profit margin associated with a given
dollar of revenue. Based on our understanding of the context and subsequent
analyses, we believe the direct cost of these stances to be near zero
on average, and we estimate mostly statistically insignificant impacts
on non-consumer outcomes that might matter for profit margins (e.g.,
prices or employee responses).} This is equivalent to maximizing the overall revenue growth induced
by its social stance decision, which for estimation purposes can be
split into the product of three terms that summarize our empirical
targets:

\begin{equation}
\underset{s\in\{\mathbf{F},\mathbf{A},\mathbf{N}\}}{\max}\underset{\text{\ensuremath{\substack{\text{Overall}\\
 \text{Revenue Growth} 
}
} }}{\underbrace{\frac{y(s)-y(\mathbf{N})}{y(\mathbf{N})}}}=\sum_{g}\underset{\text{Baseline Share}}{\underbrace{\frac{y_{g}(\mathbf{N})}{y(\mathbf{N})}}}\times\underset{\text{Awareness}}{\underbrace{\tau}}\times\underset{\substack{\text{Consumption Responsiveness}\\
\text{ (Conditional on Awareness)}
}
}{\underbrace{\frac{[y_{g}(s)-y_{g}(\mathbf{N})]\tau^{-1}}{y_{g}(\mathbf{N})}}}\label{eq:firm_problem_exact}
\end{equation}
The overall revenue growth induced by its stance is a weighted average
of group-specific responses, with weights given by the share of consumption
dollars a firm would receive from a given group if it did not take
a stance (which we refer to as baseline shares). The induced consumption
growth of a given group can be split into the product of two terms:
the share of consumers aware of a firm's stance, and the group's consumption
response conditional on awareness. This split is useful when comparing
responses to social stance events with varying levels of consumer
awareness, as induced consumption growth scales linearly with the
share of consumers aware of the firm's stance.\footnote{To see this linearity in $\tau$, define $\tilde{y}_{g}(s)$ as the
consumption by group $g$ at the firm that would occur if all group
members were aware of the firm's stance, thus $y_{g}(s)=\tau\times\tilde{y}_{g}(s)+(1-\tau)\times y_{g}(\mathbf{N})$.
Then $[y_{g}(s)-y_{g}(\mathbf{N})]/y_{g}(\mathbf{N})=\tau\times[\tilde{y}_{g}(s)-y_{g}(\mathbf{N})]/y_{g}(\mathbf{N})$.} We can therefore think of $\tau$ as a measure of treatment intensity
or event size that varies across events.

In this stylized model, firms face a potential tradeoff in catering
to the preferences of different groups when taking controversial social
stances. For example, suppose that there are two consumer groups denoted
by their social views on this issue (i.e., $G=\{\mathbf{F}or,\ \mathbf{A}gainst\}$).
The firm will prefer taking an $\mathbf{F}$ stance to no stance if
and only if the consumption increase among the aligned ($\mathbf{F}$)
group is at least as large as the decrease among the opposed group.
The net revenue impact of an $\mathbf{F}$ stance by the firm is more
positive if aligned consumers account for a larger baseline share
and/or if aligned consumers have greater consumption responsiveness
(conditional on awareness) than opposed consumers. Consumer awareness
($\tau$) affects the magnitude of revenue impacts, but does not affect
the firm's optimal stance decision given the assumption that awareness
is constant across consumer groups. This assumption is consistent
with the empirics shown in Section \ref{sec:event-selection-size}.

\subsection{Key Empirical Targets\label{subsec:key-moments}}

Estimating the terms in Equation \ref{eq:firm_problem_exact} requires:
identifying salient social stance events ($s$); measuring consumer
awareness of each firm's stance ($\tau$); separating consumers into
different groups ($g$) with likely similar social alignment; and
assembling data that allow us to measure consumption at firms by each
group and at different times ($y_{g}(s)$). We can then reasonably
proxy for baseline shares ($y_{g}(\mathbf{N})/y(\mathbf{N})$) using
consumption shares during the year preceding the firm's social stance
event. The final term left to then be estimated in the equation above
is $y_{g}(\mathbf{N})$, the counterfactual consumption that would
have occurred if the firm had not taken its stance. We can predict
this value based on contemporaneous consumption at other firms as
well as the firm's historical seasonal patterns. We estimate each
of these targets in the subsequent sections. These parameter estimates
allow us to quantify the typical strength of consumer demand responses,
to test the optimality of existing firm social stances, and to analyze
the optimal behavior of a firm facing consumers with arbitrary baseline
shares.

\section{Data\label{sec:data}}

In this section, we summarize the data sources we use to estimate
the key empirical targets highlighted by our conceptual framework.

\subsection{Transaction Data\label{subsec:data-transactions}}

We primarily use credit- and debit-card data from a large payment
card company, which allow us to measure individuals' actual consumption
at particular firms over time. The dataset contains transactions in
the U.S.\ from 2008 through March 2023, and covers approximately
20\% of all U.S.\ consumption. The dataset is longitudinal and transactions
can be linked at the card-level. For each transaction, we observe
the date, dollar amount, and merchant (along with other information).
The transaction data are depersonalized, so name, address, and other
personal information about the cardholder is not observable, other
than what can be inferred given the card's transaction history. The
data also do not specify which goods or services were purchased from
a particular merchant, nor the prices of those items. Transaction
data, in the aggregate, may be combined with depersonalized demographic
data from consumer credit reports.\footnote{Demographic data from consumer credit reports are not available for
cards that were only active during earlier years or for debit cards.
These data instead cover only credit cards that were active in recent
years, representing $\intupct$ percent of cards in our sample.} This demographic information includes the cardholder's home census
block, gender, age, and estimated household income.\footnote{At no point do we analyze consumption responses at the level of an
individual or card, instead aggregating cards into large groups prior
to our analysis of consumption responses. Per the terms of our data
agreement, all analyses also aggregate over at least five firms, with
no one firm accounting for more than 50 percent of the total.} We use this transaction dataset to impute cardholders' likely alignment
on social issues (forming groups $g$ from our conceptual framework),
to measure a firm's baseline shares across these groups ($y_{g}(\mathbf{N})/y(\mathbf{N})$),
to predict the counterfactual consumption that would have occurred
had a firm not taken a social stance ($y_{g}(\mathbf{N})$), and to
measure actual consumption by these groups at the firm over time ($y_{g}(s)$).

\subsection{Other Complementary Data Sources\label{subsec:data-other}}

Our analysis also relies on several other complementary data sources,
which we use to identify social stance events and measure consumer
awareness of each event ($s$ and $\tau$), and to analyze related
outcomes associated with these events. 

Our primary measure of consumer awareness comes from YouGov's \href{https://business.yougov.com/product/brandindex}{BrandIndex}
dataset of contemporaneous brand perception surveys of consumers,
in which YouGov surveys a nationally representative sample of at least
5,000 people each day (from their panel of more than four million
U.S.\ respondents) about their perceptions of more than 2,000 brands
operating in the U.S. Importantly for our analysis, YouGov has been
collecting the data continuously since 2007, allowing us to analyze
changes in respondents' contemporaneous perceptions of firms during
the period surrounding their social stance event. YouGov also collects
a large number of profile variables for each respondent (including
information about their demographics, party affiliation, location,
attitudes, and behaviors), allowing us to separately analyze responses
among consumers with likely different social alignment starting in
November 2012. In addition to measuring consumer awareness of firms'
social stances, we also analyze respondents' interpretation of social
stance news and their self-reported consumption responses.\footnote{For additional information regarding BrandIndex data, including sampling
methodology and complete text for all survey questions used, see Appendix
Section \ref{subsec:other-variable-construction-detail}.}

We use data from Google Trends and ProQuest's U.S.\ Newsstream primarily
to identify salient firm social stances, and to construct alternative
proxies of events' salience to consumers. Google Trends data consist
of daily relative search frequencies for given keywords on Google,
which can be compared over time, across search terms, or across geographies.
ProQuest's U.S.\ Newsstream dataset contains full-text news articles
published by more than 350 U.S.\ print and online newspapers, and
is intended as a comprehensive collection of U.S.\ news that is available
throughout our analysis period. For each news article, we observe
the full text, the publication date and outlet, and additional metadata
including the names of firms mentioned as subjects in the article.

We also use several additional data sources to study other outcomes
and behaviors for our event-study firms. We source stock prices from
CRSP. Using data from Revelio Labs, we analyze: LinkedIn employment
histories (to construct worker inflows and outflows at our event-study
firms); job postings from LinkUp, LinkedIn, and job aggregator websites;
and Glassdoor employee reviews. We use receipt-captured information
from Numerator's omni-channel consumer panel to analyze firm-level
price indices based on online and in-store purchases.\footnote{The Numerator data's coverage begins in 2017. The availability of
the job postings from LinkedIn and aggregator websites begins in 2012
and 2016, respectively. All other datasets used cover our entire analysis
period, unless otherwise noted.} We also use Nielsen's Ad Intel data to study firm-level advertising
expenditures across a variety of media types (e.g., television, radio,
internet). We also use data from D\&B Hoovers to construct a list
of the largest U.S.\ firms by revenue, and make use of their definition
of the 0--5 closest competitors of our event-study firms. We also
source data on the social alignments of a firm's non-consumer stakeholders
from \href{https://www.opensecrets.org/}{OpenSecrets} and \citet{bonica2016avenues}.
For additional detail on these different data sources and the precise
construction of variables used in our analyses, see Appendix Section
\ref{sec:data-detail}.

\section{Event Selection and Consumer Awareness\label{sec:event-selection-size}}

In this section, we describe how we systematically identify events
in which firms took controversial and salient social stances ($s$)
and measure consumer awareness ($\tau$) of each event based on Google
Trends searches, news reports, contemporaneous surveys, and queries
to a large language model.

\subsection{Identifying Candidate Social Stance Events}

We construct a dataset of $\nsocialevents$ salient social stance
events that were associated with particular firms, had a clear event
date, and which were likely to affect consumer perceptions of a firm's
social values.\footnote{By analyzing social stance events that were salient to consumers,
our event selection may exclude firm actions that the media did not
report on and that consumers did not learn about. Because we would
predict zero consumption response when consumers do not learn of a
firm's action, this restriction to salient stances is appropriate
for analyses of individuals' consumption decisions and their implications
for firms' incentives (especially coupled with our perception of near-zero
direct costs for firms of taking these stances). Our results, however,
should ultimately be understood as studying consumer responses and
firm incentives around controversial and \textit{salient} social stances.} We restrict our analysis to events that occurred between 2011 and
2022Q1, inclusive, to align with the coverage of our transaction
data and empirical methods.\footnote{The transaction dataset we use covers 2008--2023Q1, and our empirical
method requires data three years prior to and one year after the event
date.} Examples of the social stance events we identify include a controversial
advertising campaign related to racial justice, stances on widely
debated LGBTQ rights and legislation, corporate policy regarding the
provision of contraceptives or abortion access, and salient stances
on gun control issues or voting legislation.

We combine several different methods to identify these candidate corporate
social stance events, which we overview in this section and describe
in more detail in Appendix Section \ref{subsec:event-selection-detail}. 

We first implement a procedure to identify candidate events by searching
systematically for spikes in daily Google Trends searches for a given
firm name and for the firm name and keywords indicative of social
stances, using keywords like ``\emph{transgender}'' or ``\emph{gun
control}'' and repeating this search for each of the 10,000 largest
U.S.\ firms by revenue.\footnote{Additional detail on this procedure can be found in Appendix Section
\ref{subsec:event-selection-detail}, including a complete list of
searched keywords and a description of how keywords were chosen.}

We also implement a similar approach to identify candidate events
based on news coverage in ProQuest's U.S.\ Newsstream, looking for
unusual spikes in the number of news articles that mention firm names
together with keywords indicative of social stances. We complement
our news-based approach using an existing list of firm stances from
\citet{klostermann2021effect}, which identifies events by searching
for any individual news articles that contain their own set of keywords
indicative of corporate stances.

While the vast majority of events we analyze are selected by these
keyword-based Google Trends and news methods, we complement these
methods with two additional approaches based on brand perception surveys
and queries to a large language model to ensure that we have not omitted
salient events due to our choice of keywords.\footnote{Appendix Section \ref{subsec:event-selection-detail} lists the share
of events identified by each of our event selection methods. In practice,
$\sharesocialeventsgptbiadded$\% of events in our final sample were
not identified by our keyword-based methods and were added by our
brand perception and/or large language model methods.} In the first such complementary method, we identify candidate social
stance events based on discrete shifts in favorability toward a firm
among two groups of respondents who hold likely opposite views (based
on their party affiliations) in contemporaneous brand perception surveys
from the BrandIndex dataset. We further extend our list of candidate
social stance events by querying OpenAI's GPT-4 large language model
for a list of the most salient events in which firms took stances
on controversial social issues in the U.S., considering the top 50
most salient events returned by GPT-4 as candidate social stance events.\footnote{The full text of the prompt provided to GPT-4 via ChatGPT can be found
in Appendix Section \ref{subsec:event-selection-detail}.}

Taking the union of candidate firm-dates generated by the four methods
above, we then manually filter this list by using news coverage to
confirm the existence and exact timing of a social stance event while
removing false positives. We also filter to firms that we can track
consistently in our transaction data, excluding smaller firms and
firms that are purchased by consumers only via other retailers.\footnote{Our main results are robust to excluding firms with a significant
resale component.} Using the consumer awareness event-size measure defined below, we
also exclude rare candidate firm-dates that occur within two years
of a larger event at the same firm, as well as three candidate events
that were estimated to have a weakly negative event size. We typically
choose the ultimate event date based on the earliest news coverage
of a given event. We note that events are often each selected by multiple
methods, and that our main results are robust to dropping any one
method from our event selection procedure.

We provide generic descriptions for each of the $\nsocialevents$
selected firm social stance events (we denote this set $J$) in Table
\ref{tab:event-list-All}, also providing for each event the year,
direction of alignment with our consumer clusters (as defined in Section
\ref{subsec:social-preference-predictions}), and the share of consumers
we estimate were aware of the firm's stance.\footnote{These events represent social stances taken by $\nsocialfirms$ unique
firms. Results are similar if we restrict to the first or most salient
event for each firm.}

\subsection{Quantifying Event Size Based on Awareness in Contemporaneous Consumer
Surveys\label{subsec:brandindex-results}}

Having selected a set of social stance events, we use contemporaneous
surveys from YouGov's BrandIndex dataset to measure consumer awareness
of each stance. To do so, we first define the intermediate series
$a_{jt}$ as the share of BrandIndex respondents in event-time month
$t$ who report having heard something positive and/or negative about
firm $j$ in the past two weeks.\footnote{When analyzing variation over time around firm events, we define event-time
``months'' as four-week periods relative to the firm's event. Month
$t=0$ is defined as the four-week period starting with the day of
the firm's event, with month $t=-1$ then denoting the four weeks
directly preceding the firm's event.} We then define our BrandIndex-based estimate of consumer awareness
as $\hat{\tau}_{j}:=\frac{a_{j0}-a_{j,-1}}{1-a_{j,-1}}$, i.e., the
pre- vs.\ post-month change in the\emph{ }share of respondents who
have heard good or bad news about the firm, scaled by the share of
respondents who were not already reporting having heard news about
the firm.\footnote{This scaling accounts for the fact that responses will not change
among respondents who would have already reported hearing news about
the firm in the absence of its stance. It is theoretically justified
if news about the firm other than its social stance is constant over
time and independent of news about the firm's social stance (see Appendix
Section \ref{subsec:event-size-detail} for detail). This scaling
is also closely related to the literature on persuasion rates (see
\citealp{dellavigna2010persuasion}), which similarly scales changes
by the share of respondents who could possibly be converted by a treatment.} For the $\sharetauimputed$ percent of events that are not covered
by the BrandIndex dataset, we impute consumer awareness based on changes
in news coverage and Google Trends searches for the firm.\footnote{See Appendix Section \ref{subsec:event-size-detail} for detail on
this imputation procedure. Our results are robust to excluding events
for which consumer awareness was imputed by news coverage and Google
Trends, or to quantifying awareness using only these alternative proxies
(see Figure \ref{fig:consumption-responses-awareness-proxies-level}).}

In Figure \ref{fig:social_stance_awareness_brandindex} Panel A, we
show variation in average consumer awareness over time for our social
stance firms. BrandIndex respondents report hearing good and/or bad
news about firms at fairly constant rates in the months before their
social stance. In the month of the firm's social stance, consumer
awareness increases by $\taumean$ percentage points on average. This
consumer awareness measure varies significantly across events as shown
in Figure \ref{fig:social_stance_awareness_brandindex} Panel B, with
consumer awareness around $\taupnineninerounded$ percent for the
most salient stances, whereas the 75th percentile and median values
are much smaller at $\taupsevenfive$ percent and $\taumedian$ percent,
respectively. This variation highlights the need to scale observed
consumption responses relative to consumer awareness, as discussed
in Section \ref{sec:conceptual-framework}.

In Figure \ref{fig:social_stance_awareness_brandindex_byalignment},
we report consumer awareness separately among respondents depending
on their likely alignment with the firm's stance (based on their self-reported
party affiliations). Panel A shows similar magnitude spikes in consumer
awareness among consumers who are likely aligned with the firm's stance,
opposed to the firm's stance, or who are less strongly socially aligned/opposed.
These spikes in the month of the firm's social stance are slightly
larger in magnitude on average among opposed vs.\ aligned consumers,
but this difference is not statistically significant at standard significance
levels ($\taumeanopposed$\% vs.\ $\taumeanaligned$\% awareness,
respectively, p-value=$\taumeanalignedlessopposedpval$). Similarly
plotting the mean of $a_{jt}$ itself over time by respondent alignment
in Panel B, we note that aligned and opposed respondents report hearing
good and/or bad news about the firm at nearly identical rates on average
in the month of the firm's stance ($\attentionshposmeanopposed$\%
vs.\ $\attentionshposmeanaligned$\% among aligned and opposed respondents,
respectively).\footnote{The insignificant difference in estimated consumer awareness is driven
by the fact that aligned consumers are slightly more likely to report
having heard good or bad news about the firm throughout most of the
ten months \emph{prior} to the firm's event. See Appendix Section
\ref{subsec:other-variable-construction-detail} for detail on the
construction of these series.} In the empirical analysis below, we assume that awareness does not
vary with alignment.

\section{Measurement of Consumer Social Alignment and Baseline Shares\label{sec:social-preferences-baseline-shares}}

In this section, we describe how we use transaction data to impute
cardholders' likely social alignment with firm stances, to aggregate
consumers into groups with similar imputed social alignment (groups
$g$), and to measure the baseline consumption share each of our social
stance firms receives from these different groups ($y_{g}(\mathbf{N})/y(\mathbf{N})$).
We will use these imputed social alignment groups and baseline shares
in our analysis of consumption responses in subsequent sections.

\subsection{Imputing Social Alignment and Consumer Groups\label{subsec:social-preference-predictions}}

The longitudinal nature of the transaction dataset allows us to impute
a cardholder's likelihood of alignment with the firm's social stance
based on the other transactions they make throughout the card's history,
as well as their demographics (when available). Prior work has demonstrated
how consumption histories can predict a myriad of demographic characteristics
including income, education, gender, race, and ideology \citep{bertrand2023coming}.
To impute social alignment in our context, we start with a subset
of consumers with donations to PACs, charitable organizations, and
other non-profits that clearly indicate that these donors are likely
socially aligned with or opposed to one or more of the $\nsocialevents$
social stances in our analysis. For computational purposes, we then
partition these donations into two clusters, which we arbitrarily
label ``For'' and ``Against.'' All causes that are associated
with a position direction on a given social issue are included in
the same cluster, and we group together position directions across
distinct issues when there is a higher relative co-occurrence of donations
to those associated causes than to causes associated with the opposite
position on this social issue (e.g., clustering ``pro-LGBTQ+'' donations
with ``support for stricter gun control'' donations in the ``For''
cluster). We define a donor as aligned with a stance if they donate
to a cause that is associated with a similar position to the firm's
social stance (or to other causes in the same cluster as these aligned
causes), and as opposed if they donate to a cause that is associated
with or shares a cluster with an opposing position on this issue.\footnote{We would ideally like to estimate the likelihood of a cardholder's
alignment or opposition to each firm social stance separately, but
doing so would increase the computation burden of these predictions
by a factor of $\nsocialevents$ (the number of social stance events).
We implement our clustering approach to minimize this computational
burden. While this clustering is motivated primarily by computational
constraints, this can be justified by the fact that individuals' views
are strongly and increasingly correlated across distinct social issues
\citep{fiorina2016political}. We exclude from our definition of donors
a small share ($\doubledonorpct$ percent) of would-be donors who
donate to causes in both clusters.}

We use these donors as a labeled dataset (including more than $\donorcountfloor$
million cards) on which to train a machine learning model to predict
social alignment with firm stances, defined as the probability of
likely sharing the same For/Against position. Our predictors consist
of: indicators for ever purchasing at each of the 1,000 merchants
in the data most predictive of donor alignment on their own by $\chi^{2}$
(excluding the donations directly used to tag donor social preferences,
as well as firms with social stance events and their closest competitors);\footnote{This $\chi^{2}$ statistic is highest for firms that have particularly
skewed consumption shares from donors by cluster (relative to their
consumption shares in the entire economy) and high overall transaction
counts by donors, which together make them useful for differentiating
between a large number of donors by cluster.} the demographics of inferred home counties;\footnote{We infer an individual's home county as the modal county of their
in-person transactions throughout time. County characteristics include
population distributions across age groups, race, voting, urbanicity,
and other demographics.} and other general demographics (when available from credit reports).

In our prediction exercise, we first randomly split the dataset into
a training sample (70 percent of cards) and a holdout sample (30 percent).
XGBoost \citep{chen2016xgboost}, a tree-based ensemble method, is
used to classify donors. We fit XGBoost to the full 70 percent training
sample of donors using parameters selected by cross-validation, and
make predictions for our holdout sample to evaluate the model's out-of-sample
performance.\footnote{Additional detail on our predictive social preference exercise can
be found in Appendix Section \ref{subsec:social-alignment-detail}.}

\begin{figure}[H]
\begin{centering}
\caption{Donor Social Alignment Predictions\label{fig:donor-ideology-predictions}}
\subfloat[Panel A: Social Alignment Prediction Densities, by True Alignment
Status]{\begin{centering}
\par\end{centering}
\centering{}\includegraphics[width=0.84\textwidth]{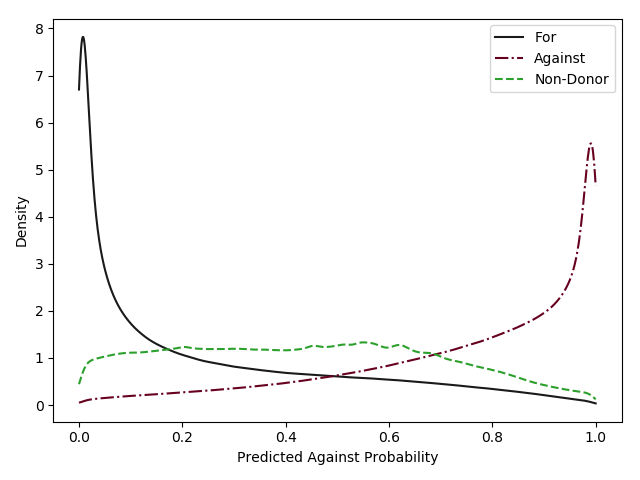}}
\par\end{centering}
\begin{centering}
\subfloat[Panel B: Median Predicted Social Alignment Decile, among All Cards
by State]{\begin{centering}
\par\end{centering}
\centering{}\vspace*{-1cm}\includegraphics[width=0.84\textwidth]{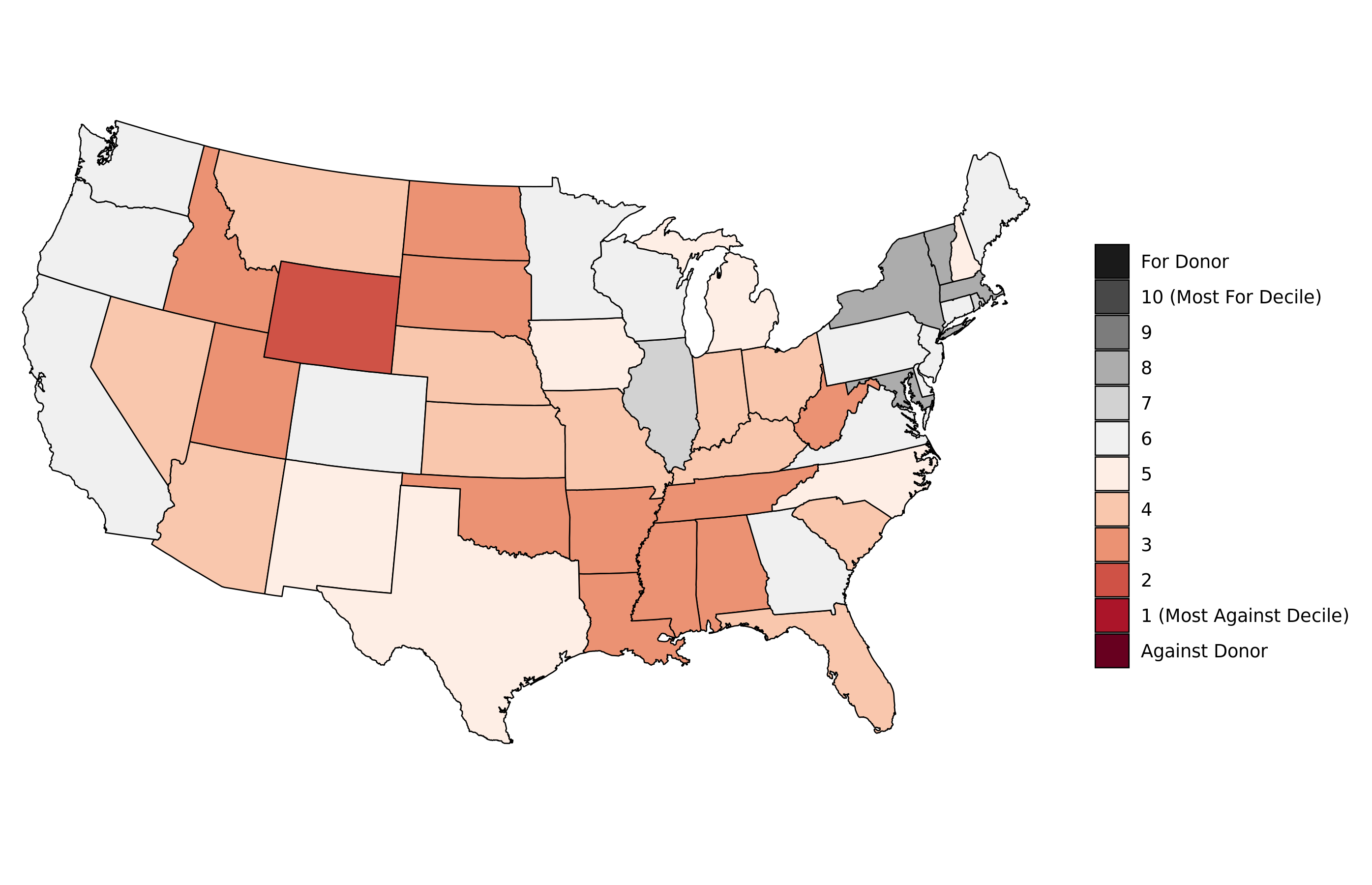}}
\par\end{centering}
{\footnotesize\vspace*{-1cm}Note: Figure summarizes social alignments
predicted by transactions and demographics. Panel A shows predicted
social alignment density distributions by sample (i.e., for both of
our holdout donor clusters, and for non-donors). Among the sample
of cards with observed donations indicative of clear social alignments,
densities are plotted separately for donors to causes in the (arbitrarily
labeled) ``For'' vs.\ ``Against'' donation clusters. Panel B
maps (for each state) the median predicted decile of alignment with
causes in the ``For'' cluster among all cards in that state, with
deciles 10 and 1 denoting non-donors most likely to be aligned with
vs.\ opposed to causes in this cluster, respectively.}{\footnotesize\par}
\end{figure}

We evaluate the performance of our predictive model among our holdout
donor sample in Figure \ref{fig:donor-ideology-predictions} Panel
A, which shows the density of predicted alignment probabilities by
true alignment status. These predicted class probabilities effectively
separate donors socially aligned with For vs.\ Against positions,
in that our predicted probabilities of being aligned with For positions
are high among consumers truly aligned with For stances (based on
their observed true donations, not used as predictors) and are low
for consumers truly aligned with Against positions on those stance
issues.\footnote{Figure \ref{fig:donor-ideology-predictions} Panel A also shows the
density of our predicted alignment probabilities among non-donors,
which are more diffuse than among our donor groups.} When making out-of-sample predictions on our holdout sample, we achieve
$\donorpredictionsbalancedaccuracyoos$ balanced accuracy, which can
be interpreted as the probability that a randomly drawn donor would
be assigned to the correct class (for which a random coin flip would
be expected to produce $0.5$). The model achieves an area under the
ROC-curve (i.e., ROC-AUC) of $\donorpredictionsrocaucoos$, which
can be interpreted as the probability that a randomly drawn consumer
truly aligned with a given class will receive a higher predicted probability
of membership for that class than would a randomly drawn donor truly
aligned with the other class (for which a random coin flip would be
expected to produce $0.5$). Our predictive accuracy is primarily
driven by the transaction indicators, rather than by home county or
credit report demographics.\footnote{To illustrate the drivers of our baseline out-of-sample performance
($\donorpredictionsbalancedaccuracyoosthreedigits$ balanced accuracy;
$\donorpredictionsrocaucoosthreedigits$ ROC-AUC), we refit our predictive
model separately for each of our three predictor subsets (transaction
indicators; county-level demographics; and demographics from credit
reports, when available). The out-of-sample balanced accuracy measures
for these subsets are $\donorpredictionsbalancedaccuracyoostransactionsthreedigits$,
$\donorpredictionsbalancedaccuracyooscountygeothreedigits$, and $\donorpredictionsbalancedaccuracyoostudemothreedigits$,
respectively. ROC-AUC values are $\donorpredictionsrocaucoostransactionsthreedigits$,
$\donorpredictionsrocaucooscountygeothreedigits$, and $\donorpredictionsrocaucoostudemothreedigits$.} We show a generic list of the most influential predictors selected
by this algorithm (as defined by the gain in predictive accuracy from
including this predictor) in Table \ref{tab:most-influential-predictors},
noting that these predictors are interpretable and intuitive. While
we cannot identify individual merchants under our data agreement,
these include media subscriptions, donations to other non-profit organizations
associated with clear social alignments, purchases at merchants with
industries and geographic distributions that are particularly correlated
with likely social alignment, and other similar transactions and demographics
that are plausibly predictive of an individual's alignment on social
issues.

Having estimated this predictive model on our training dataset of
donors, we now have a function mapping an individual's transactions
and demographics to a measure of their social alignment (likelihood
of alignment with a given For vs.\ Against stance among donors),
which we apply to the transactions and demographics of non-donors
to impute their individual social alignment. In doing so, we assume
that the relationships between an individual's social alignment and
their transactions and demographics among observed donors are similar
to the relationships among non-donors. We are more confident of this
external validity when analyzing the relative ordering of consumers'
social alignment as opposed to the levels of predicted class probabilities
due to the fact that our training data contain more donors from one
of the two clusters and due to our computationally-motivated clustering
approach (which ignores that true alignment levels likely vary across
stances and issues). As a result, we focus on the relative ordering
of consumers in terms of social alignment (rather than the predicted
likelihood of alignment as levels) in our subsequent analysis.

We partition consumers into the following 12 groups ordered by likely
social alignment: donors aligned with For causes (and likely opposed
to Against positions); card-weighted deciles among non-donors decreasingly
ordered by their likelihood of alignment with For causes; and donors
aligned with Against causes. To give an intuition for how these predicted
social alignments vary, in Figure \ref{fig:donor-ideology-predictions}
Panel B we map each state's median imputed social preference decile
across cards with imputed home counties in that state.\footnote{See Figure \ref{fig:donor-ideology-predictions-bycounty} for an analogous
map by county.}

\subsection{Measuring Baseline Consumption Shares by Group}

To quantify the relative importance of each group to a given firm,
we would ideally like to know the share of consumption (in \$) that
the firm would receive from that group (among all groups) if it did
not take a stance (i.e., $y_{g}(\mathbf{N})/y(\mathbf{N})$). While
we cannot directly observe this counterfactual quantity given that
the firms in our analysis did take stances, we can reasonably proxy
for this value as the share of consumption that the firm received
from a group in the year \emph{preceding} its stance.

\begin{figure}[H]
\begin{centering}
\caption{Group Shares of Pre-Existing Consumption at Event-Study Firms (Baseline
Shares)\label{fig:baseline_shares}}
\includegraphics[width=1\textwidth]{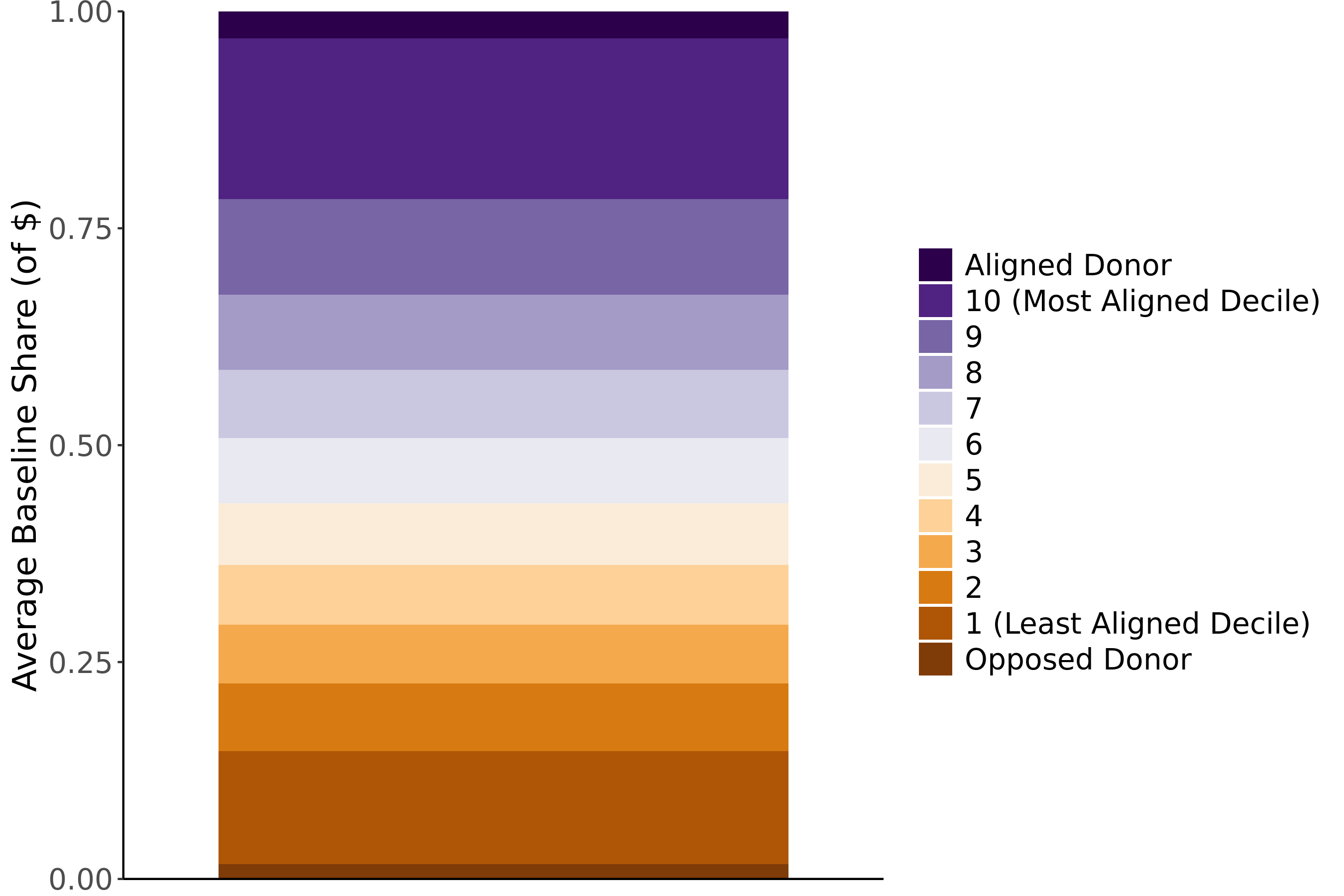}
\par\end{centering}
{\footnotesize Note: Figure shows shares of consumption (in \$) at
social stance firms by alignment group in the year preceding these
stances. Consumer groups are defined as described in Section \ref{sec:social-preferences-baseline-shares},
ordering consumers based on their predicted alignment with firm social
stances. These baseline shares are averaged across social stance events,
weighting events by consumer awareness of the firm's stance ($\tau_{j}$,
as defined in Section \ref{sec:event-selection-size}).}{\footnotesize\par}
\end{figure}

We show these estimated baseline shares in Figure \ref{fig:baseline_shares},
taking a $\tau_{j}$-weighted average of these baseline shares across
events.\footnote{We show these shares by position direction in Figure \ref{fig:baseline_shares-bydirection},
as well as each group's share of consumption (in \$) when aggregating
across all U.S.\ firms in the transaction data throughout the period
studied (2008-2023Q1).} We see that on average firms take stances that are aligned with the
social preferences of their existing customer base. Firms that take
stances received more pre-existing consumption from groups that are
likely socially aligned with the direction of their stance than from
groups that are likely opposed to the firm's stance.

\section{Consumption Responses\label{sec:consumption-reponses}}

We now use our consumer awareness measures, alignment groups, and
transaction data to estimate potentially heterogeneous consumption
responses to firms' social stances. Our empirical target throughout
this section is consumption responsiveness conditional on awareness,
i.e., $\frac{[y_{g}(s)-y_{g}(\mathbf{N})]\tau^{-1}}{y_{g}(\mathbf{N})}\approx\frac{log(y_{g}(s))-log(y_{g}(\mathbf{N}))}{\tau}$,
and we use this log approximation in our empirical estimation.\footnote{We make this log approximation to consumption growth (coming from
a first-order Taylor expansion) in our empirical estimation due to
log consumption's increased robustness to outliers and for increased
tractability, as this avoids the appearance of the unknown $y_{g}(\mathbf{N})$
in the denominator.} We can think of estimating consumption responses to the firm's stance
as an imputation problem, in that we observe the actual consumption
that occurred after the firm's stance ($log(y_{g}(s))$) but do not
directly observe the consumption that would have occurred if the firm
had not taken this stance ($log(y_{g}(\mathbf{N}))$). We first show
consumption changes by group, normalized by a simple counterfactual:
each group's consumption change at all other firms in the economy.
We then provide our preferred causal estimates of consumption responsiveness
by imputing our no-event consumption counterfactual via a synthetic
difference-in-differences design that uses contemporaneous consumption
at related firms and past historical patterns at the social stance
firm.

\subsection{Consumption Changes by Group\label{subsec:consumption-changes-bygroup}}

Before incorporating our synthetic counterfactual, we first show changes
in consumption by each group at the event-study firm in the months
surrounding the firm's social stance event, relative to changes at
all other firms. More precisely, we show $\frac{[\Delta log(y_{gjt}(s))-\Delta log(y_{g,-j,t})]}{\tau}$,
where $y_{gjt}(s)$ denotes observed consumption in dollars by group
$g$ at firm $j$ in event-time month $t$, $y_{g,-j,t}(s)$ similarly
denotes consumption for this group and month at all other firms in
the economy, $\Delta$ denotes changes from month $-1$ to month $t$,
and $\tau$ scales by our measure of consumer awareness. Taking changes
over time controls for pre-existing differences across firms and groups
(i.e., removing any group$\times$firm effects) and removing group-specific
trends in consumption at all other firms controls for group-specific
trends in consumption that affect all firms (i.e., removing any group$\times$time
effects). This simple counterfactual does not control for group time-trends
that are specific to the social stance firm or shocks to consumption
at the social stance firm itself that vary over time.

We plot these values in Figure \ref{fig:consumption-levels-bygroup},
taking a weighted-average across events.\footnote{For comparability with Figure \ref{fig:consumption-responses-bygroup},
we use the same precision weights used in that analysis. These weights
are proportional to $\tau^{2}_{j}$, but also scale inversely with
the estimated variance of our estimates of no-event counterfactual
consumption. See Section \ref{subsec:event-study-counterfactual}
for detail on the construction of and motivation for these weights.} We note that in the ten months preceding the firm's event, the normalized
consumption of relatively aligned and opposed consumer groups varies
over time but generally moves similarly. This suggests that any time-specific
shocks (such as seasonality) generally have similar effects across
these groups.

In sharp contrast, in the month of a firm's social stance event we
see large and sharply diverging changes in normalized consumption
across groups, consistent with a demand for social alignment with
firms. Donors aligned with the firm's stance see a sharp increase
in consumption of about $\coefaligneddonorlevels$ percent (per 25
percent consumer awareness) in the month of the stance, while donors
opposed to the firm's stance see a sharp decline in consumption of
about $\negcoefopposeddonorlevels$ percent in this same month. We
see similar divergent sorting among non-donors by imputed social alignment
decile; non-donors predicted to be most aligned with the firm's stance
increase their consumption by about $\coefaligneddecilelevels$ percent
while the most opposed non-donors decrease their consumption by about
$\negcoefopposeddecilelevels$ percent. All 12 groups' consumption
changes are ordered exactly as predicted by a preference for social
alignment, with smaller change magnitudes among non-donors less clearly
aligned with or opposed to the firm's social stance. These differences
attenuate in magnitude (especially for the large responses among donors)
but largely persist even a year after the firm's stance. The sharp
timing and striking divergence in these consumption changes provide
clear evidence of consumer demand for social alignment with firms. 

However, these consumption responses also reflect other shocks to
the firm (e.g., seasonality) that affect consumption, unrelated to
the firm's social stance. Indeed, we see some month-to-month changes
that similarly affect all consumers during both pre-event and post-event
months and are therefore most consistent with these confounds. We
need to account for these potential confounds in order to quantify
the causal effects of the firm's stance on consumption (for each group
and overall) and to analyze optimal behavior for revenue-maximizing
firms.

\subsection{Imputing No-Event Counterfactual Consumption\label{subsec:event-study-counterfactual}}

To control for these other shocks at the firm, we impute the counterfactual
overall consumption that would have occurred at each event-study firm
had it not taken a social stance, using a synthetic difference-in-differences
(henceforth synthetic DiD) design \citep{arkhangelsky2021synthetic}.
We train this counterfactual by predicting, for each event, the weekly
consumption series for the social stance firm ($log(y_{j\tilde{t}}$))
in the two years before the firm's event ($-104\leq\tilde{t}<0$,
where $\tilde{t}$ denotes weeks relative to the date of the firm's
social stance).\footnote{The higher frequency of week-level data (relative to monthly data
shown elsewhere) helps to inform the weights placed on different control
units in our synthetic DiD counterfactual by ensuring that this counterfactual
moves similarly to the firm week-by-week during the pre-event training
period, while still aggregating across day-of-week patterns that are
less important for our synthetic DiD to match.} We then use this model to forecast consumption at the firm in the
absence of an event.

Our synthetic DiD estimator can be expressed as follows:

\[
\underset{\omega_{0},\omega}{\min}\sum^{-1}_{\tilde{t}=-104}\Bigg[\underset{\widehat{log(y_{j\tilde{t}})}}{\underbrace{\left(\omega_{0}+\sum_{k}\omega_{k}log(y_{k\tilde{t}})\right)}}-log(y_{j\tilde{t}})\Bigg]^{2}+\zeta\left\Vert \omega\right\Vert ^{2}_{2}
\]

\[
\text{s.t. }\sum_{k}\omega_{k}=1;\;\omega_{k}\geq0
\]
In words, for a given event-firm's weekly consumption series ($log(y_{j\tilde{t}}$))
we seek to form a synthetic series ($\widehat{log(y_{j\tilde{t}})}$)
as an $\omega_{k}$-weighted average of control units ($k$) such
that this synthetic series moves in parallel with the target social
stance firm, while allowing for a fixed difference ($\omega_{0}$)
between the two.\footnote{Our intercept $\omega_{0}$ is chosen to normalize average consumption
in the pre-event month to zero.} We use the following as a superset of possible control units: consumption
at the firm in the same week of the previous year ($log(y_{j,\tilde{t}-52})$);
consumption at each of the thousands of other U.S.-based firms in
the economy for the same week ($log(y_{k\tilde{t}})\;\forall k\neq j$);
and contemporaneous total consumption across all other firms in the
same $n$-digit NAICS industry as event-firm $j$ ($log(\sum_{j'\in F_{n}(j)\setminus j}y_{j'\tilde{t}})\;\forall n\in\{0,1,\cdots,6\}$,
where $F_{n}(j)$ denotes the set of firms with the same $n$-digit
industry as our event-study firm). We exclude a firm's closest competitors\footnote{Our synthetic DiD estimates are very similar when including or excluding
the closest competitors of our event-study firms as potential control
units.} (as defined by D\&B Hoovers) and other firms with a social stance
event, both as possible control units and when defining industry-level
consumption. When making predictions after the firm's event, this
choice of possible control units ensures that we only use pre-event
data at the social stance firm and/or contemporaneous data at other
firms that did not take a social stance. As in synthetic controls,
the weights placed on these control units are constrained to be non-negative
and sum to 1, which helps to avoid regularization bias. The synthetic
DiD estimator also penalizes the sum of squared weights ($\left\Vert \omega\right\Vert ^{2}_{2}$)
in order to spread out weights across units, with the amount of this
regularization determined by the hyperparameter $\zeta$.

We tune hyperparameters governing the set of possible control units\footnote{We consider restricting to the set of possible control units based
on revenue, industry, or a first-stage Lasso regression (regressing
the social stance firm's weekly consumption series on all potential
controls). The associated hyperparameters we tune are: the L1 penalty
$\lambda$ in a first stage Lasso regression used to select predictors
by restricting to firms with non-zero Lasso coefficients; the share
of firms $\nu$ to keep when selecting the largest firms by revenue
as predictors; and the number of digits $m$ to use when restricting
to firms in the same $m$-digit industry as predictors. We include
values of $\nu$ and $m$ that allow for no filtering on these size
and industry characteristics.} and the regularization hyperparameter $\zeta$ in a data-driven way
for each firm-event by maximizing our out-of-sample forecast accuracy
in pre-event data. More specifically, we look at a series of three-year
periods that occur entirely before the firm's social-stance event.
For each three-year period and possible combination of hyperparameters,
we use the first two years as training data on which we fit a synthetic
DiD estimator, and then use this synthetic control to forecast (out-of-sample)
weekly consumption at the event-study firm. We choose hyperparameters
that minimize the mean squared error of these pre-event out-of-sample
forecasts, as in these pre-event windows we directly observe ``no
stance'' consumption (as the firm has not yet taken a stance) and
want to forecast this series as accurately as possible. We also use
this mean squared error as an estimate of the average variance of
our estimator for a given firm, which we denote $\hat{\sigma}^{2}_{j}:=\widehat{Var}[log(y_{j\tilde{t}})-\widehat{log(y_{j\tilde{t}})}]$.

Having thus selected our model's hyperparameters in a data-driven
way, we fit our synthetic DiD estimator to the two years of data preceding
a firm's social stance event. This fit determines our choice of weights
and intercept ($\omega_{k}$ and $\omega_{0}$), which determines
our synthetic control series ($\widehat{log(y_{j\tilde{t}})}$) and
estimated consumption responsiveness ($[log(y_{j\tilde{t}})-\widehat{log(y_{j\tilde{t}})}]/\tau_{j}$)
in our training period ($\tilde{t}\in[-104,-1]$) and in our out-of-sample
forecasts following the firm's event ($\tilde{t}\in[0,51]$). We aggregate
these treatment effect estimates of consumption responsiveness by
taking a precision-weighted average of event-specific estimates, with
precision weights given by $w_{j}:=\widehat{Var}([log(y_{j\tilde{t}})-\widehat{log(y_{j\tilde{t}})}]/\tau_{j})^{-1}=\frac{\tau^{2}_{j}}{\hat{\sigma}^{2}_{j}}$.
We plot our actual and predicted consumption series by event-week
in Figure \ref{fig:counterfactual-levels-weekly}.

\subsection{Estimates of Consumption Responsiveness\label{subsec:event-study-counterfactual-bygroup}}

Comparing actual scaled consumption at the firm ($log(y_{j\tilde{t}})/\tau_{j}$)
relative to our synthetic DiD control ($\widehat{log(y_{j\tilde{t}})}/\tau_{j}$)
allows us to estimate causal effects of the firm's stance on consumption
at the firm, which we plot in Figure \ref{fig:consumption-responses-bygroup}.
Panel A shows estimated overall consumption responsiveness along with
a 95\% confidence interval.\footnote{When performing statistical inference on our consumption response
estimates, it is important to account for uncertainty in our synthetic
DiD control ($\widehat{log(y_{j\tilde{t}})}/\tau_{j}$). We do so
using a wild cluster bootstrap approach that incorporates residuals
from our past forecasts on pre-event data. See Appendix Section \ref{subsec:bootstrap-detail}
for detail.} We estimate a statistically significant increase in overall consumption
of about $\coefoverall$ percent (per 25 percent consumer awareness)
on average in the month of the firm's event. This decreases on average
in the following months to values which are generally weakly positive
but not statistically significant (at the 95\% level).

To provide group-specific estimates of consumption responsiveness,
we assume that the group-specific consumption series in Figure \ref{fig:consumption-levels-bygroup}
are on parallel trends except for the firm's event (after having already
differenced out groups' consumption trends at all other firms and
initial differences in consumption levels at $t=-1$). We therefore
shift each group's series for a given firm by the same amount in a
given month such that their average (weighted by baseline shares)
aggregates up to our estimated treatment effect on overall consumption
for that social stance firm (based on the synthetic DiD estimator
above).\footnote{More formally, we shift each group-level response for a given firm-event-month
by the constant $c_{jt}=\left[log(y_{jt})-\widehat{log(y_{jt})}\right]-\sum_{g}\widehat{\frac{y_{gj}(\mathbf{N})}{y_{j}(\mathbf{N})}}\left[log(y_{gjt})-log(\tilde{y}_{gjt})\right]$,
where $\widehat{log(y_{jt})}$ is our synthetic DiD counterfactual
for overall consumption, $\widehat{\frac{y_{gj}(\mathbf{N})}{y_{j}(\mathbf{N})}}$
are our estimated baseline shares, and $\left[log(y_{gjt})-log(\tilde{y}_{gjt})\right]$
is the group-specific consumption responses shown in Figure \ref{fig:consumption-levels-bygroup}
(prior to this synthetic DiD adjustment).} We then take a ($w_{j}$) precision-weighted average across events.
This shift incorporates our synthetic DiD counterfactual to control
for seasonality and other confounding shocks unrelated to the firm's
social stance, allowing us to estimate consumption responses under
our parallel-trends assumption.\footnote{Figure \ref{fig:consumption-responses-alt-groupspecific} shows alternative
consumption responses shifted by group-specific synthetic DiD counterfactuals,
relaxing this parallel trends assumption. Each group's counterfactual
is estimated using analogous group-specific potential control units
(i.e., the group's consumption at other U.S.\ firms, at $n$-digit
NAICS industry aggregates, and at the event-firm lagged by one year).
Hyperparameters and synthetic weights are allowed to vary by group.
Results are generally similar when adjusting relative to these group-specific
counterfactuals, in that we again observe sharply diverging consumption
responses by alignment. We prefer our baseline aggregate synthetic
DiD adjustment over the group-specific version to ensure that our
group-specific effects are consistent with our overall estimates,
and because group-specific counterfactuals add significant noise to
our relative responses over longer time horizons due to the need to
compare different series that each have their own distinct forecast
errors.}

We show these shifted, group-specific estimates of consumption responsiveness
in Figure \ref{fig:consumption-responses-bygroup} Panel B. We again
observe sharply diverging consumption responses, with $\coefaligneddonormatchtotal$
percent and $-\negcoefopposeddonormatchtotal$ percent effects among
aligned and opposed donors, respectively, during the month of the
firm's social stance event. As expected given this constant shift,
we again see ordering of consumption responses according to social
value alignment. Responses are generally positive among aligned donors
and most non-donor deciles, and are mostly negative among opposed
donors and the deciles of non-donors most opposed to the firm's stance.
Consumption responses are remarkably persistent, with some gradual
decreases in magnitudes. Figure \ref{fig:consumption-responses-awareness-proxies-level}
shows that our group-specific estimates of consumption responsiveness
are robust to alternative proxies for consumer awareness.\footnote{These alternative proxies for consumer awareness include changes in
news coverage about the firm, changes in news coverage specifically
about a social stance taken by the firm (as classified by an LLM),
and changes in the firm's Google Trends search index. Figure \ref{fig:consumption-responses-awareness-proxies-log}
shows robustness to changes in logs.}

In Figure \ref{fig:consumption-responses-vsalignmeans}, we compare
each group's estimated consumption response to the mean probability
of alignment with the firm's stance among all cards in that group,
using the prediction probabilities estimated in Section \ref{subsec:social-preference-predictions}.
We refer to this mean probability as the estimated share of that group
aligned with the firm's stance ($ShareAligned_{gj}$). In Panel A
of this figure, we plot each group's estimated consumption response
in the event-month against its share aligned. In addition to the ordering
of consumption responses by alignment and the weakly positive response
point estimates among most non-donors noted previously, this figure
shows how the consumption response gradient with respect to the share
aligned changes at different points in the $ShareAligned_{gj}$ distribution.
Aligned and opposed donor responses (and to a lesser extent the responses
of the most aligned and opposed non-donor deciles) are more extreme
than we might expect were we to linearly extrapolate how consumption
changes with $ShareAligned_{gj}$ based on the non-donor deciles.
The second most extreme deciles (i.e., the 10th--20th and 80th--90th
percentiles) differ in their share aligned by about $\seconddiffalignmentshare$
and in their consumption response by about $\seconddiff$ percent.
The aligned and opposed donor groups themselves presumably differ
in their aligned share by 1 (less than doubling the difference vs.\ these
deciles) but differ in their consumption responses by $\donordiff$
percent (roughly quintupling the difference vs.\ these deciles).
This suggests that the consumption of aligned and opposed donors is
likely more responsive to social alignment than that of non-donors
even after accounting for the fact that they hold a greater share
of aligned/opposed consumers, which makes intuitive sense as these
donors have already shown a willingness to pay financially for their
social views through their donations.\footnote{The weaker relationship between consumption responses and the share
aligned among non-donors vs.\ donors could also be driven in part
by attenuation given that the share aligned is measured with noise
among non-donor groups. Given that the share aligned is calculated
fairly precisely when aggregating across the many millions of cards
within each group as we do here, attenuation is unlikely to be the
primary driver of this result.}

In Figure \ref{fig:consumption-responses-vsalignmeans} Panel B, we
estimate the gradient of consumption responses with respect to a group's
share aligned, comparing across groups within a month. We measure
these gradients as the coefficients $\beta_{t}$ in the following
regression specification:{\small
\[
\frac{log(y_{gjt})-\widehat{log(y_{gjt})}}{\tau_{j}}=\gamma_{t}+\beta_{t}ShareAligned_{gj}+\varepsilon_{gjt}
\]
}We regress our estimated consumption response for a given group $g$,
event $j$, and event-month $t$ on the alignment share of that group
(allowed to vary flexibly by month), controlling for month fixed effects
and using the same precision weights described in Section \ref{subsec:event-study-counterfactual}.
Our estimated consumption response differences out our synthetic difference-in-differences
no-stance counterfactual ($\widehat{log(y_{gjt})}$), and is scaled
by consumer awareness ($\tau_{j}$), as described in Sections \ref{subsec:event-study-counterfactual}
and \ref{sec:event-selection-size}, respectively. We initially see
a gradient of about $\coefalignedsharemonthzero$ percent in the event-month,
so that going from 0\% to 100\% alignment is associated with a $\coefalignedsharemonthzero$
percentage point increase in consumption response. This gradually
attenuates to about $\coefalignedsharemonthnine$ percent after ten
months. This gradient estimate is statistically significant at the
95\% level in each post-event month when clustering standard errors
by event.

In Figure \ref{fig:consumption-responses-differences-cis}, we also
visualize and provide 95\% confidence intervals for the difference
in consumption responses between the aligned and opposed donors (Panel
A) and between the most aligned and most opposed non-donor deciles
(Panel B). We similarly see a sharp and statistically significant
jump in this consumption response in the month of the firm's social
stance, which gradually attenuates by about half after ten months
while remaining statistically significant through this endpoint of
our analysis.

Figure \ref{fig:consumption-responses-placebo} shows analogous results
by group from a placebo exercise in which we shift actual social stance
event dates one year earlier and rerun all analysis (including synthetic
DiD training and estimation) using these one-year-earlier placebo
dates. In this placebo exercise we do not see the same patterns of
a temporary spike in consumption during the month of the event, nor
do we see sharply diverging consumption responses by alignment with
the firm's stance.

We note that our estimates rely on our parallel-trends assumption,
which we evaluate during the pre-period by reproducing our group-level
consumption responses with three years of pre-event data in Figure
\ref{fig:consumption-responses-bygroup-longpre}. Having already differenced
out group-specific trends at all other firms in the economy, we see
generally similar trends across groups in consumption at the event-study
firms throughout the three years preceding an event. In particular,
we never observe sharply diverging responses among consumers with
different social preferences. This lends support to our parallel trends
assumption, which we also relax in Figure \ref{fig:consumption-responses-alt-groupspecific}.

We similarly show consumption responses over a longer 2-year post-event
horizon in Figure \ref{fig:consumption-responses-bygroup-longpost}.\footnote{Panel A shows these responses in levels, while Panel B normalizes
the average consumption response of the middle two deciles of non-donors
to zero in each period. When analyzing this longer two-year post period,
we normalize relative to these middle deciles rather than relative
to our synthetic control because the latter relies on one-year-lagged
consumption data at the firm taking a social stance. The responses
in Panel B should thus be interpreted as relative rather than absolute
consumption responses.} Among non-donors, consumption by alignment deciles reconverges after
roughly one year, with somewhat greater persistence among the most
opposed decile until about two years after the firm's stance. In contrast,
aligned$-$opposed donor differences persist strongly even two years
later and show little sign of abating.

Figure \ref{fig:consumption-responses-by-stancetype} shows our overall
and group-specific estimates of consumption responsiveness when separately
aggregating events in which ``For'' cluster donors are likely aligned
with vs.\ opposed to the firm's stance. We observe similar divergent
responses for both cluster For-aligned and -opposed stances, with
aligned consumption increases and opposed consumption decreases in
both cases on average. On average, stances aligned with Against donors
seem to induce somewhat larger consumption responses (both positive
and negative) for a given level of consumer awareness, and these events
drive the increase in overall consumption during the event-month.

\subsection{Stance Impacts on Related Outcomes\label{sec:related-outcomes}}

We also use BrandIndex data to analyze respondents' interpretation
of social stance news, as well as their self-reported consumption
responses.\footnote{See Appendix \ref{subsec:other-variable-construction-detail} for
detail regarding the construction and analysis of these related outcomes.} Figure \ref{fig:brandindex_favorability} shows that respondents
who are socially aligned with the firm's stance (based on the respondents'
self-reported demographics) interpret this news positively and increase
their favorability toward the firm,\footnote{Figure \ref{fig:brandindex_favorability_levels} shows that prior
to the firm's stance, different respondent alignment groups had similar
favorability toward the firm. This similarity in pre-event favorability
levels is consistent with the idea that these firms were on average
viewed relatively neutrally prior to their social stance event.} while socially opposed respondents feel more negatively about this
news and about the firm following its social stance event.\footnote{We note that we used divergence in favorability toward the firm or
in interpretation of news about the firm as one of several criteria
for selecting possible events. We observe similarly diverging BrandIndex
favorability and news interpretation responses when restricting to
events chosen by our other event selection methods, alleviating the
potential concern that this result could mechanically reflect our
BrandIndex-based event selection procedure, and suggesting that this
is instead a real effect typical of controversial social stances taken
by firms.} This change in favorability translates into self-reported purchase
behavior in Figure \ref{fig:brandindex_purchasebehavior}, as socially
aligned consumers more frequently say that they would consider purchasing
from or that they intend to purchase from the firm, while fewer socially
opposed consumers do so following the firm's stance. These favorability
and self-reported consumption responses persist even a year after
the firm's stance. This finding corroborates and complements the more
incentivized responses we estimate in our analysis of transaction
data.\footnote{Relative to our analysis of transaction data, our self-reported responses
in consumer surveys have the benefit that they do not rely on imputed
social preferences or no-stance counterfactuals, as we observe less
seasonality in BrandIndex respondents' consideration or purchase intent
and we directly observe self-reported demographics/social views in
these consumer surveys. However, these self-reported BrandIndex responses
could be ``cheap talk'' in that they are not reflected in consumers'
actual purchase decisions. They also allow for only coarse measures
of respondents' social views and only capture the extensive margin,
which leaves us unable to fully quantify impacts on the firms' revenues
(which depend on both extensive and intensive margin responses). This
motivates our analysis of transaction data, through which we are able
to address each of these limitations and to more fully quantify consumer
responses and firms' revenue-maximizing behavior.}

In Figure \ref{fig:brandindex_learningchannels}, we also analyze
changes in respondents' exposure to information about the firm across
different potential learning channels. We see that per 25 percent
consumer awareness of a firm's stance, there is a sharp $\womexposureyzero$
percentage point increase on average in the share of consumers who
report having recently talked with someone about the brand (i.e.,
Word-of-Mouth exposure, Panel A) and a $\adawarenessyzero$ percentage
point increase in the share who report having recently seen advertising
for the brand (Panel B). This suggests that consumers learn about
firms' stances through multiple channels, with most of this learning
coming from channels other than direct advertising by the firm.\footnote{In Figure \ref{fig:adspend}, we analyze our event-study firms' advertising
spending in Nielsen Ad Intel data, finding that firms' total advertising
expenditures do not change significantly (either statistically or
relative to pre-period variation) in the months following their social
stance event. This rules out increased advertising spending as a likely
driver of the estimated overall consumption increases, and is consistent
with the idea that the direct cost of taking social stances is small
relative to the revenue responses we estimate. This lack of increased
advertising expenditures also suggests that the observed increase
in the share of BrandIndex respondents reporting having seen an advertisement
for the firm might partially reflect an increase in memorability or
earned (rather than paid) impressions.}

In Figure \ref{fig:priceindex}, we use Numerator's receipt-captured
data to analyze how our event-study firms change prices within the
same products at different points in time. We see some mixed evidence
weakly suggestive of a temporary increase in prices in the month of
a firm's social stance, with stronger increases for firm-branded products
than for other products sold under the firm's banner (i.e., in its
stores or on its website) without the firm's brand. This could reflect
a profit-maximizing response to increased demand and a potential increase
in profit margins per unit of revenue.

We also analyze stock price responses in Figures \ref{fig:stock-prices}
and \ref{fig:stock-prices-altrisk}, both in raw returns and as abnormal
returns in excess of various standard risk models.\footnote{We show abnormal returns defined in excess of the following benchmarks:
CRSP's value-weighted market return; CAPM; a Fama-French 3-Factor
Model; and a Fama-French 3-Factor Model with Momentum.} We do not find clear evidence on average of immediate impacts from
the firms' social stances on their stock prices among publicly-owned
firms in the month of the firm's stance or in subsequent months. Any
such impacts are difficult to separate from typical trends and variation
in the firms' stock prices, and we cannot rule out magnitudes consistent
with our estimated temporary consumption increase.\footnote{In our context, stock price analyses face limitations in that they
are unavailable for privately-owned firms, are sensitive to the exact
event-start date and horizon, and do not reveal the composition of
a firm's consumer base.}

\section{Supply-Side Implications and Predictors of Firm Behavior\label{sec:supply-side}}

Having now estimated group-specific consumption responsiveness, we
return to our stylized model and discuss when our estimates imply
that taking social stances maximizes revenue as a function of a firm's
baseline shares across consumer groups. We then analyze the extent
to which the preferences of firms' different stakeholder groups and
their ownership structure predict the direction of a firm's stance.

\subsection{Net Revenue Impact by Counterfactual Baseline Shares}

We analyze how the revenue impacts of the stances firms took would
differ if they counterfactually faced different distributions of consumers.
As described in Sections \ref{sec:conceptual-framework} and \ref{sec:consumption-reponses},
the net revenue impact of a firm's stance depends on the direction
of the firm's stance and on its baseline shares across consumer alignment
groups, with more positive impacts when firm stances are better aligned
with the firm's consumer base. Figure \ref{fig:cumulativeimpacts-by-stancetype}
estimates the cumulative revenue impact implied by our average stance
effects, weighting these same responses by two sets of baseline shares:
the actual $\tau_{j}$-weighted averages among firms that took social
stances, and the reversed baseline shares they would have faced had
they taken the opposite For/Against stance. Because firms' social
stances are better aligned with their actual existing consumers than
are these opposite stance counterfactuals, we see that estimated cumulative
net revenue impacts would be lower if firms faced consumer groups
with these reversed baseline shares. 

Figure \ref{fig:baseline-share-examples} shows that baseline shares
can vary even more dramatically in different contexts, such as across
states or across industries. In Figure \ref{fig:impacts-bystate},
we map states by the estimated cumulative net sales impact after five
months induced by taking a stance, using our average estimates of
consumption effects (as shown in Figure \ref{fig:consumption-responses-bygroup}
Panel B) and assuming that firms face the baseline shares of overall
consumer expenditures within that state. Taking a stance aligned with
the Against cluster is estimated to have positive revenue impacts
in the South, Midwest, and Southwest, and negative impacts in urban
and coastal states, while the reverse is generally true if an average
stance were instead taken in the opposite For direction. We thus estimate
that a firm could, on average, benefit from or be hurt by taking either
a For or Against stance on social issues given plausible distributions
of consumer baseline shares, and that net revenue impacts will likely
depend on the firm's geographic distribution of consumers and on its
industry.

\subsection{Predictors of Firm Behavior}

We next analyze the extent to which the alignments of our social stance
firms' different stakeholders predict stance directions (For vs.\ Against).
We control for a firm's ownership structure as an indicator for whether
the firm is publicly (rather than privately) owned. We measure consumer
social alignments as the baseline-share-weighted average of $100\times ShareAligned_{gj}$
(for alignment with For cluster positions) across groups, i.e., the
percentage of pre-existing consumption at the firm that we estimate
comes from For-aligned vs.\ Against-aligned consumers. We measure
the preferences of employees, CEOs, and boards of directors as the
percentage of donations going to recipients aligned with the For direction,
sourcing this data for employees from \href{https://www.opensecrets.org/}{OpenSecrets}
and for the CEO and board members from \citet{bonica2016avenues}.
We weight events in these regressions by $\tau_{j}$.

Table \ref{tab:reg_is_foraligned} shows results from regressing an indicator
for firms having taken a stance in this direction (rather than in
the opposite direction) on these firm characteristics among our set
of firm social stances. Stances taken by publicly-owned firms are
more likely to be in the For direction than stances taken by privately
held firms. Firms take stances that are aligned with their employees,
consumers, and to some extent their CEO, but the social preferences
expressed in board donations are not correlated with the direction
of firms' social stances. Because firms take stances that strongly
align with their employees' social preferences, the positive revenue
impacts of these stances may underestimate their full profit benefits,
as the latter would include any cost reductions that result from the
labor supply or productivity responses of well-aligned employees.\footnote{Motivated by the correlation between firm stances and employees' social
preferences, we also use data from Revelio Labs to analyze firm behavior
toward and potential impacts on employees around the time of these
stances. We find little evidence of clear changes. Figure \ref{fig:jobpostings}
shows that the number of new job postings and the average salary of
these postings for our event-study firms do not differ significantly
from U.S.\ trends in the months surrounding firms' stances. Figure
\ref{fig:workerflows} analyzes worker flows to and from the firm
(based on LinkedIn employment histories), which also do not change
significantly. Figure \ref{fig:glassdoorreviews} shows no clear impacts
on average Glassdoor employee reviews of the firm overall, on a ``Culture
and Values'' dimension, or when splitting reviews by county-level
alignment terciles. Taken together, these results provide little evidence
of employee-side impacts of these firm social stances.} 

\section{Conclusion\label{sec:conclusion}}

Our results demonstrate that social alignment between firms and consumers
is an economically meaningful dimension of demand in our setting,
not merely a stated preference. When firms take controversial social
stances, consumers aligned with the stance increase their spending
significantly, while opposed consumers reduce theirs. These diverging
responses persist throughout the distribution of consumers' social
views and well beyond the initial event.

These findings speak to ongoing debates over corporate social engagement
and the firm's objective function. In our sample, firm stances can
harm revenues when poorly aligned with consumers' social views, but
on average cause positive (but small) overall revenue impacts. This
occurs both because aligned buycotts are larger than opposed boycotts
on average, and because firms tend to take stances aligned with the
social preferences of their existing consumers (and employees). This
highlights how stakeholder rhetoric, which is frequently interpreted
as evidence of a shift away from shareholder value, can in some cases
be aligned with more traditional objectives through the market responses
of these stakeholders.

Several directions for future research follow naturally. One direction
would study the mechanisms underlying these responses, including the
relative importance of extensive- and intensive-margin adjustments,
consumer switching across firms, competitive interactions, and social-image
concerns. Future research might also analyze consumers' and other
stakeholders' preferences over firms' social values outside the context
of salient social stances, or might examine the causes of firm behavior.
A related question is whether stakeholder responses, firms\textquoteright{}
incentives to engage with social issues, and firms\textquoteright{}
realized behavior have changed over time.\footnote{Since our study period, there have been anecdotal reports of changing
consumer behavior and shifts in the frequency or direction of firms'
controversial social stances, alongside changes in the regulatory,
labor-market, and investor environments.} Additional research might analyze how firms empirically aggregate
the potentially distinct preferences of their different stakeholders,
studying whose preferences matter for determining firm behavior and
how this depends on institutions, such as the firm's governance structure
and regulatory environment.

\pagebreak{}

\leftskip=2em 
\parindent=-2em
\begin{spacing}{1.2}
\setlength{\bibsep}{2pt} 

\noindent{\small\bibliographystyle{chicago}
\bibliography{ConsumingValues}
}{\small\par}

\end{spacing}
\parindent=2em
\leftskip=0em

\pagebreak{}

\noindent{}
\begin{figure}[H]
\begin{centering}
\caption{Consumer Awareness of Events, Based on BrandIndex Responses\label{fig:social_stance_awareness_brandindex}}
\subfloat[Panel A: Unusual Awareness of News About Firm, Averaged Across Events ]{\begin{centering}
\par\end{centering}
\centering{}\includegraphics[width=0.75\textwidth]{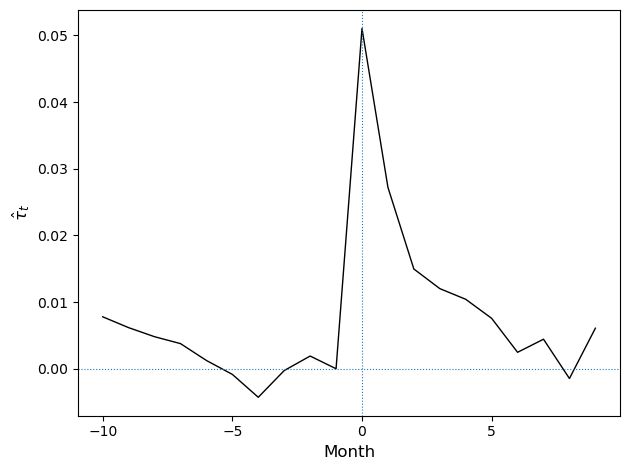}}
\par\end{centering}
\begin{centering}
\subfloat[Panel B: Consumer Awareness Distribution Across Events (Histogram)]{\begin{centering}
\par\end{centering}
\centering{}\includegraphics[width=0.75\textwidth]{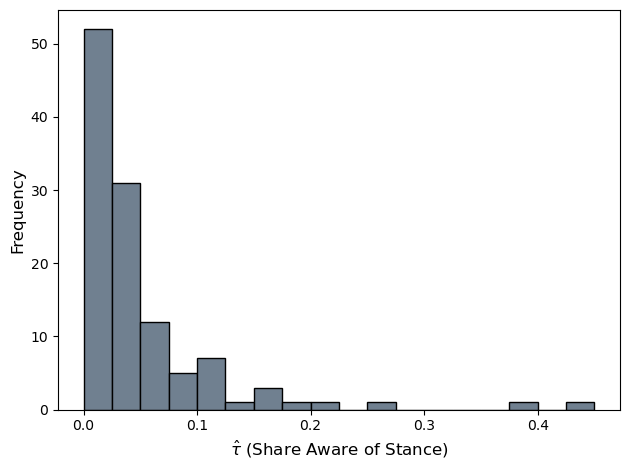}}
\par\end{centering}
{\small}{\footnotesize Note: Figure shows changes in consumer awareness
of firms, based on contemporaneous BrandIndex responses to the questions:
}{\footnotesize\emph{``Over the past two weeks, which of the following
brands have you heard something POSITIVE/NEGATIVE about (whether in
the news, through advertising, or talking to friends and family)?}}{\footnotesize ''
Defining $a_{t}$ as the share who report having heard positive and/or
negative news about the brand among respondents in month $t$, Panel
A shows $\hat{\tau}_{t}:=\frac{a_{t}-a_{-1}}{1-a_{-1}}$, averaged
by month across event-study firms. Months are defined as 4-week periods
relative to the firm's event. Panel B shows a histogram summarizing
across events our measure of consumer awareness ($\hat{\tau}:=\frac{a_{0}-a_{-1}}{1-a_{-1}}$).}{\footnotesize\par}
\end{figure}
\pagebreak{}
\pagebreak{}
\pagebreak{}
\begin{figure}[H]
\begin{centering}
\caption{Estimated Causal Effects of Social Stances on Consumption at Firm\label{fig:consumption-responses-bygroup}}
\subfloat[Panel A: On Overall Consumption]{\centering{}\includegraphics[width=0.77\textwidth]{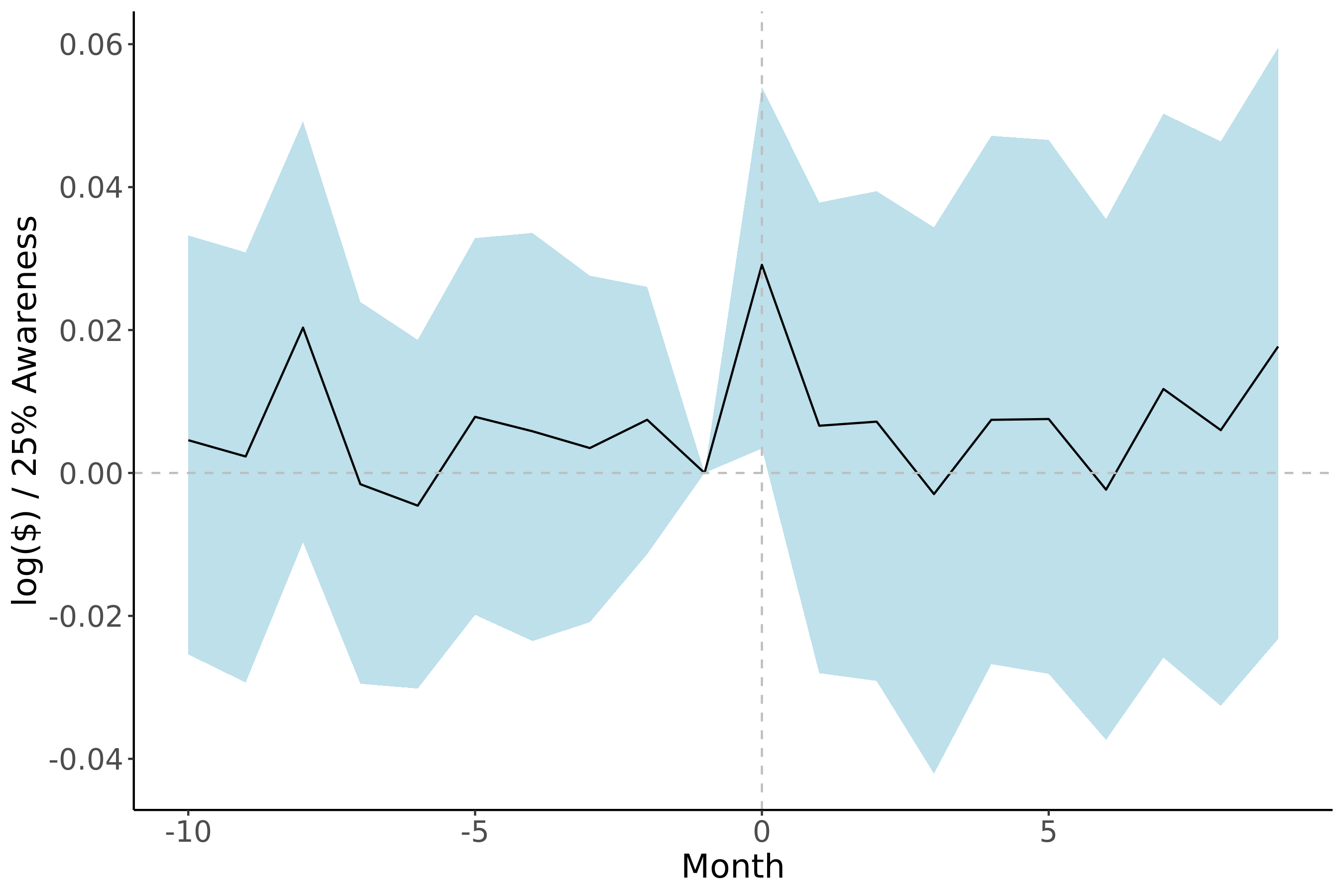}}
\par\end{centering}
\begin{centering}
\subfloat[Panel B: On Consumption by Consumer Social View Group]{\begin{centering}
\par\end{centering}
\centering{}\includegraphics[width=0.9\textwidth]{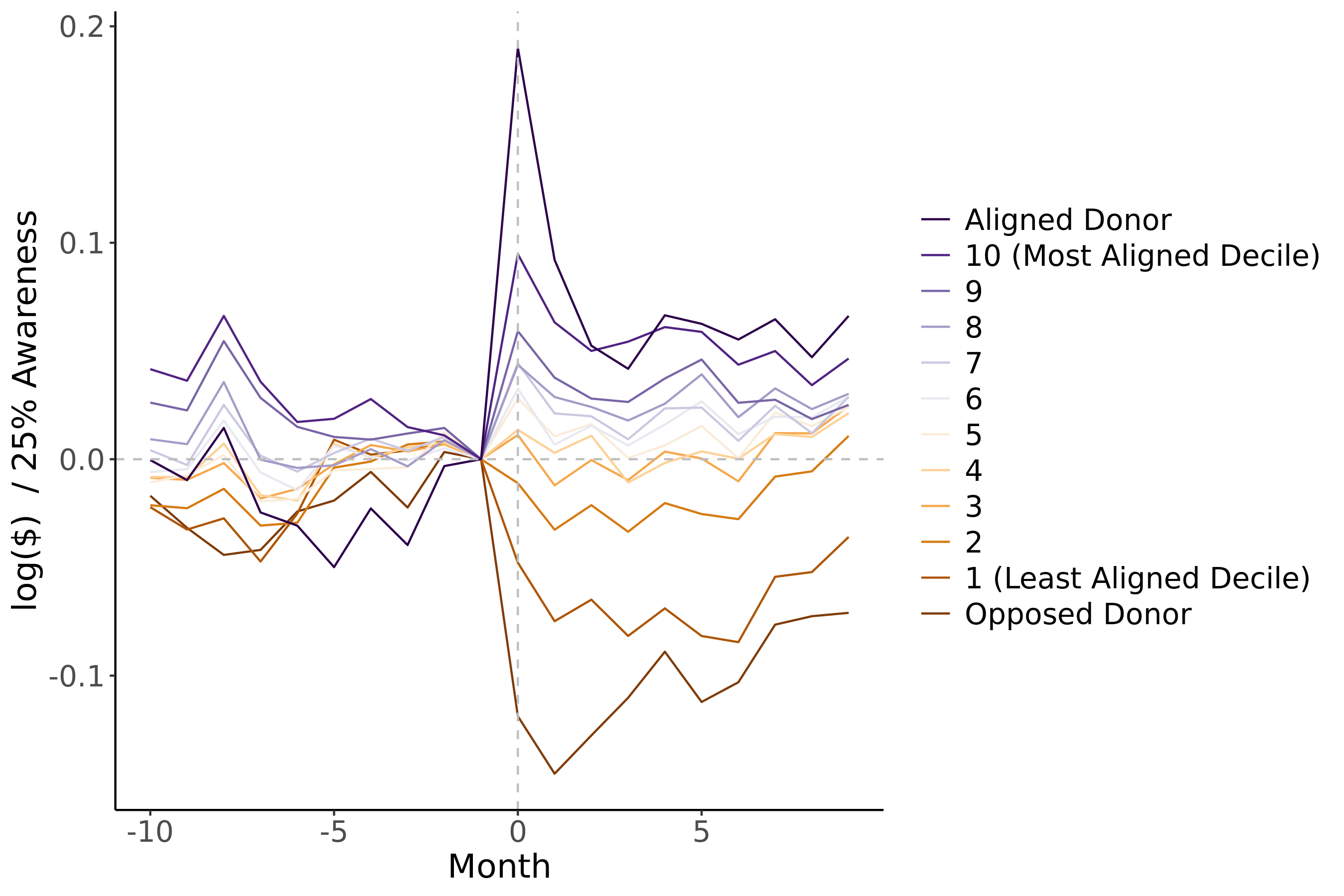}}
\par\end{centering}
{\footnotesize Note: Figure shows estimated causal effects of the firm's
stance on log consumption in the months surrounding firms' social
stances, overall and by consumer group. Panel A shows overall effects,
calculated as the difference between observed consumption and a no-stance
counterfactual predicted using a synthetic difference-in-differences
design (see Section \ref{subsec:event-study-counterfactual}). Effects
are scaled relative to consumer awareness and are averaged across
firms using a precision-weighted average. 95\% confidence intervals
are constructed using a wild cluster bootstrap approach that accounts
for uncertainty in our synthetic difference-in-differences counterfactuals.
Panel B similarly provides causal estimates for the impact of the
firm's social stance on consumption for each group (see Section \ref{sec:consumption-reponses}).
}{\footnotesize\par}
\end{figure}
\pagebreak{}
\begin{figure}[H]
\begin{centering}
\caption{Consumption Responses by Group, Using Alternative Awareness Proxies\label{fig:consumption-responses-awareness-proxies-level}}
\includegraphics[width=0.95\textwidth]{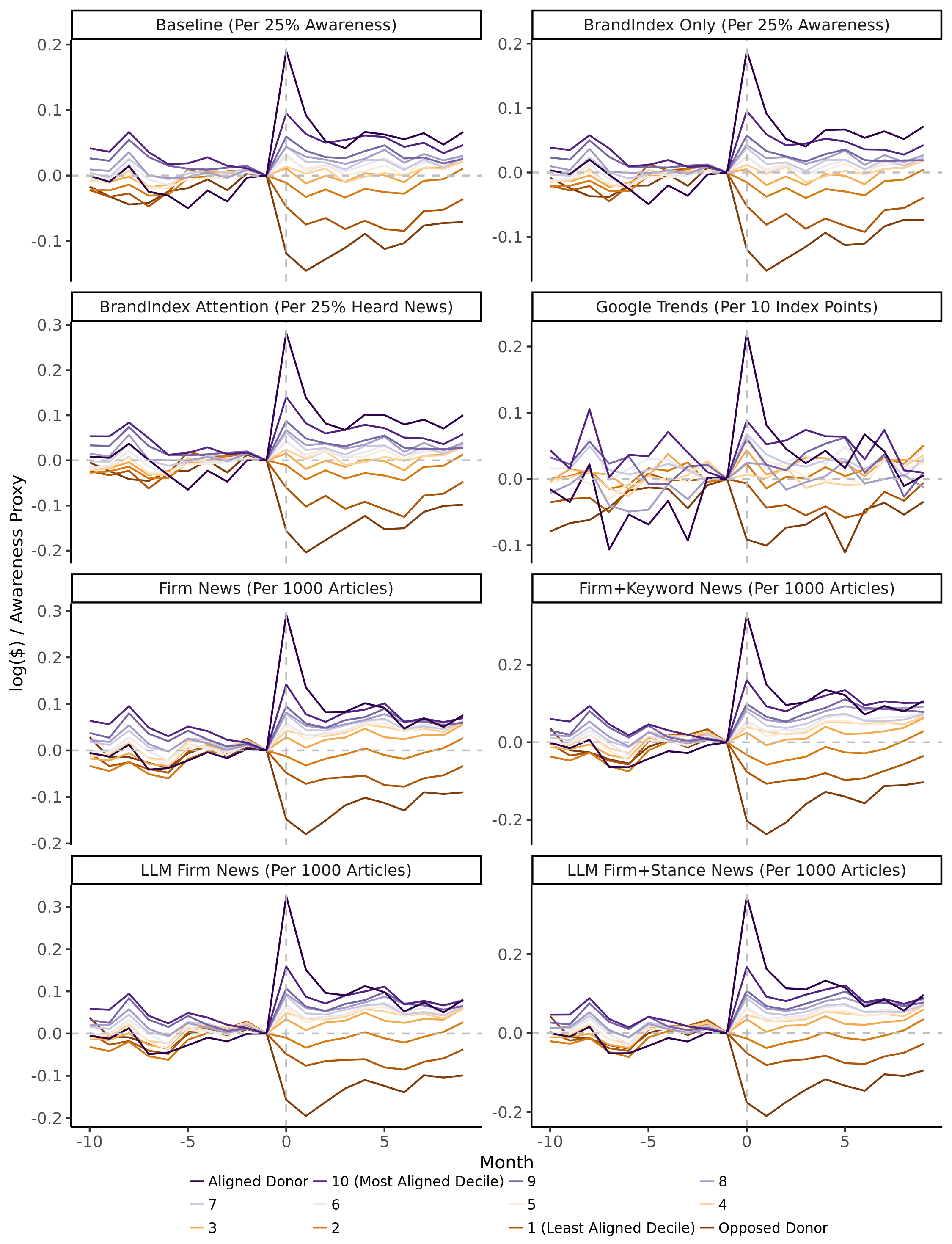}
\par\end{centering}
{\footnotesize Note: Figure analyzes group-specific consumption responses
across awareness proxies. The first subpanel repeats Figure \ref{fig:consumption-responses-bygroup}
Panel B, while other subpanels follow the same specification except
that they use the event-month change in the subpanel title as their
proxy for consumer awareness. This affects both the consumption response
scaling and the precision weights used to aggregate across events,
as described in Sections \ref{sec:event-selection-size} and \ref{subsec:event-study-counterfactual}.}{\footnotesize\par}
\end{figure}
\pagebreak{}
\begin{figure}[H]
\begin{centering}
\caption{Consumption Responses vs. Share Aligned, by Group \label{fig:consumption-responses-vsalignmeans}}
\subfloat[Panel A: Event-Month Consumption Response vs.\ Share Aligned, by
Group]{\begin{centering}
\par\end{centering}
\centering{}\includegraphics[width=0.8\textwidth]{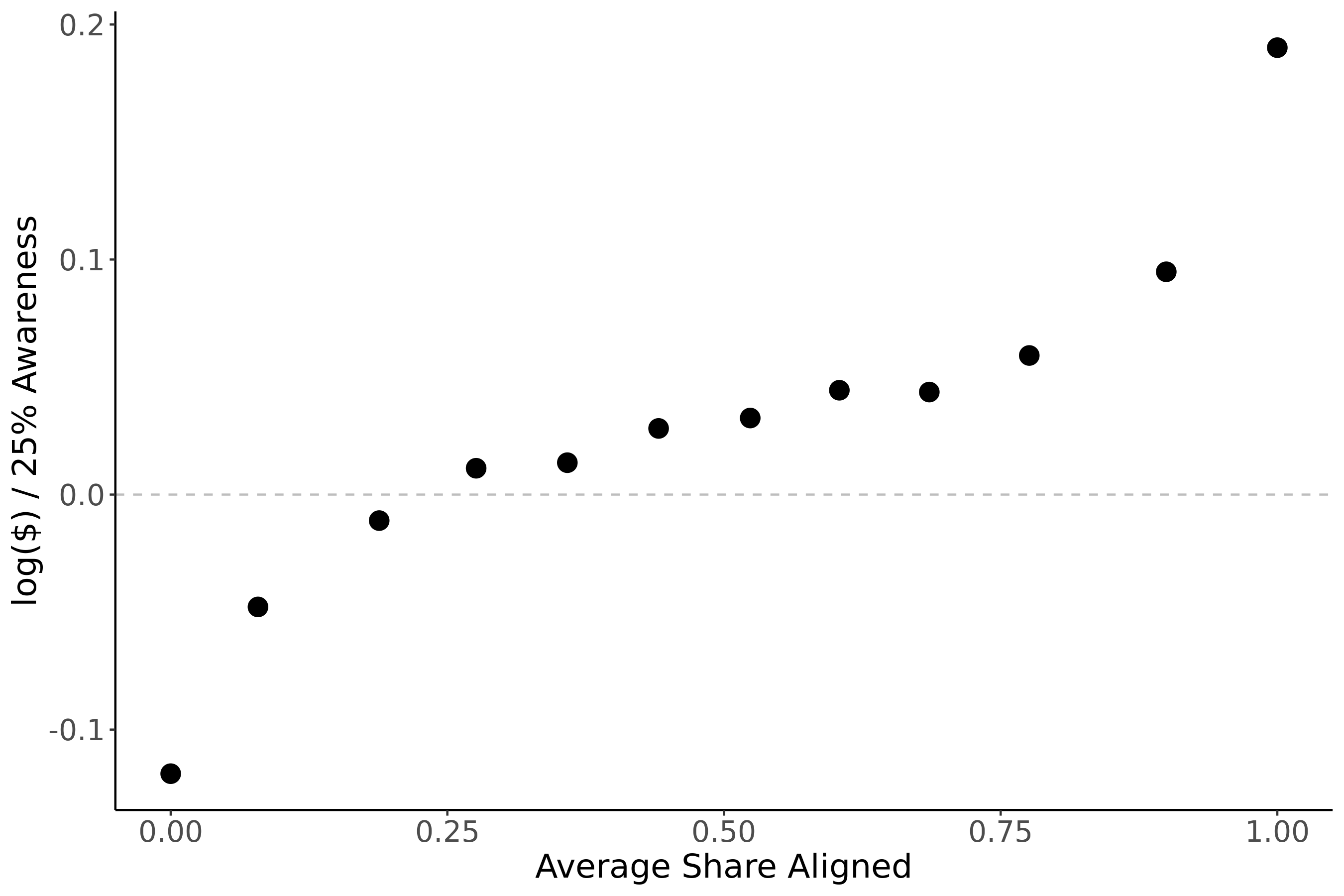}}
\par\end{centering}
\begin{centering}
\subfloat[Panel B: Consumption Response Gradient vs.\ Share Aligned, Across
Groups]{\begin{centering}
\par\end{centering}
\centering{}\includegraphics[width=0.8\textwidth]{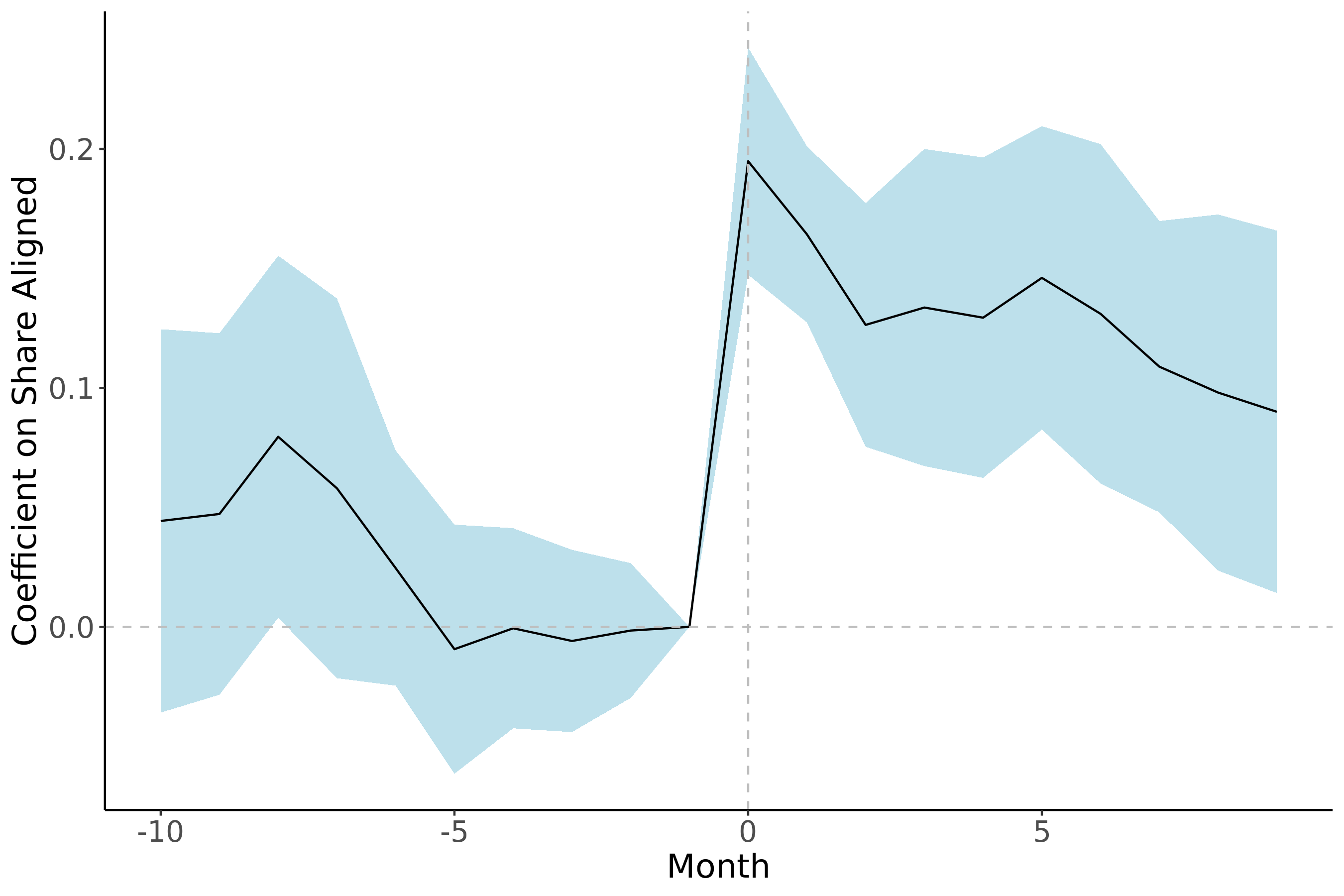}}
\par\end{centering}
{\footnotesize Note: Figure compares consumption responses (as shown
in Figure \ref{fig:consumption-responses-bygroup} Panel B) by group
to the mean share of individuals in that group predicted to be aligned
with the firm's stance. Panel A shows for each alignment group the
average estimated consumption response in the month of the firm's
social stance event (y-axis, matching the $t=0$ value in Figure \ref{fig:consumption-responses-bygroup}
Panel B) vs.\ the average share of consumers predicted to be aligned
with the firm's stance (x-axis). Alignment shares are constructed
as the mean probability of alignment among all non-donors by decile
group and direction (see Section \ref{subsec:social-preference-predictions}
for detail), as 1 for the Aligned Donor group, and as 0 for the Opposed
Donor group. Averages are then taken across firm-events and use the
same precision weights described in Section \ref{subsec:event-study-counterfactual}.
Panel B plots the coefficients $\beta_{t}$ and 95\% confidence intervals
from the regression of consumption responses on share aligned, as
described in Section \ref{subsec:event-study-counterfactual-bygroup},
clustering standard errors by event.}{\footnotesize\par}
\end{figure}

\pagebreak{}
\begin{table}[H]
\caption{Most Influential Social Alignment Predictors\label{tab:most-influential-predictors}}

\begin{centering}
{\footnotesize \setstretch{0.95}\begin{tabular}{ccc}
\toprule
Direction &                      Type &                                                               Name \\
\midrule
  Against &      Demographic (County) &                                              Republican Vote Share \\
      For &               Transaction &            \emph{[National Online$+$Print Newspaper, Based in NY]} \\
  Against &               Transaction &                                      Other Political Organizations \\
  Against &               Transaction &                                    \emph{[Pro-Gun Membership Org]} \\
  Against &               Transaction &                                      Other Religious Organizations \\
      For &               Transaction &                                                  Other Restaurants \\
      For &               Transaction &                    \emph{[Non-Profit Reproductive Healthcare Org]} \\
  Against &      Demographic (County) &                                             Share Commutes by Auto \\
  Against &               Transaction &                                        Other Sporting Goods Stores \\
      For &               Transaction &                                         Other Theatrical Producers \\
      For &               Transaction &                                    Other Parking Lots,Meters,Garag \\
      For &               Transaction &                               \emph{[Non-Profit Human Rights Org]} \\
  Against &               Transaction &                            \emph{[Christian Humanitarian Aid Org]} \\
      For &               Transaction &                                                  Other Book Stores \\
      For &               Transaction &                       \emph{[National Grocery Chain, Based in CA]} \\
      For &               Transaction &                                     Other Local Commuter Transport \\
      For &               Transaction &                                      Other Motion Picture Theatres \\
      For &               Transaction &                     \emph{[Online Marketplace for Handmade Goods]} \\
      For &               Transaction &                                    Other Eating Places And Restaur \\
      For &               Transaction &                                    Other Colleges/Univ/Jc/Professi \\
  Against &               Transaction &               \emph{[Farm Supplies and Home Improvement Retailer]} \\
      For &               Transaction &                                    Other Bars/Taverns/Lounges/Disc \\
  Against &               Transaction &  \emph{[Online e-Commerce and Affiliate Marketplace, Based in ID]} \\
  Against &               Transaction &                                     \emph{[Cable$+$Media Company]} \\
  Against &               Transaction &                             \emph{[Hunting and Outdoors Retailer]} \\
  Against &               Transaction &                                      Other Automotive Parts Stores \\
  Against &               Transaction &                  \emph{[Steakhouse Restaurant Chain, Based in TX]} \\
      For &               Transaction &                         \emph{[Audio Streaming and Media Company]} \\
  Against &               Transaction &                          \emph{[Identity Theft Software Provider]} \\
  Against &               Transaction &                   \emph{[Christian Catalog and Internet Retailer]} \\
      For &               Transaction &                                    \emph{[Food Delivery Platform]} \\
  Against &               Transaction &                               \emph{[Tool and Equipment Retailer]} \\
      For &               Transaction &                    \emph{[Newspaper, Magazine, and Media Company]} \\
      For &               Transaction &                                 \emph{[National Railroad Company]} \\
      For &      Demographic (County) &                                                   Share Race Black \\
      For &               Transaction &                                        \emph{[Furniture Retailer]} \\
      For &      Demographic (County) &                                             Share Commutes by Walk \\
  Against &               Transaction &                                          Other Public Golf Courses \\
  Against &               Transaction &                                       Other Religious Goods Stores \\
  Against &               Transaction &            \emph{[National Online$+$Print Newspaper, Based in NY]} \\
      For &               Transaction &                        \emph{[Child-Focused Humanitarian Aid Org]} \\
  Against &  Demographic (Individual) &                                                       Age in Years \\
  Against &               Transaction &                                      \emph{[Veterans Service Org]} \\
      For &               Transaction &                                               Wikimedia Foundation \\
  Against &               Transaction &                                       \emph{[Pillow Manufacturer]} \\
\bottomrule
\end{tabular}
}
\par\end{centering}
\begin{singlespace}
{\footnotesize Note: Table shows a list of the 45 most influential
transaction and demographic predictors of social alignment, as defined
by the gain in predictive accuracy from including this predictor in
our XGBoost prediction model. Predictors are in decreasing order of
influence. For each predictor, we list the univariate direction it
would predict for alignment with the For vs.\ Against donation clusters,
its type, and the name of this predictor. ``Other {*}'' predictors
are groups of merchants created by the payment card company, aggregating
all merchants in a given industry that aren't consistently identifiable
elsewhere as individual merchants. We replace the names of individual
merchants with generic descriptions (in brackets), as we are unable
to identify individual merchants under our data agreement.}{\footnotesize\par}
\end{singlespace}
\end{table}

\pagebreak{}
\begin{table}[H]
\caption{Predicting Stance Direction ($\mathds{1}_{\{\text{Stance is For}\}}$),
by Stakeholder Preferences\label{tab:reg_is_foraligned}}

\begin{centering}

\begingroup
\centering
\begin{tabular}{lccccc}
   \tabularnewline \midrule \midrule
   Model:                   & (1)          & (2)         & (3)           & (4)         & (5)\\  
   \midrule
   Is Public                & 0.397$^{**}$ &             &               &             &   \\   
                            & (0.157)      &             &               &             &   \\   
   Consumer For \%          &              & 0.011$^{*}$ &               &             &   \\   
                            &              & (0.007)     &               &             &   \\   
   Employees For Donate \%  &              &             & 0.009$^{***}$ &             &   \\   
                            &              &             & (0.001)       &             &   \\   
   CEO For Donate \%        &              &             &               & 0.003$^{*}$ &   \\   
                            &              &             &               & (0.002)     &   \\   
   Board For Donate \%      &              &             &               &             & -0.001\\   
                            &              &             &               &             & (0.003)\\   
   \midrule
   Observations             & 116          & 116         & 101           & 36          & 42\\  
   R$^2$                    & 0.168        & 0.068       & 0.449         & 0.128       & 0.011\\  
   \midrule \midrule
\end{tabular}
\par\endgroup

\par\end{centering}
{\footnotesize Note: Signif. Codes: {*}{*}{*}: 0.01, {*}{*}: 0.05,
{*}: 0.1. Table shows stance-level regressions in which the stance's
direction is predicted by the preferences of different stakeholder
groups. The dependent variable is an indicator for the stance being
aligned with causes in the (arbitrarily labeled) ``For'' donation
cluster. Predictors include an indicator for being a publicly-owned
(rather than privately held) firm, the percentage of consumers estimated
to be similarly aligned with positions in the ``For'' donation cluster,
and the percentage of donations that went to similarly aligned groups
from a firm's employees, CEO, and board of directors. See Section
\ref{sec:supply-side} for detail on these measures. Events are weighted
by estimated consumer awareness ($\tau_{j}$). Heteroskedasticity-robust
standard errors in parentheses.}{\footnotesize\par}
\end{table}

\pagebreak{}

\appendix

\section{Detail on Data and Measurement\label{sec:data-detail}}

\setcounter{figure}{0}
\renewcommand\thefigure{\thesection\arabic{figure}}  
\setcounter{table}{0}
\renewcommand\thetable{\thesection\arabic{table}}

\subsection{Detail on Event Selection Procedures\label{subsec:event-selection-detail}}

\subsubsection{Selecting Events Based on Google Trends:}

We identify candidate events based on Google Trends searches for
firm names in combination with particular keywords indicative of possible
social stances. We generated a list of firm names over which to search
by pulling the 10,000 largest U.S.\ firms from D\&B Hoovers by 2021
revenues, excluding subsidiaries, public sector organizations, and
non-profits. We pulled the name and tradestyle for each such firm
from D\&B Hoovers, and standardized these names (e.g., by removing
common firm suffixes like Inc.). We then programmatically queried
Google Trends searches for each firm name and keyword using the \emph{pytrends}
Python package. If no data was returned (due to insufficient searches)
using the firm name, we repeated this pull using the company's tradestyle.
We pulled search data for each firm and keyword at the month-level
(which indexed the largest value in a month to 100). We also separately
pulled this series together with a common reference search across
firms (so that searches for different firms could be compared in levels
with a common reference index). We then created a composite index
of these series by standardizing each series based on its mean and
standard deviations over 6, 12, or 24-month backward-looking rolling
windows, and then took an unweighted average of these values across
these three different windows as well as our self-indexed vs.\ commonly-indexed
series. By averaging these self-indexed and commonly-indexed series,
we valued events with unusual increases relative to typical patterns
for a given firm+keyword combination, and which showed a significant
increase in absolute (rather than just relative) searches through
this inclusion of a common reference search. We then selected the
2,500 largest firm-months by this index as candidate months. For each
candidate month, we then pulled daily Google Trends searches for the
firm+keyword combination in a 270-day window around the firm's event
(which is the maximum window length over which Google Trends returns
daily searches). We then returned candidate event firm-dates as those
with the largest increase from a 28-day backward- to a 28-day forward-looking
moving average for this daily series.

The precise keywords used to identify potential Google Trends searches
were: \emph{gay,} \emph{transgender, immigration, political contributions,
political, voting, controversy, buycott, boycott, gun control, abortion.}
These keywords were selected as common keywords and topics present
in discussions around a set of firm social stances that we first identified
through a manual search. Our choice of keywords was also guided by
the stance topics identified in \citet{klostermann2021effect}, and
by querying OpenAI's ChatGPT for suggested keywords that could be
used to identify social stances taken by firms based on Google Trends
searches and news coverage.

\subsubsection{Selecting Events Based on News Coverage:}

We selected candidate events based on news coverage by initially pulling
all articles from ProQuest's U.S.\ Newsstream that mention at least
one of a union of social keywords anywhere in their text and which
mention a firm as a subject (as identified by the ORG metadata field).
In addition to the full list of social keywords used when analyzing
Google Trends, we also include the following additional keywords when
analyzing news articles: \emph{racial,} \emph{social issue,} \emph{lesbian,}
\emph{queer,} \emph{lgbtq,} \emph{lgbtq+,} \emph{lgbtqia,} \emph{daca,}
\emph{gun}s\emph{,} \emph{second amendment,} \emph{reproductive rights}.
The smaller set of keywords used when analyzing Google Trends was
motivated by computational constraints, as querying Google Trends
for each additional keyword carried with it a higher computational
cost than including additional keywords when filtering news articles
from ProQuest's U.S.\ Newsstream.

For each firm mentioned as a subject in an article containing these
keywords, we produced a daily time series counting the number of such
articles mentioning the firm, and selected candidate event firm-dates
as those with the largest difference between a 28-day forward- and
a 28-day backward-looking moving average for this daily series. We
also supplement this list of events based on news coverage by adding
events identified in \citet{klostermann2021effect}, which identifies
events by searching for any individual news articles that contain
their own set of keywords indicative of corporate stances.

\subsubsection{Selecting Events Based on BrandIndex Favorability Responses:}

We selected candidate events based on BrandIndex surveys by first
calculating for each firm the difference in net favorability toward
the firm (calculated as the share with a positive impression minus
the share with a negative impression) between Democratic vs.\ Republican
respondents. We similarly calculated this difference in the net share
who reported having heard good minus bad news about the firm in the
last two weeks. We then selected candidate firm-dates as those with
the largest magnitude difference between a 28-day forward- and a 28-day
backward-looking moving average for either of these daily series.

\subsubsection{Selecting Events Based on Queries to GPT-4:}

To generate a list of candidate social stance events from GPT-4, we
provided the following prompt to OpenAI's ChatGPT on April 22nd, 2023
(at which point ChatGPT was on its ``Mar 23'' version):

\emph{``List 40 of the most notable and widely covered events in
which individual companies took partisan stances on controversial
social/political issues in the U.S.}\ \emph{between 2010 and 2022,
inclusive. For each event, provide the following variables:}

\emph{- company: company name,}

\emph{- date: the start date of the company's stance event (MM/DD/YY
format)}

\emph{- ideology: the ideological direction (conservative or liberal)
of the firm's stance}

\emph{- description: a brief (2-6 word) description of the company's
stance}

\emph{Order events according to their notability (descending), and
output this list in csv format.''}\\

Querying ChatGPT for more than 40 events at a time typically exceeded
limits to its output length. As a result, we extended this list beyond
40 events through the following follow-up prompt:

\emph{``Extend this list by adding another 40 of the most notable
and widely covered events in which individual companies took partisan
stances on controversial social/political issues in the U.S.}\ \emph{between
2010 and 2022, inclusive. For each event, provide the following variables:}

\emph{- company: company name,}

\emph{- date: the start date of the company's stance event (MM/DD/YY
format)}

\emph{- ideology: the ideological direction (conservative or liberal)
of the firm's stance}

\emph{- description: a brief (2-6 word) description of the company's
stance}

\emph{Order events according to their notability (descending), and
output this list in csv format.}

\emph{Avoid duplicates by not choosing any events which have the same
company name and month as an event you've already suggested.''}\\
We then selected the first fifty suggestions from this combined list
as candidate social stance events.

\subsubsection{Combining and Filtering Candidate Events}

For each of the candidate firm-event dates selected by the automated
Google Trends, news coverage, BrandIndex, and GPT-4-based methods
described above, we then manually checked for the existence of a social
stance taken by the firm around this date. We did so by searching
the internet and querying news articles about the firm, and we filtered
out candidate events that were not associated with a social stance
taken by the firm. This included filtering out candidates for which
we were unable to identify any salient event for the firm around this
date. This also included dropping events that were falsely flagged
by our automated methods as a social stance, for example dropping
candidates that were associated with a spike in social keyword activity
because a shooting occurred at one of the firm's stores (without the
firm taking a clear social stance in response) or because an executive
used a racial slur. Having identified a clear social stance by the
firm for each remaining candidate event, we also manually categorized
the stances into topics based on discussions in the news articles
and other online materials reviewed in our search. We also assigned
a tentative start date for each social stance event. We also filtered
out stances taken by firms that were not consistently identifiable
as merchants in our transaction data, typically because they exclusively
sell their products through other merchants or because they are particularly
small firms.

After implementing this filtering procedure, we then took the union
of candidates selected by our four different automated methods. We
grouped together, as a single candidate event with multiple possible
dates, candidates that occurred within 28 days of each other for the
same firm, using fuzzy string matching to identify different names
that correspond to the same firm. We then finalized the single start
date for each event as the date on which the firm initially took its
social stance. In rare cases for which this date was not itself directly
reported, we used the earliest publication date of news articles or
other online materials that mention the firm's social stance.

Having identified a set of actual social stances taken by firms, we
then used BrandIndex responses and other data to quantify consumer
awareness of each firm's stance, as described in Appendix Section
\ref{subsec:event-size-detail}. We then dropped a small number of
events by restricting our list to events which are the largest social
stance events (in terms of consumer awareness) for that firm within
a $\pm$2 year window. We then dropped three events for which we estimated
that a non-positive share of consumers were aware of the firm's stance
(because the share of consumers who reported hearing recent good or
bad news about the firm decreased in the month following the firm's
stance).

This procedure ultimately selected $\nsocialevents$ social stance
events, which were taken by $\nsocialfirms$ unique firms. We list
each event's year, direction, estimated consumer awareness, and generic
description in Table \ref{tab:event-list-All}. Of these $\nsocialevents$
social stance events, some are selected only by our Google Trends
method ($\sharesocialeventsgtonly$\%), by our news coverage method
($\sharesocialeventsnewsonly$\%), by our BrandIndex method ($\sharesocialeventsbionly$\%),
and by our GPT-4-based method ($\sharesocialeventsgptonly$\%). The
remaining $\sharesocialeventsmultiplesources$\% of events are selected
by multiple methods. The share of events selected by multiple methods
increases to $\sharesocialeventsmultiplesourcestauweighted$\% when
weighting events by their estimated consumer awareness, as events
that were more salient to consumers are more frequently identified
by multiple methods.

\subsection{Detail on Event Size Measurement\label{subsec:event-size-detail}}

\subsubsection{Quantifying Consumer Awareness Based on BrandIndex Responses}

As described in Section \ref{sec:event-selection-size}, for most
events, we use contemporaneous brand perception surveys from BrandIndex
to quantify the share of consumers who were likely aware of the firm's
stance. We refer to this share as ``consumer awareness.'' We identify
this share based on responses to the following question asked by BrandIndex:
``\emph{Over the past two weeks, which of the following brands have
you heard something {[}positive/negative{]} about (whether in the
news, through advertising, or talking to friends and family)?}''
Defining $a_{jt}$ as the share of BrandIndex respondents in event-time
month $t$ who report having heard something positive or negative
about firm $j$ in the past two weeks, we then define our BrandIndex-based
estimate of consumer awareness as $\hat{\tau}_{j}:=\frac{a_{j0}-a_{j,-1}}{1-a_{j,-1}}$.
This share represents the pre- vs. post-event-month change in the
share of respondents who report having recently heard good or bad
news about the firm, scaled by the share of respondents who were not
already reporting having heard recent news about the firm.

This denominator scaling accounts for the fact that the numerator
will undercount awareness among respondents who hear about the firm's
stance but who would already have reported hearing other news about
the firm unrelated to its social stance. As one potential justification
for this metric, suppose that a continuum of respondents become aware
of the firm's social stance with i.i.d.\ probability $\tau_{j}$
and independently become aware of other news about the firm with i.i.d.\ probability
$\gamma_{j}$, fixed over time. We can then see that our empirical
measure provides an estimate of consumer awareness under these assumptions
as follows:
\begin{align*}
a_{j,-1} & =\gamma_{j}\\
a_{j0} & =\mathds{P}[AwareStance]+\mathds{P}[AwareOther]-\mathds{P}[AwareStance\cap AwareOther]=\tau_{j}+\gamma_{j}-\tau_{j}\gamma_{j}\\
\tau_{j} & =\frac{\tau_{j}(1-\gamma_{j})}{1-\gamma_{j}}=\frac{a_{j0}-a_{j,-1}}{1-a_{j,-1}}
\end{align*}
The extent of this undercounting will, in practice, depend on the
correlation between social stance awareness and awareness of other
news about the firm. Here we assume independence between these two
events, following calculations of persuasion rates in the literature
(e.g., \citealp{dellavigna2010persuasion}) in using this $1-a_{j,-1}$
scaling as our adjustment factor.

In Figure \ref{fig:awareness-lines}, we plot $a_{j,t}$ (Panel A)
and $\hat{\tau}_{jt}:=\frac{a_{jt}-a_{j,-1}}{1-a_{j,-1}}$ (Panel
B) by event. Panel A shows that firms vary substantially in the share
of consumers reporting good/bad news about the firm in the pre-period
(values range from near zero to shares around 0.5), highlighting the
need for this scaling adjustment in order to avoid differential undercounting
across events. Panel A also shows reasonable consistency in $a_{jt}$
within firms over pre-event months, consistent with $\gamma_{j}$
being similar over time and suggesting that $a_{j,-1}$ is likely
a reasonable proxy for the share of consumers that would hear about
news other than the firm's social stance in month $t=0$.

\subsubsection{Quantifying Event Salience Based on News Coverage and Google Trends}

As an alternative to our BrandIndex-based quantification of consumer
awareness, we also proxy for the relative salience of each candidate
event using data from ProQuest\textquoteright s U.S.\ Newsstream.
To quantify the salience of each event, we define an event's size
as the increase in the number of news articles mentioning the social
stance firm ($j$) as a subject in the month following the event relative
to the preceding month, i.e., $EventSize_{j}:=\#articles_{j0}-\#articles_{j,-1}$.

We visualize the salience of these social stances to consumers in
Figure \ref{fig:social_stance_news}. Panel A shows variation in news
coverage of social stance firms by month around their social stance
event, relative to the month preceding the firm's event and averaged
across firm-events (i.e., showing $[\sum_{j\in J}(\#articles_{jt}-\#articles_{j,-1})]/|J|$
for months $t\in[-10,9,\cdots,8,9]$). We see that on average across
events, the firm taking a social stance is covered by $\articlediffmean$
additional news articles in the month following relative to the month
preceding its social stance event. This represents an unusual $\articlediffmeanpct$
percent increase in news coverage relative to the average number of
articles covering the firm in the month preceding its stance (i.e.,
$[\sum_{j\in J}\#articles_{j,-1}]/|J|$=$\articlepremonthavg$). We
have confirmed by looking at the text of these news articles that
this increase in news coverage is primarily driven by the firm's social
stance event itself, rather than by news covering some other aspect
of the firm. News coverage is relatively constant during the months
preceding the firm's event and is somewhat elevated in subsequent
months following the firm's event month. This sharp spike in news
coverage is consistent with the occurrence of an event (the firm's
social stance) which is likely to be salient to consumers and to affect
their perceptions of the social values associated with the firm, thus
enabling our analysis of how individuals' consumption responds to
these changes in perceptions.

In Panel B of this same figure, we use a histogram to show heterogeneity
in the event-month news coverage increases across events. We see that
events vary in their induced news coverage. A handful of the largest
social stance events see more than 1,000 article increases in news
coverage in the month following the firm's event, suggesting that
consumers are most likely to be aware of these events. Many of these
events are much less salient to consumers, as the 75th percentile
and median values are $\articlediffpsevenfive$ and $\articlediffmedian$
news article increases, respectively.

In Figure \ref{fig:social_stance_news_llm}, we show an analogous
figure after using an LLM (gpt-4o-mini) to restrict to news articles
that specifically discuss a controversial social stance taken by the
event-study firm. On average, we see $\llmStanceOrgHighEventSizeMean$
additional news articles covering controversial social stances taken
by the firm in the month following its stance relative to the preceding
month, an $\llmStanceOrgHighPctDiff$ percent increase relative to
the average number of such articles ($\llmStanceOrgOrProductHighPreMean$)
in that preceding month. This suggests that most of the increased
news coverage of the firm is indeed discussing its controversial social
stance.

We also show changes in log Google Trends searches for the firm around
its social stance event in Figure \ref{fig:google-trends}, observing
a sharp spike in Google searches in the month of the firm's social
stance. Relative to our BrandIndex-based measure of awareness, we
also observe greater month-to-month variation in both firm news coverage
and Google searches in months far removed from the firm's social stance.
This suggests that these alternative measures are likely noisier proxies
for consumer awareness.

Figures \ref{fig:consumption-responses-awareness-proxies-level} and
\ref{fig:consumption-responses-awareness-proxies-log} re-estimate
our grouped consumption responses after scaling by these alternative
awareness proxies.\footnote{These figures also include a panel in which an LLM (gpt-4o-mini) identifies
and filters to articles about the firm, dropping cases in which the
firm name in the ORG metadata field was a false positive.} These alternatives generate response gradients that are similar to
our preferred baseline, with consumption rising most among groups
aligned with the firm's stance and falling most among opposed groups.

\subsubsection{Imputing Consumer Awareness for Events Not Covered by BrandIndex}

We prefer BrandIndex-based measures of consumer awareness (when available)
to measures based on news coverage and Google Trends searches, as
this preferred measure is most closely related to the empirical target
$\tau$ highlighted by our conceptual framework. It also appears more
stable over time in the absence of an event (e.g., exhibits less seasonality
and noise) than alternative measures, and avoids potential issues
when comparing the news mentions or searches of firms that vary in
their name's commonality or potential for variants. Estimates based
on Google Trends also face a concern that searches for a firm could
in part reflect purchase intent (i.e., searching for their website
in order to buy a product) and might therefore directly reflect changes
in consumer demand.

As mentioned in Section \ref{sec:event-selection-size}, $\sharetauimputed$
percent of firm-events in our sample are not covered by the BrandIndex
dataset in the months around the firm's event. This means that we
cannot quantify consumer awareness of these firm-events directly from
BrandIndex responses and must instead impute consumer awareness of
these events using other data sources. We choose our imputation method
via cross-validation. We include the following as potential predictors
in this imputation exercise: the (pre- vs. post-month) change in Google
Trends searches for the firm (without specifying additional keywords);
the change in the number of news articles mentioning the firm as a
subject (without specifying additional keywords); and the change in
the number of news articles mentioning the firm as a subject \emph{and}
that also include at least one of our news coverage social stance
keywords. We also include as predictors changes in logs for each of
the three metrics above. When any of these six variables are missing,
we replace this missing value with the average value of this predictor
across events, and include as a potential predictor an indicator for
whether the firm had a missing value for this variable.

To make these imputation predictions, we consider the following four
methods: an elastic net; random forest; XGBoost; and stepwise selection.
We select our preferred model by minimizing the five-fold cross-validated
RMSE when predicting consumer awareness among the set of firms covered
by BrandIndex data. This cross-validation procedure selects stepwise
selection with nine features as our preferred imputation method, including
the following as controls in a linear regression (in addition to a
constant term): the six changes in levels and in logs mentioned above,
as well as missingness indicators for changes in Google Trends searches,
for changes in log Google Trends searches, and for changes in the
log number of news articles mentioning the firm in conjunction with
social keywords. The cross-validated RMSE of these predictions was
$\awarenessimputationRMSE$.

\subsection{Detail on Imputing Social Preferences and Alignment\label{subsec:social-alignment-detail}}

As described in Section \ref{subsec:social-preference-predictions},
we use a machine learning model (XGBoost) to predict an individual's
social preferences and alignment based on her transactions and demographics.
To do so, we form a labeled dataset of consumers with donations to
PACs, charitable organizations, and other non-profits that clearly
indicate that these donors are likely socially aligned with or opposed
to one or more of the $\nsocialevents$ social stances. After partitioning
these donors into two arbitrarily-labeled ``For'' and ``Against''
clusters of correlated social views, we split donors into a training
sample (70 percent of cards) and a holdout sample (30 percent). We
include the following as predictors: indicators for ever purchasing
at each of the 1,000 merchants in the data most predictive of donor
alignment on their own by $\chi^{2}_{j}$ (excluding the donations
directly used to tag donor social preferences, as well as firms with
social stance events and their closest competitors);\footnote{For a given firm $j$, $\chi^{2}_{j}\propto\frac{(N_{jA}N_{\sim jF}-N_{jF}N_{\sim jA})^{2}}{(N_{jF}+N_{jA})(N_{jA}+N_{\sim jA})(N_{jF}+N_{\sim jF})(N_{\sim jA}+N_{\sim jF})}$,
where $N_{jg}$ is defined as the number of transactions by cluster
$g$ donors at firm $j$ at any point in time, and $N_{\sim jg}$
denotes the number of transactions by cluster $g$ at all firms in
the economy other than $j$ (excluding the merchants used to define
our clusters).} the demographics of inferred home counties;\footnote{We infer an individual\textquoteright s home county as the modal county
of their in-person transactions throughout time. This agrees with
their home county as listed in a credit report snapshot (when available)
$\countyagreementpct$ percent of the time, with much of the disagreement
due to changing home locations over time. We exclude from our analysis
cards with zero in-person transactions. } and other general demographics (when available from credit reports).
We fit this model via weighted maximum likelihood estimation. We observe
a smaller number of donors in the Against cluster, and so we uniformly
upweight donors from this cluster so that both clusters are given
the same total weight when evaluating the likelihood of our predictions.
We empirically tune the parameters of the XGBoost algorithm using
five-fold cross-validation on the 70 percent training sample.\footnote{We gradually tighten the grid of tuning parameters over which we search
as we increase the sample size used for training. These tuned parameters
take the following final values: \textit{learning\_rate}=0.12, \textit{gamma}=1,
\textit{max\_depth}=6, \textit{subsample}=0.8, \textit{colsample\_bytree}=0.8,
\textit{reg\_lambda}=10, \textit{reg\_alpha}=0.1. All other parameters
are kept at their defaults unless otherwise noted.} We then fit XGBoost to the full 70 percent training sample of donors
using the parameters selected by this cross-validation, and we make
predictions for our holdout sample to evaluate the model\textquoteright s
out-of-sample performance.\footnote{We exclude 10 percent of the training sample as an evaluation set
to determine when adding trees to our XGBoost ensemble no longer improves
our evaluation-set predictions (i.e., early stopping). Based on this
early stopping criterion, our final fit uses $\xgboostntrees$ trees.
All out-of-sample predictive measures refer to the 30 percent of the
data not used during training, and do not include the 7 percent of
the overall sample used as an evaluation dataset to determine early
stopping during training.} These performance metrics and the outputs of our predictions are
described in Section \ref{subsec:social-preference-predictions}.
This trained model gives us a mapping from individuals' transactions
and demographics to a measure of their social alignment (likelihood
of alignment with a given For vs. Against stance among donors). Applying
this mapping to the transactions and demographics of non-donors allows
us to impute their individual social alignment and to partition them
into social-alignment deciles based on these predictions.

\subsection{Detail on Wild Cluster Bootstrap\label{subsec:bootstrap-detail}}

When performing statistical inference on the overall consumption response
estimates (see Section \ref{subsec:event-study-counterfactual}),
it is important to account for uncertainty in our synthetic DiD control
($\widehat{log(y_{j\tilde{t}})}/\tau_{j}$). We do so using a wild
cluster bootstrap approach that incorporates residuals from our past
forecasts on pre-event data. Recall that in these past forecasts,
we look at a series of three-year periods that occur entirely before
the firm's social-stance event. For each three-year period, we use
the first two years as training data on which we fit a synthetic DiD
estimator, and then use this synthetic control to forecast (out-of-sample)
weekly consumption at the event-study firm. We treat these residuals
as forecast errors, as in these pre-event windows we directly observe
``no stance'' consumption (as the firm has not yet taken a stance)
and want to forecast this series accurately.

In each bootstrap iteration, we randomly sample firm-events with sampling
probabilities proportional to their precision weights. Once we have
drawn the time-series for a given firm-event within a bootstrap iteration,
we then uniformly sample one past forecast series from the set of
all possible past forecasts for that firm-event, and add the residuals
from the drawn past forecast series multiplied by a Rademacher weight
($\pm1$ each with 50\% probability) to the estimated overall consumption
for that firm-event. We then average across the sampled firm-event+residual
series within a bootstrap iteration to produce an estimated overall
consumption time series for that bootstrap iteration. We conduct \bootstrapiterations\ such
independent bootstrap iterations. We then construct a 95\% confidence
interval for our overall consumption response estimates in Figure
\ref{fig:consumption-responses-bygroup} Panel A by using as our bounds
the 2.5\% and 97.5\% quantiles across bootstrap iterations for each
month.

\subsection{Detail on Construction of Other Variables\label{subsec:other-variable-construction-detail}}

\subsubsection{YouGov BrandIndex Survey Details and Question Text}

To produce its BrandIndex dataset, YouGov owns and operates a syndicated
global panel of more than 17 million respondents. More than four million
of these respondents are located in the U.S., and we restrict our
analysis only to this subset for consistency with the other data sources
we analyze. Panel members sign up through a double opt-in process
through which they register to join YouGov's panel, validate their
email address, and start by sharing demographics about themselves
(e.g., age, gender, and race). Panelists are then invited to complete
brand preference surveys, in which they answer questions about multiple
brands from a single product category (e.g., ``Grocery Stores''
or ``Skin Care and Cosmetics'') on a given day.\footnote{In their U.S.\ subsample, YouGov elicits preferences regarding 2,000+
brands spread over 45+ product categories. Both numbers have varied
over time.} Panelists can only complete a particular survey on a given day if
they receive an invitation to do so from YouGov, and YouGov employs
a lock-out period following the completion of a survey to ensure that
a given respondent does not complete multiple surveys within a short
time window. Panelists are randomly assigned to product categories
using a quota system to ensure that responses for each product-category$\times$day
are in expectation nationally representative based on race, income,
gender, and region (relative to U.S.\ Census data). YouGov also uses
weights when aggregating responses to account for unexpected variation
in completion rates, thereby ensuring that responses are also nationally
representative ex-post. YouGov respondents receive points for their
survey completion, which they can exchange for rewards like Amazon
gift cards or movie tickets. YouGov collects responses from at least
5,000 U.S.\ respondents each day, collecting these data since June
3rd, 2007.

Within a given survey, respondents first select the brands that they
are aware of within the product category (from a list of up to 40
brands). They then answer the remaining questions in the survey only
for the brands of which they said they were aware. BrandIndex produces
two kinds of metrics: 2-point metrics (e.g., Yes/No responses), and
3-point metrics (e.g., Positive, Negative, or Neutral). The exact
wording of questions varies by product category to reflect the product
category name, type of good, and typical purchase frequency. Here
we provide the questions seen by YouGov BrandIndex respondents for
each of the BrandIndex-based metrics used in our analysis, using as
an example the exact question text for the ``Dining: Fast Food''
product category. For each question, we list the name given to this
metric by YouGov, and specify whether the metric is on a 2-point or
3-point response scale.
\begin{itemize}
\item ``Aided Brand Awareness'' (2-point, initial question): \emph{Which
of the following restaurant chains have you {*}ever{*} heard of? Please
select all that apply.}
\item ``Buzz'' (3-point): \emph{Over the PAST TWO WEEKS, which of the
following restaurant chains have you heard something POSITIVE about
(whether in the news, through advertising, or talking to friends and
family)? / Now which of the following have you heard something NEGATIVE
about over the PAST TWO WEEKS?}
\item ``Attention'' (2-point): {[}Yes if respondent reported Positive
and/or Negative ``Buzz''{]}
\item ``Consideration'' (2-point): \emph{When you are in the market next
to purchase food or drink, from which of the following would you consider
purchasing?}
\item ``Purchase Intent'' (2-point): \emph{From which of these would you
be most likely to purchase? }{[}Follow-up to ``Consideration''{]}
\item ``Current Customer'' (2-point): \emph{Have you purchased food or
drink from any of the following restaurant chains in the past 30 days?}
\item ``Former Customer'' (2-point): \emph{Have you ever purchased food
or drink from any of the following restaurant chains? }{[}Excludes
``Current Customers''{]}
\item ``Impression'' (3-point): \emph{Overall, of which of the following
restaurant chains do you have a POSITIVE impression? / Now which of
the following restaurant chains do you have an overall NEGATIVE impression?}
\item ``Word-of-Mouth Exposure'' (2-point): \emph{Which of the following
restaurant chains have you talked about with friends and family in
the PAST TWO WEEKS (whether in-person, online, or through social media)?}
\item ``Advertising Awareness'' (2-point): \emph{Which of the following
restaurant chains have you seen an advertisement for in the PAST TWO
WEEKS?}
\end{itemize}
We construct the variables used in our analysis from these questions
as follows. As described in Section \ref{sec:event-selection-size},
we define $a_{jt}=\frac{\sum_{i\in I_{jt}}w_{i}\mathds{1}\{\text{Reported Positive and/or Negative Buzz}\}}{\sum_{i\in I_{jt}}w_{i}}$
as the share reporting having heard something positive and/or negative
about the firm in the past two weeks, among all responses $I_{jt}$
that asked about $j$'s firm in event-month $t$.\footnote{We do \emph{not} exclude individuals who were unaware of the firm
(in the ``Aided Brand Awareness'' question) from the denominator,
although they did not answer subsequent questions about the brand.} Responses $i$ are weighted by the survey weights $w_{i}$ provided
by YouGov to make responses nationally-representative for that product-category
and day. Constructing $\hat{\tau}_{jt}=\frac{a_{jt}-a_{j,-1}}{1-a_{j,-1}}$,
Figure \ref{fig:social_stance_awareness_brandindex} Panel A then
plots the average of $\hat{\tau}_{jt}$ across all event-firms in
a given event-month, among events covered by BrandIndex. Constructing
$\hat{\tau}_{j}$ as equal to $\hat{\tau}_{j0}$ for events covered
by BrandIndex and imputing this value from Google Trends and news
coverage when not covered by BrandIndex (as described in Appendix
Section \ref{subsec:event-size-detail}), Figure \ref{fig:social_stance_awareness_brandindex}
Panel B then plots a histogram of this consumer awareness measure
across all $\nsocialevents$ events.

Figure \ref{fig:social_stance_awareness_brandindex_byalignment} shows
averages of $a_{\tilde{g}jt}$ and $\hat{\tau}_{\tilde{g}jt}$ across
firms, adding an alignment group dimension $\tilde{g}$ by calculating
these metrics for each firm-event separately among respondents ($I_{\tilde{g}jt}$)
split by their social alignment.\footnote{We define social alignment among BrandIndex survey respondents based
on their self-reported party affiliation, oriented relative to our
donation clusters and stances based on related donations in those
clusters. This demographic question was answered previously and asks
``\emph{Generally speaking, do you think of yourself as a...? {[}Democrat,
Republican, Independent, Other, Not Sure{]}}.'' We drop the $\otheraffiliationpct$
percent of respondents answering ``Other'' or ``Not Sure'' to
this question from all analyses of splits by alignment group in the
BrandIndex data.} YouGov's 2-point metrics (with the exception of ``Attention'')
and the splits by party affiliation are also only available starting
in 11/13/2012. When producing any given time-series figure, we balance
our panel by dropping events that aren't covered throughout the period
for the metric shown in that figure (e.g., a hypothetical event with
date 1/1/2013 would be dropped from any figure showing splits by party
affiliation ten months prior).

When producing Figure \ref{fig:brandindex_favorability} Panel A,
we first calculate for each group, firm-event, and month $buzz_{\tilde{g}jt}=\frac{\sum_{I_{\tilde{g}jt}}w_{i}(\mathds{1}\{\text{Reported Positive Buzz}\}-\mathds{1}\{\text{Reported Negative Buzz}\})}{\sum_{I_{\tilde{g}jt}}w_{i}}$,
i.e., the share of respondents in a given alignment group reporting
positive news about the firm minus the share reporting negative news.
We then define our outcome metric for each group, firm-event, and
month as $\frac{buzz_{\tilde{g}jt}-buzz_{\tilde{g}j,-1}}{\hat{\tau}_{j}}$
and calculate a precision-weighted average of this series across firm-events
for a given month and group (i.e., weighting firm-events by $\hat{\tau}^{2}_{j}$).\footnote{In all plots showing how responses scale with awareness, we actually
divide by $4\times\hat{\tau}_{j}$, so that responses are scaled relative
to 25 percent awareness rather than 100 percent awareness. We do so
because the latter is not within the range of observed awareness for
actual social stance events. For expositional simplicity, we sometimes
abuse notation by also referring to this denominator as $\hat{\tau}_{j}$.} In Panel B, we similarly calculate favorability toward the firm using
responses to ``Impression'' rather than ``Buzz.'' We plot $\hat{\tau}_{j}$-weighted
averages of the levels $buzz_{\tilde{g}jt}$ and $impression_{\tilde{g}jt}$
in Figure \ref{fig:brandindex_favorability_levels}.

To produce Figure \ref{fig:brandindex_purchasebehavior} Panel A,
we first calculate for each group, firm-event, and month $considers_{\tilde{g}jt}=\frac{\sum_{I_{\tilde{g}jt}}w_{i}(\mathds{1}\{\text{Included firm among answers to Consideration}\})}{\sum_{I_{\tilde{g}jt}}w_{i}}$,
i.e., the share of respondents who report that they would consider
purchasing from that firm when next in the market for its product
category. We then similarly define our outcome metric as $\frac{considers_{\tilde{g}jt}-considers_{\tilde{g}j,-1}}{\hat{\tau}_{j}}$
and calculate a precision-weighted average of this series across firm-events
for a given month and group (i.e., weighting firm-events by $\hat{\tau}^{2}_{j}$).
We similarly produce Panel B by calculating purchase intent (or more
precisely, that the respondent reports being most likely to purchase
from the firm) using responses to ``Purchase Intent'' rather than
``Consideration''. We similarly produce Panels A and B of Figure
\ref{fig:brandindex_learningchannels} using responses to ``Word-of-Mouth
Exposure'' and ``Ad Awareness,'' respectively.

\subsubsection{Nielsen Ad Intel Data on Advertising Expenditures}

We source data on advertising expenditures from the Nielsen Ad Intel
dataset at the Kilts Center. We linked our event-study firms to Nielsen
identifiers based on a combination of fuzzy name matching and manual
linking. We then use the Occurrence data files to identify advertisements
in which each event-study firm is listed as an associated brand, and
sum the dollars spent across all such advertisements within a firm-month
to get the firm's total spending. We include all media types available
through Nielsen's Ad Intel dataset, except for Spot TV ads, which
we exclude throughout our analysis due to inconsistencies in the availability
of spending over time. Measuring this quantity ($spend$) in thousands
of dollars, we then analyze $log(spend+1)$ as our outcome variable
in Figure \ref{fig:adspend}.

\subsubsection{Numerator Receipt-Captured Transaction Data}

We source item-level data on quantities sold and their prices from
Numerator's omni-channel consumer panel (accessed via the Kilts Center).
We link our event-study firms to Numerator data using fuzzy-string
matching and manual linking for each of two criteria: (1) identifying
items in which the firm is listed as the brand, parent-brand, or in
the item description; and (2) identifying items in which the firm
is listed under the banner field, indicating that the product was
sold in the firm's store or on its website, but isn't necessarily
branded as the firm's product. We use Numerator's \textit{ITEM\_ID}
to identify when two items being sold represent the same product,
dropping items with \textit{ITEM\_ID} values that are zero or missing.
We also drop purchases in which an item's unit price is listed as
zero, balance our panel of items, and filter to event-firms with at
least five distinct items.

In order to construct price indices for each firm, we then define
the following variables. Let $q_{uit}$ denote the quantity of item
$i$ purchased by user $u$ in period $t$, let $p_{uit}$ denote
the unit price of this transaction, and let $w_{ut}$ denote the user
weights provided by Numerator to make their panel representative of
the U.S.\ population. We then define the weighted quantity $Q_{it}=\underset{u}{\sum}w_{ut}q_{uit}$
and the weighted average unit price as $\bar{p}_{it}=\frac{\sum_{u}w_{ut}p_{uit}q_{uit}}{Q_{it}}$.
For firm $j$ with items $I_{j}$, we then define our Laspeyres and
Paasche price indices, respectively, as:

\[
P^{(L)}_{jt}=\frac{\sum_{i\in I_{j}}\bar{p}_{it}Q_{i,-1}}{\sum_{i\in I_{j}}\bar{p}_{i,-1}Q_{i,-1}};\quad P^{(P)}_{jt}=\frac{\sum_{i\in I_{j}}\bar{p}_{it}Q_{it}}{\sum_{i\in I_{j}}\bar{p}_{i,-1}Q_{it}}
\]

\subsubsection{CRSP Stock Returns}

We source daily data on stock prices from CRSP, accessed via WRDS.
We manually construct matches of our event-study firms to PERMCO and
PERMNO identifiers in CRSP. For firms with multiple PERMNOs, we retain
the identifier whose security start and end date cover the event date
and analysis window. If multiple PERMNOs satisfy this criterion, we
retain the most liquid share class. We source from CRSP the daily
total return for each of these securities, CRSP's value-weighted market
return, and daily Fama-French factors plus momentum. For one event-study
firm listed outside the U.S., we use the firm's comparable daily returns
from Compustat--Capital IQ (again accessed via WRDS), the corresponding
continental market return, and (from Kenneth French's online data
library) the corresponding continental Fama-French plus momentum factors.

We follow standard practice when constructing our market-adjusted
\citep{sharpe1964capital}, Fama-French 3-Factor \citep{fama1993common},
and Fama-French 3-Factor plus Momentum \citep{carhart1997persistence}
risk models. When constructing these risk model benchmarks, we follow
the default parameters in WRDS U.S.\ daily event-study tool in selecting:
an estimation window of 100 trading days to estimate the expected
return and residual return variance; a required minimum of 70 non-missing
return observations within this estimation window; and a gap of 50
trading days between the end of the estimation window and the beginning
of the event window (with our event window matching our plotting window).

\subsubsection{Revelio Labs Data (Job Postings, Worker Flows, Employee Reviews)}

We use data purchased directly from Revelio Labs to analyze firms'
behavior toward and impacts on employees. The data contain a firm
reference file, job postings gathered from LinkUp, LinkedIn, and job
aggregator websites, worker employment histories gathered from LinkedIn,
and employee reviews gathered from Glassdoor. We first link our firms
to identifiers in the Revelio Labs data (common across these datasets)
based on fuzzy name matching, manual linking, and searches for known
employees, for company/LinkedIn URLs, and for reviews pertaining to
the target firm. We also link to identifiers listed as subsidiaries
of our target firm. We then use this unified set of firm identifiers
throughout our analysis of Revelio Labs data.

When analyzing job postings, we count the number of job postings that
have an originally posted date in our target month. To avoid dropping
zeros when taking logs, we add one to all firm-month job posting totals
before taking logs. We also analyze the average salary (in thousands
of dollars) listed among these new job postings. In Figure \ref{fig:jobpostings}
Panels A and C, we then analyze the log of this value as our dependent
variable. In Panels B and D, we construct analogous series among all
U.S.\ job postings other than those by our target firm, and subtract
this U.S.\ aggregate from our firm-specific series in order to detrend
(with any fixed difference in levels handled by the fact that we normalize
relative to month $t=-1$).

When analyzing worker flows in Figure \ref{fig:workerflows}, we count
as inflows the number of users who are listed as working for our target
firm in month $t$ but not in $t-1$. We count as outflows the number
of users listed as working for the firm in month $t-1$ but not in
month $t$. This figure analyzes $log(x+1)$ of each of these values.

In Figure \ref{fig:glassdoorreviews}, we analyze anonymous Glassdoor
reviews of firms by their employees. We do so by taking the average
rating (initially on a 1-5 star scale) across reviews of the firm
posted in a given month. We compute metrics for two rating dimensions:
the ``Overall'' rating and the ``Culture and Values'' rating.
We plot these averages in Panels A and B. To construct Panels C and
D, we first geocode each review's raw location using the Google Maps
API to get the review's associated county. We then split counties
into social value terciles by their 2016 presidential vote shares,\footnote{Reviews from New York City are associated with the aggregate share
of its five constituent counties. Reviews from Alaska are dropped
due to data availability.} such that each tercile contains roughly the same number of U.S.\ reviews.
We then average ratings for our event-study firms by firm, month,
and alignment tercile (based on the direction of the firm's stance),
shown as the dependent variables in Panels C and D.

\pagebreak{}

\section{Appendix Exhibits}
\noindent\begin{flushleft}
\vspace*{-0.8cm}\begin{center}
\begin{figure}[H]
\begin{centering}
\caption{Consumer Awareness of Firm Social Stances, by Alignment\label{fig:social_stance_awareness_brandindex_byalignment}}
\vspace*{-0.2cm}\subfloat[Panel A: Unusual Awareness of News About Firm ($\hat{\tau}_{t}:=\frac{a_{t}-a_{-1}}{1-a_{-1}}$)]{\begin{centering}
\par\end{centering}
\centering{}\includegraphics[width=0.75\textwidth]{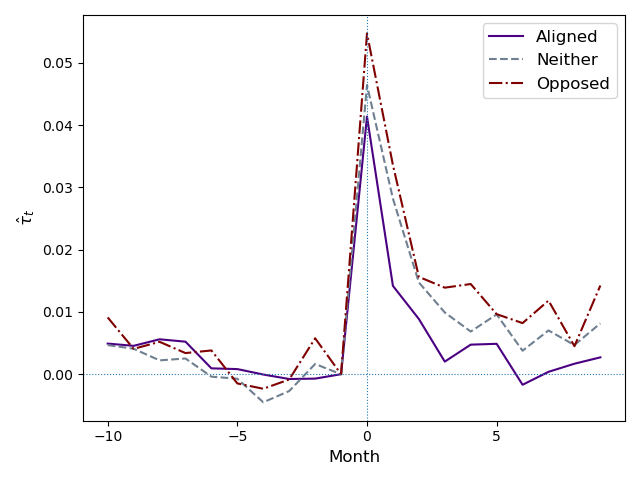}}
\par\end{centering}
\begin{centering}
\vspace*{-0.3cm}\subfloat[Panel B: Share Reporting Recent Good or Bad News ($a_{t}$)]{\begin{centering}
\par\end{centering}
\centering{}\includegraphics[width=0.75\textwidth]{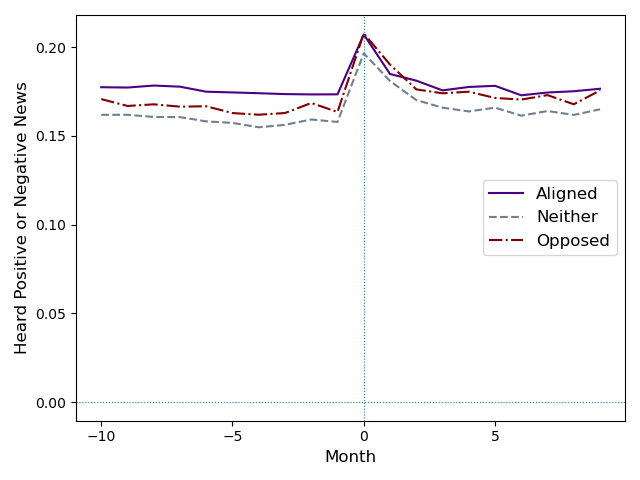}}
\par\end{centering}
{\footnotesize Note: Figure shows consumer awareness of firms' social
stance events by social alignment, based on BrandIndex survey responses.
Define $a_{t}$ as the share who report having heard positive or negative
news about the brand in the last two weeks among all respondents in
month $t$. Panel A shows our consumer awareness measure $\hat{\tau}_{t}:=\frac{a_{t}-a_{-1}}{1-a_{-1}}$,
averaged by month across event-study firms separately for respondents
by alignment. Panel B similarly shows averages of $a_{t}$.}{\footnotesize\par}
\end{figure}
\pagebreak{}
\begin{figure}[H]
\begin{centering}
\caption{Median Predicted Social Alignment Decile, among All Cards by County
\label{fig:donor-ideology-predictions-bycounty}}
\par\end{centering}
\begin{centering}
\includegraphics[width=1\textwidth]{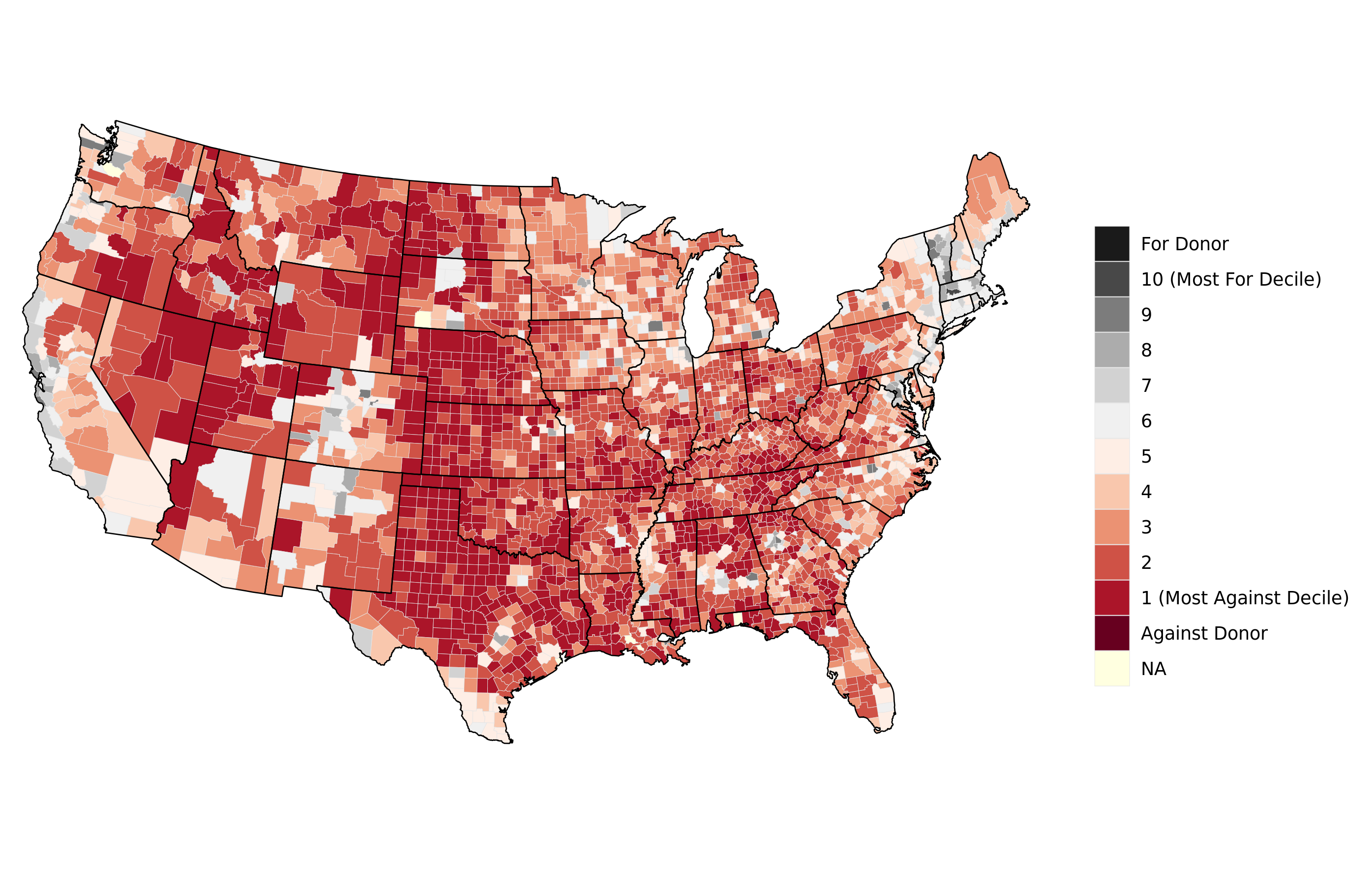}
\par\end{centering}
{\footnotesize Note: Figure maps (for each county) the median predicted
decile of alignment with causes in the (arbitrarily labeled) ``For''
donation cluster among all cards in that county, with deciles 10 and
1 denoting non-donors most likely to be aligned with vs.\ opposed
to causes in this cluster, respectively.}{\footnotesize\par}
\end{figure}
\pagebreak{}
\begin{figure}[H]
\begin{centering}
\caption{Group Shares of Pre-Existing Consumption at Event-Study Firms, by
Position\label{fig:baseline_shares-bydirection}}
\includegraphics[width=1\linewidth]{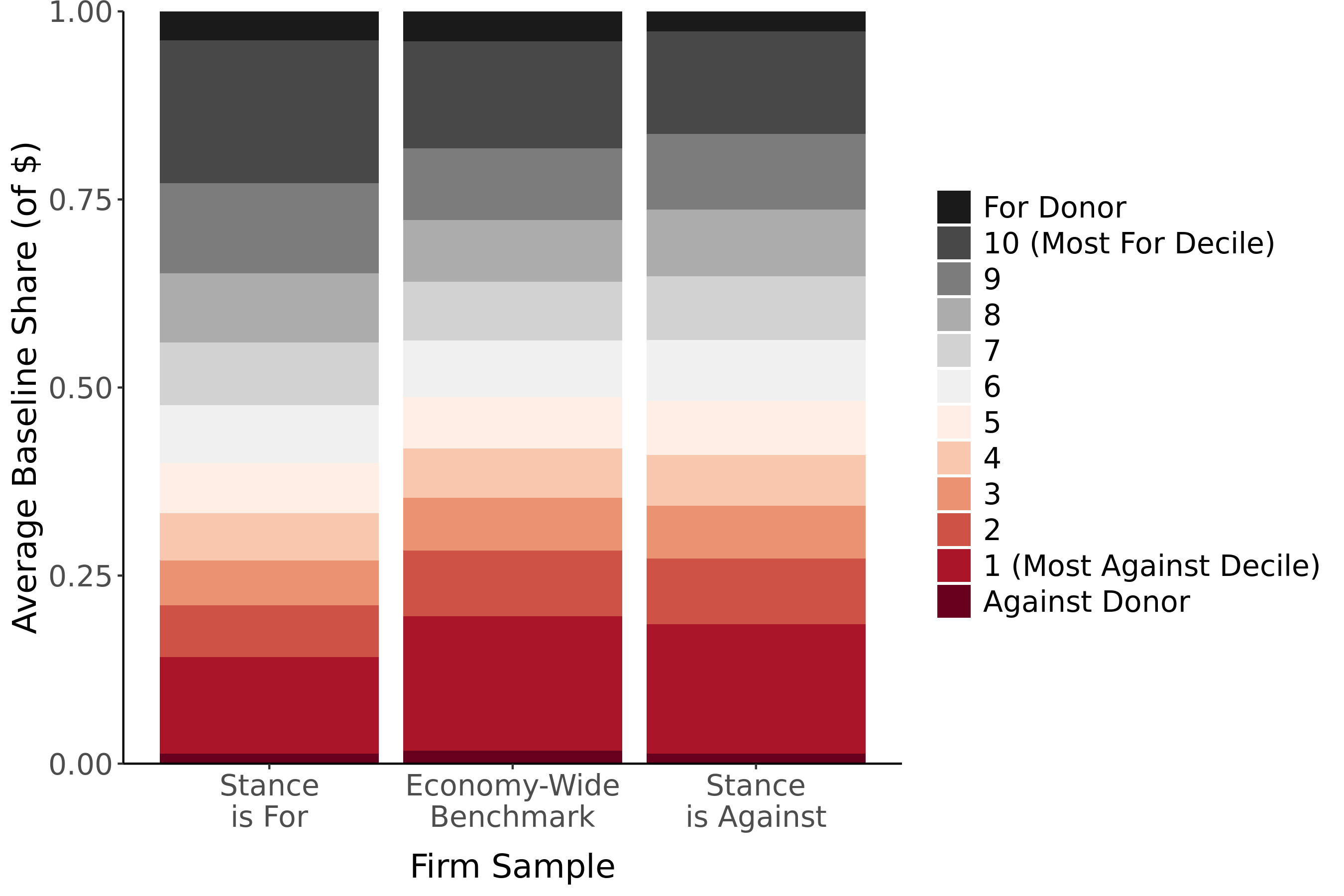}
\par\end{centering}
{\footnotesize Note: Figure shows shares of consumption (in \$) by
group. Consumer groups are defined as described in Section \ref{sec:social-preferences-baseline-shares},
ordering consumers based on their predicted social preference alignment
with the For donation cluster on social issues. The leftmost and rightmost
columns show baseline shares at firms taking social stances aligned
with vs.\ opposed to donations in this cluster, respectively. Baseline
shares refer to the share of consumption (in \$) coming from each
group at the firm in the year preceding these stances. Baseline shares
are weighted by consumer awareness of the firm's stance ($\tau_{j}$,
as defined in Section \ref{sec:event-selection-size}) when averaging
baseline shares across these firm-events. In the middle bar, we show
each group's share of consumption (in \$) aggregated across all U.S.\ firms
in the transaction data throughout the period studied (2008--2023Q1).}{\footnotesize\par}
\end{figure}
\pagebreak{}
\begin{figure}[H]
\begin{centering}
\caption{Consumption Responses by Group (vs.\ Group's Consumption at All Other
Firms)\label{fig:consumption-levels-bygroup}}
\includegraphics[width=1\linewidth]{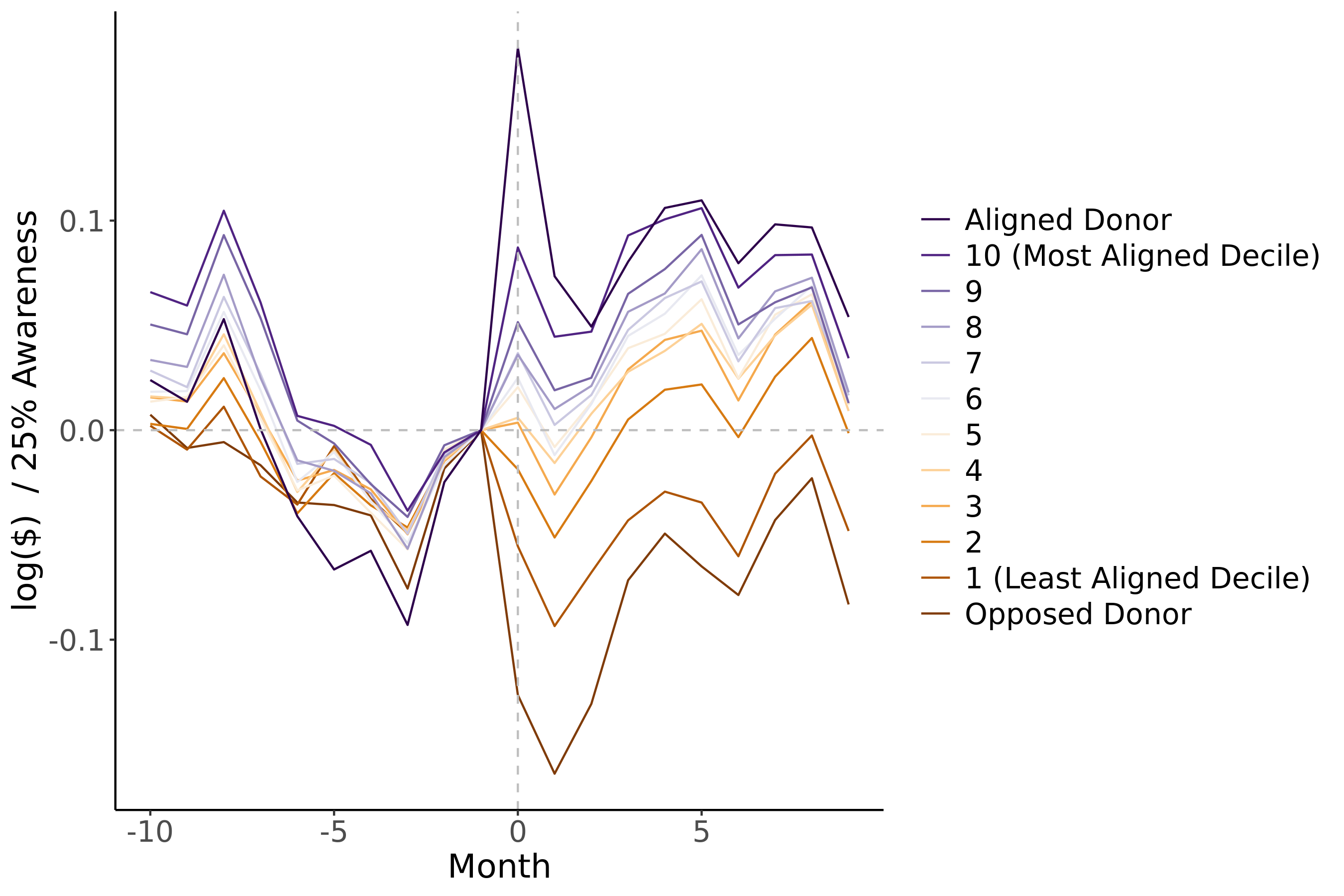}
\par\end{centering}
{\footnotesize Note: Figure shows changes in log consumption at firms
in the months surrounding their social stances, by consumer social
alignment groups. Consumer social alignment groups are constructed
as defined in Section \ref{subsec:social-preference-predictions}.
Changes in log consumption by group are normalized relative to the
month before a firm's social stance and relative to changes in that
group's consumption at all other firms in the economy. Changes are
scaled relative to consumer awareness and are averaged across firms
using a precision-weighted average, as described in Sections \ref{sec:event-selection-size}
and \ref{subsec:event-study-counterfactual}.}{\footnotesize\par}
\end{figure}
\pagebreak{}
\begin{figure}[H]
\begin{centering}
\caption{Predicting No-Event Counterfactual Consumption at Event-Study Firms\label{fig:counterfactual-levels-weekly}}
\par\end{centering}
\begin{centering}
\includegraphics[width=1\linewidth]{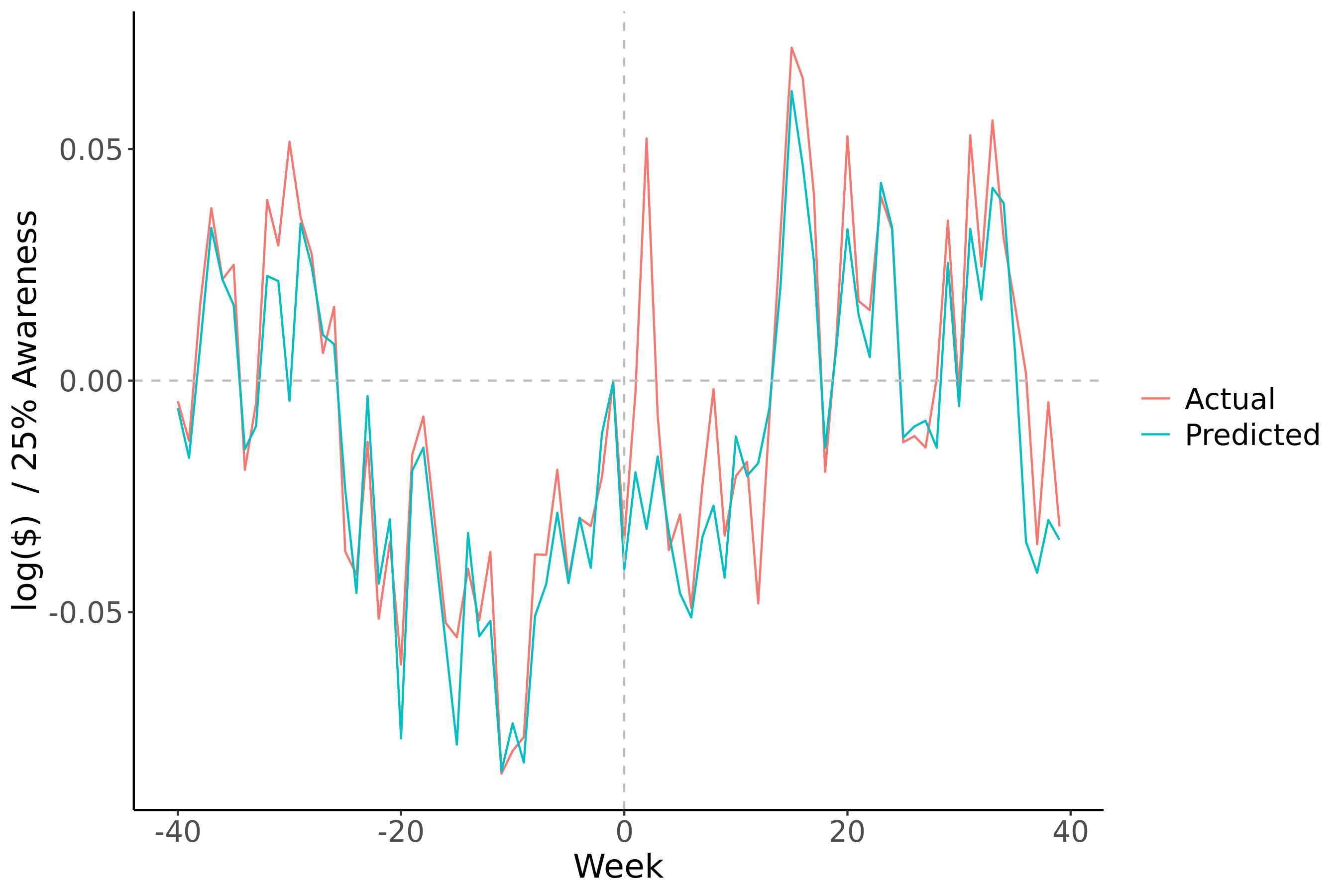}
\par\end{centering}
{\footnotesize Note: Figure shows actual and predicted log consumption
at firms taking social stances by event-week, normalized for visualization
purposes relative to the month prior to the firm's social stance event
and to changes in log consumption at all other firms in the economy.
Log consumption in the absence of a firm's social stance is predicted
using a synthetic difference-in-differences design as described in
Section \ref{subsec:event-study-counterfactual}, using as predictors
contemporaneous consumption at other firms and past consumption at
the social stance firm. Changes are scaled relative to consumer awareness
and are averaged across firms using a precision-weighted average,
as described in Sections \ref{sec:event-selection-size} and \ref{subsec:event-study-counterfactual}. }{\footnotesize\par}
\end{figure}
\pagebreak{}
\begin{figure}[H]
\begin{centering}
\caption{Consumption Responses vs.\ Group-Specific Synthetic DiD Counterfactuals\label{fig:consumption-responses-alt-groupspecific}}
\par\end{centering}
\begin{centering}
\includegraphics[width=1\linewidth]{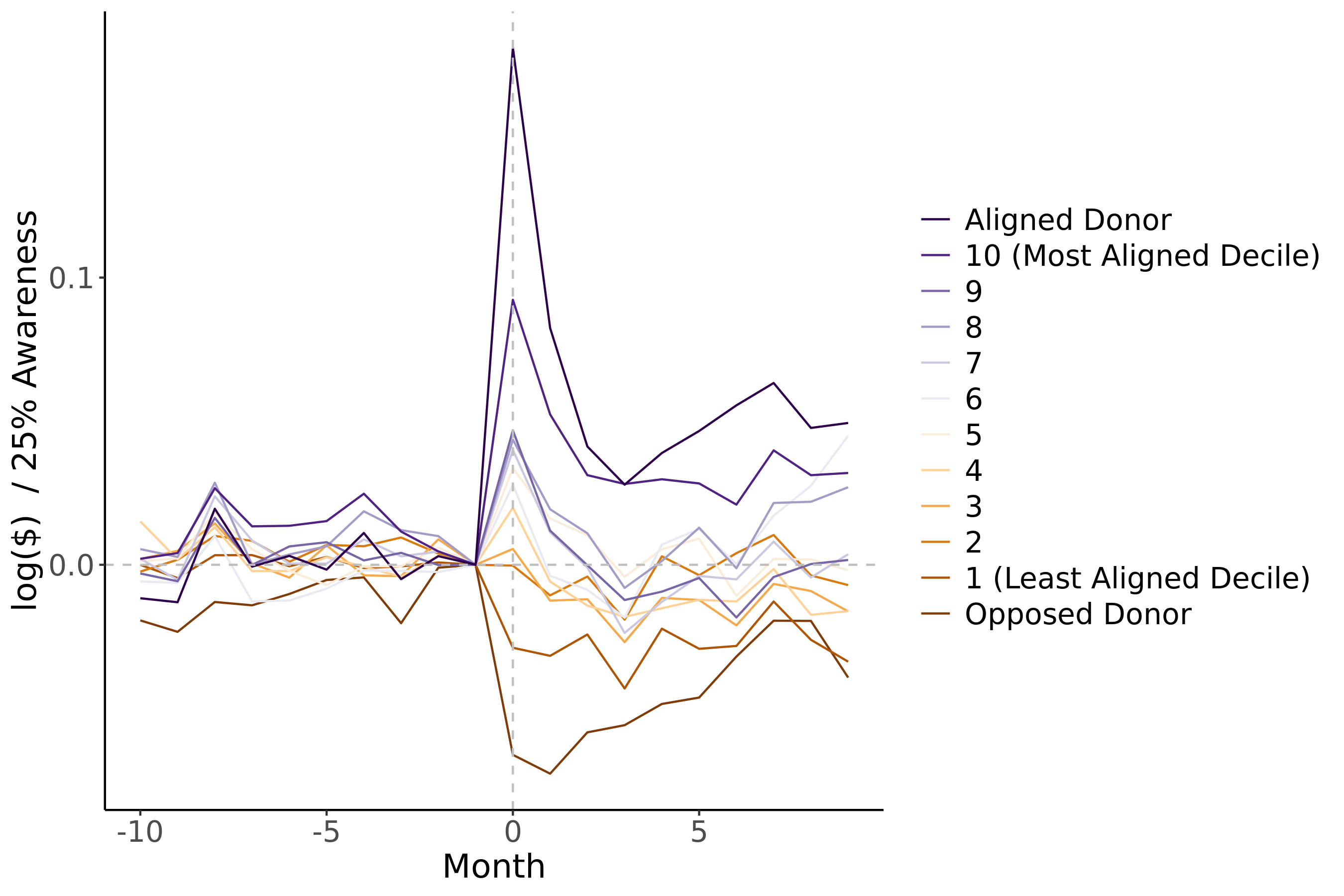}
\par\end{centering}
{\footnotesize Note: Figure follows Figure \ref{fig:consumption-responses-bygroup}
Panel B in showing estimated causal effects of the firm's stance on
log consumption in the months surrounding their social stances, by
consumer group and relative to a synthetic DiD counterfactual. Whereas
Figure \ref{fig:consumption-responses-bygroup} Panel B shifts group-level
consumption responses by the same amount within a period so that they
aggregate up to our estimated overall effect on consumption for the
firm (estimated relative to a synthetic counterfactual for firm-wide
consumption), this figure estimates causal effects by comparing each
group's observed consumption response relative to a group-specific
synthetic DiD counterfactual (based on that group's consumption at
analogous possible control units). Hyperparameters and synthetic weights
are allowed to vary by group. As in Figure \ref{fig:consumption-responses-bygroup}
Panel B, effects are scaled relative to consumer awareness and are
averaged across firms using a precision-weighted average.}{\footnotesize\par}
\end{figure}
\pagebreak{}
\begin{figure}[H]
\begin{centering}
\caption{Consumption Responses by Group, Log Difference Awareness Proxies\label{fig:consumption-responses-awareness-proxies-log}}
\includegraphics[width=0.95\textwidth]{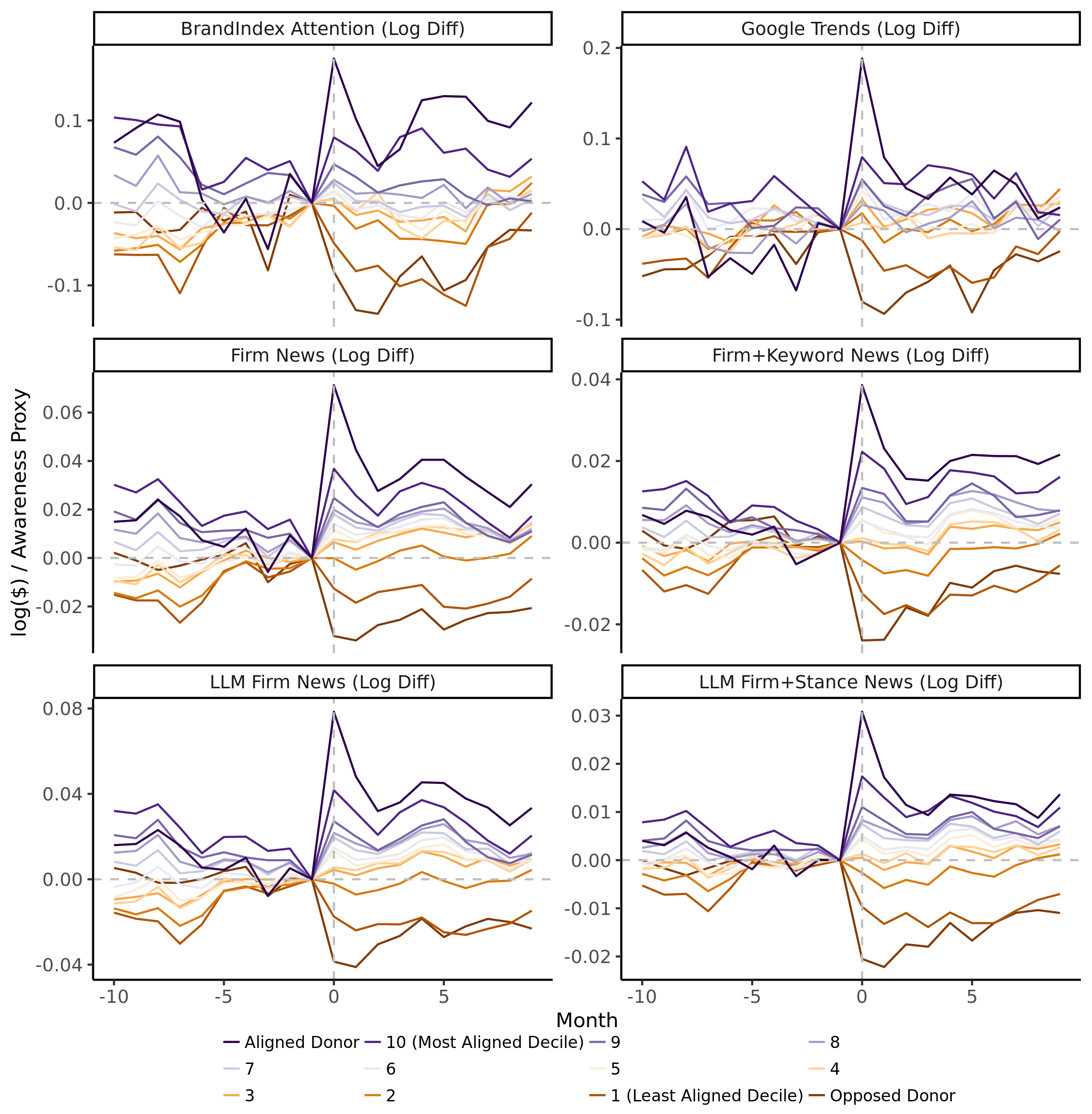}
\par\end{centering}
{\footnotesize Note: Figure follows Figure \ref{fig:consumption-responses-awareness-proxies-level}
in showing how estimated consumption responses by group vary with
our choice of awareness proxy. This figure differs from Figure \ref{fig:consumption-responses-awareness-proxies-level}
in that it proxies for awareness using the event-month change in the
log value of the variable listed in the subpanel title, rather than
change in level.}{\footnotesize\par}
\end{figure}
\pagebreak{}
\begin{figure}[H]
\begin{centering}
\caption{Differences in Consumption Response, Across Groups \label{fig:consumption-responses-differences-cis}}
\subfloat[Panel A: Aligned$-$Opposed Donors]{\begin{centering}
\par\end{centering}
\centering{}\includegraphics[width=0.87\linewidth]{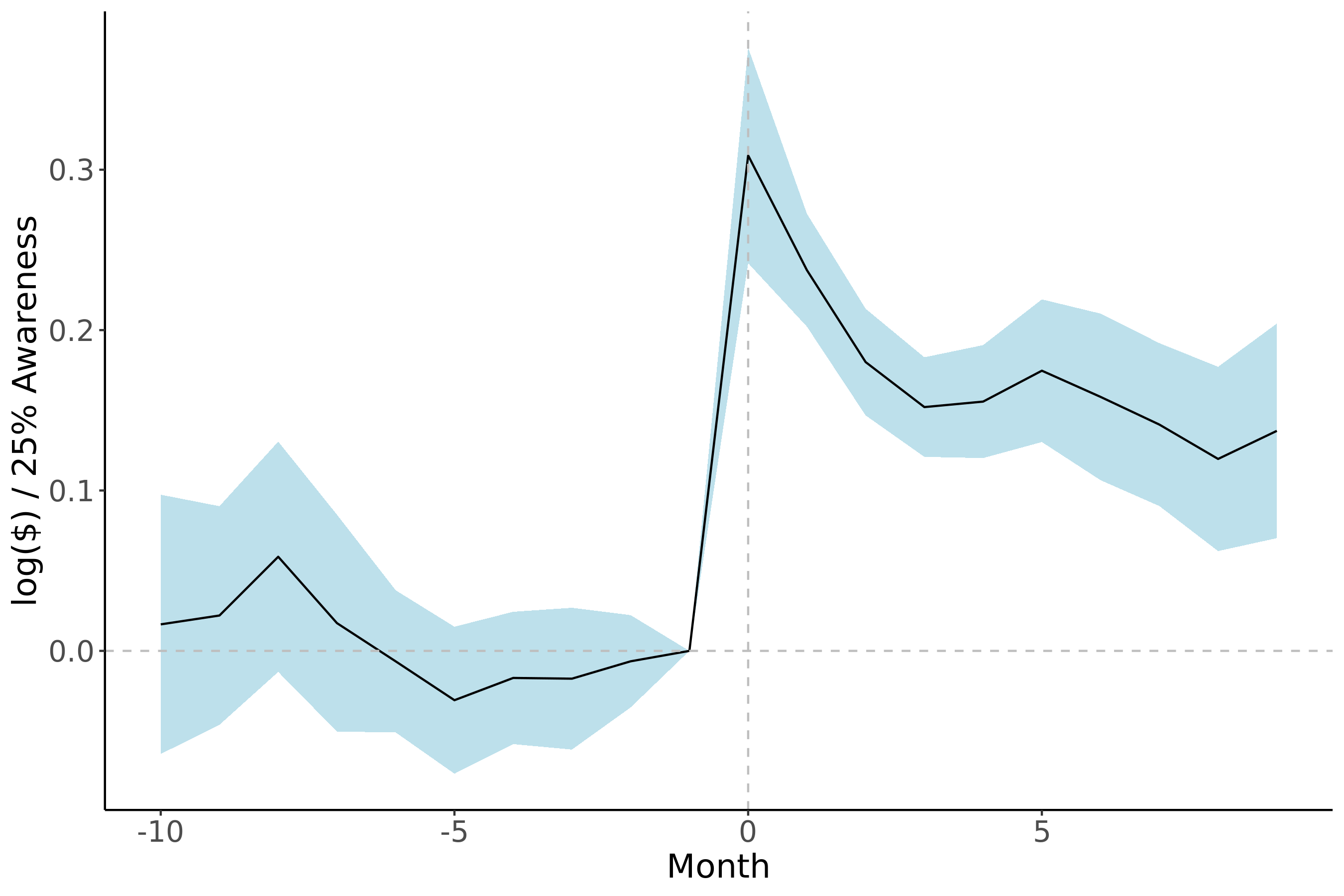}}
\par\end{centering}
\begin{centering}
\subfloat[Panel B: Most Aligned$-$Most Opposed Non-Donor Deciles]{\begin{centering}
\par\end{centering}
\centering{}\includegraphics[width=0.87\textwidth]{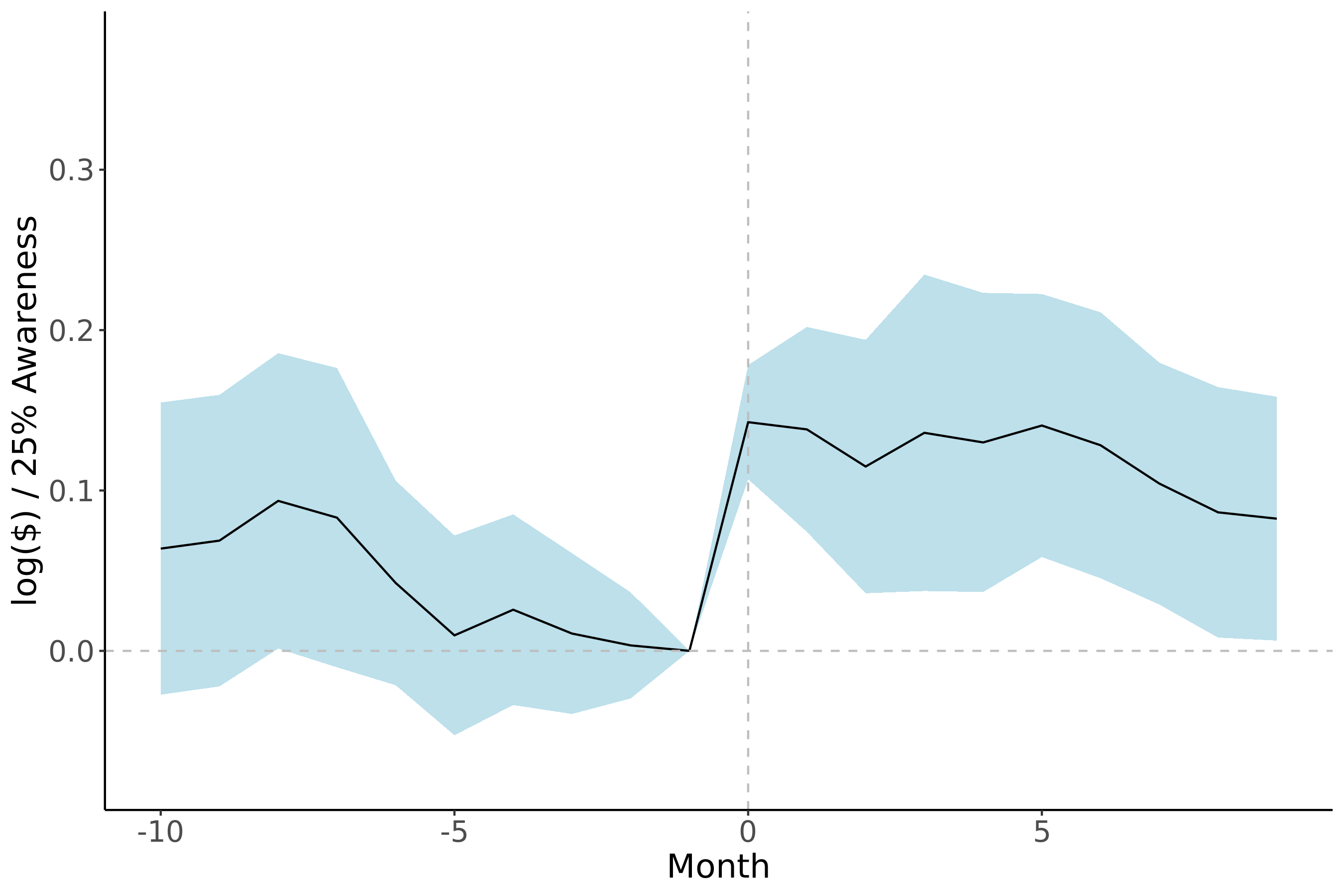}}
\par\end{centering}
{\footnotesize Note: Figure shows differences between the consumption
responses of different groups (as shown in Figure \ref{fig:consumption-responses-bygroup}),
along with a 95\% confidence interval for this difference. Panel A
shows the consumption response difference among the Aligned vs.\ Opposed
donor groups. Panel B shows the consumption response difference among
the most aligned decile vs.\ most opposed non-donor deciles. Responses
are scaled relative to consumer awareness and averaged across firms
using a precision-weighted average, as described in Sections \ref{sec:event-selection-size}
and \ref{subsec:event-study-counterfactual}. Standard errors are
clustered by event.}{\footnotesize\par}
\end{figure}
\pagebreak{}
\begin{figure}[H]
\begin{centering}
\caption{(One-Year-Prior Placebo) Changes in Consumption at Social Stance Firms,
by Group\label{fig:consumption-responses-placebo}}
\vspace*{-0.3cm}\subfloat[Panel A: Response Levels by Group]{\begin{centering}
\par\end{centering}
\centering{}\includegraphics[width=0.58\linewidth]{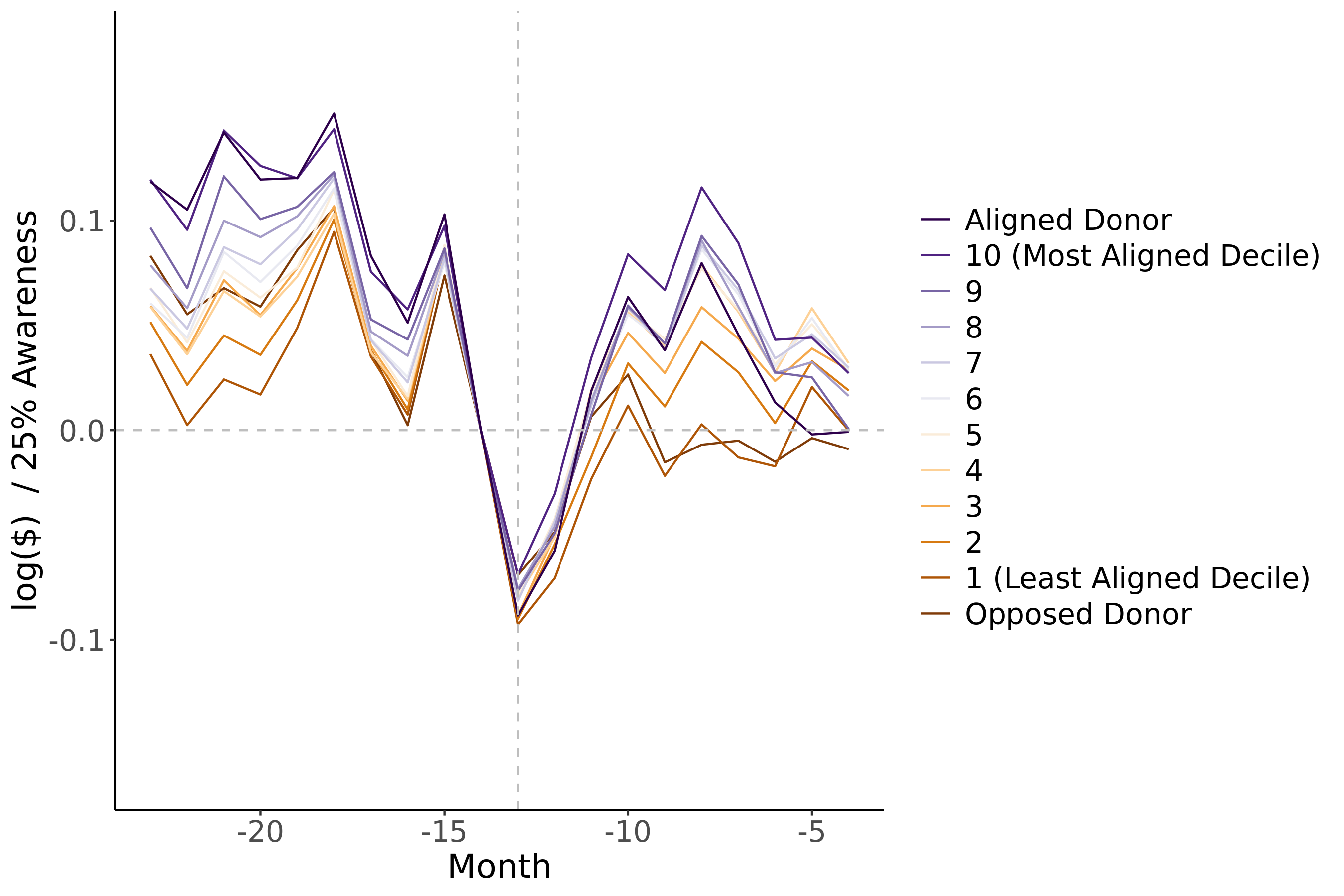}}
\par\end{centering}
\begin{centering}
\vspace*{-0.1cm}\subfloat[Panel B: vs.\ Overall Counterfactual]{\begin{centering}
\par\end{centering}
\centering{}\includegraphics[width=0.58\textwidth]{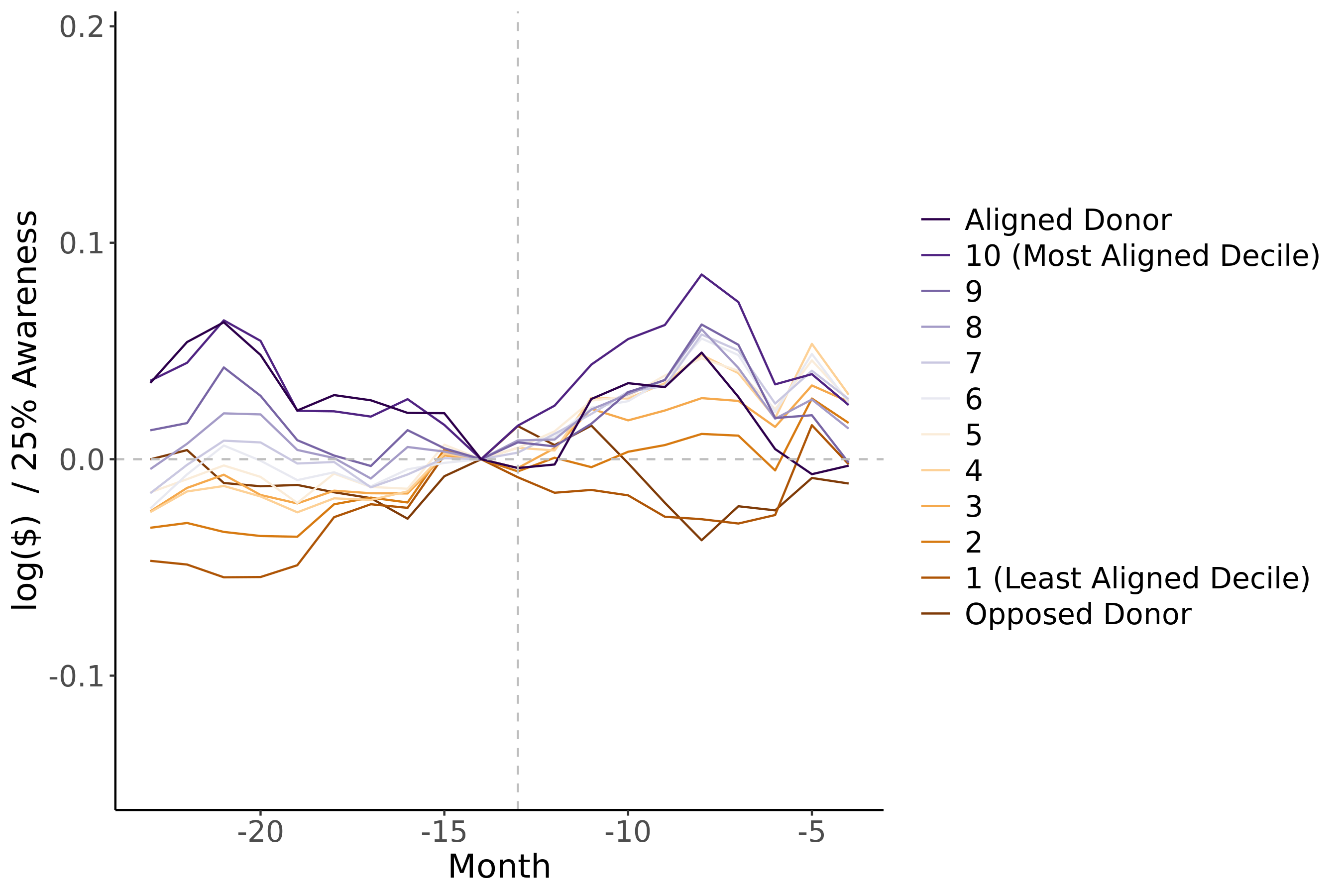}}
\par\end{centering}
\begin{centering}
\vspace*{-0.1cm}\subfloat[Panel C: vs.\ Group-Specific Counterfactual]{\begin{centering}
\par\end{centering}
\centering{}\includegraphics[width=0.58\textwidth]{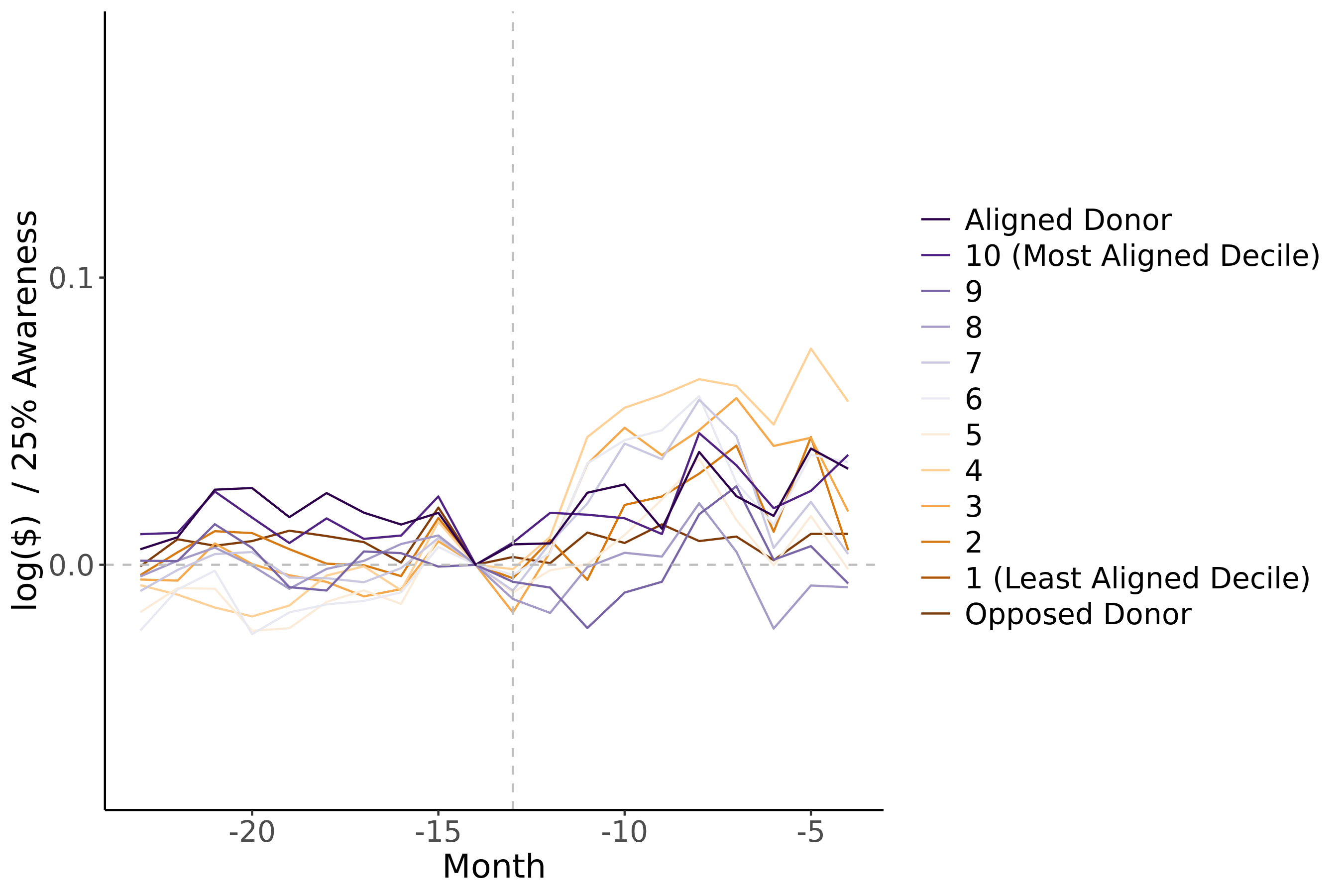}}
\par\end{centering}
{\footnotesize Note: Figure shows changes in log consumption at firms
by consumer social alignment groups in a placebo exercise, rerunning
our analysis as if social stance events occurred one year prior to
their actual date. Panel A shows consumption response in levels, Panel
B shifts these to aggregate up to an estimated overall consumption
impact, and Panel C compares consumption levels to group-specific
synthetic counterfactuals. The y-axis range and all other specifications
of Panels A-C follow Figure \ref{fig:consumption-levels-bygroup},
Figure \ref{fig:consumption-responses-bygroup} Panel B, and Figure
\ref{fig:consumption-responses-alt-groupspecific}, respectively.
}{\footnotesize\par}
\end{figure}
\pagebreak{}
\begin{figure}[H]
\begin{centering}
\caption{Changes in Consumption at Social Stance Firms, by Group (3-Year Pre-Period)
\label{fig:consumption-responses-bygroup-longpre}}
\subfloat[Panel A: Response Levels by Group (vs.\ Group's Consumption at All
Other Firms)]{\begin{centering}
\par\end{centering}
\centering{}\includegraphics[width=0.9\textwidth]{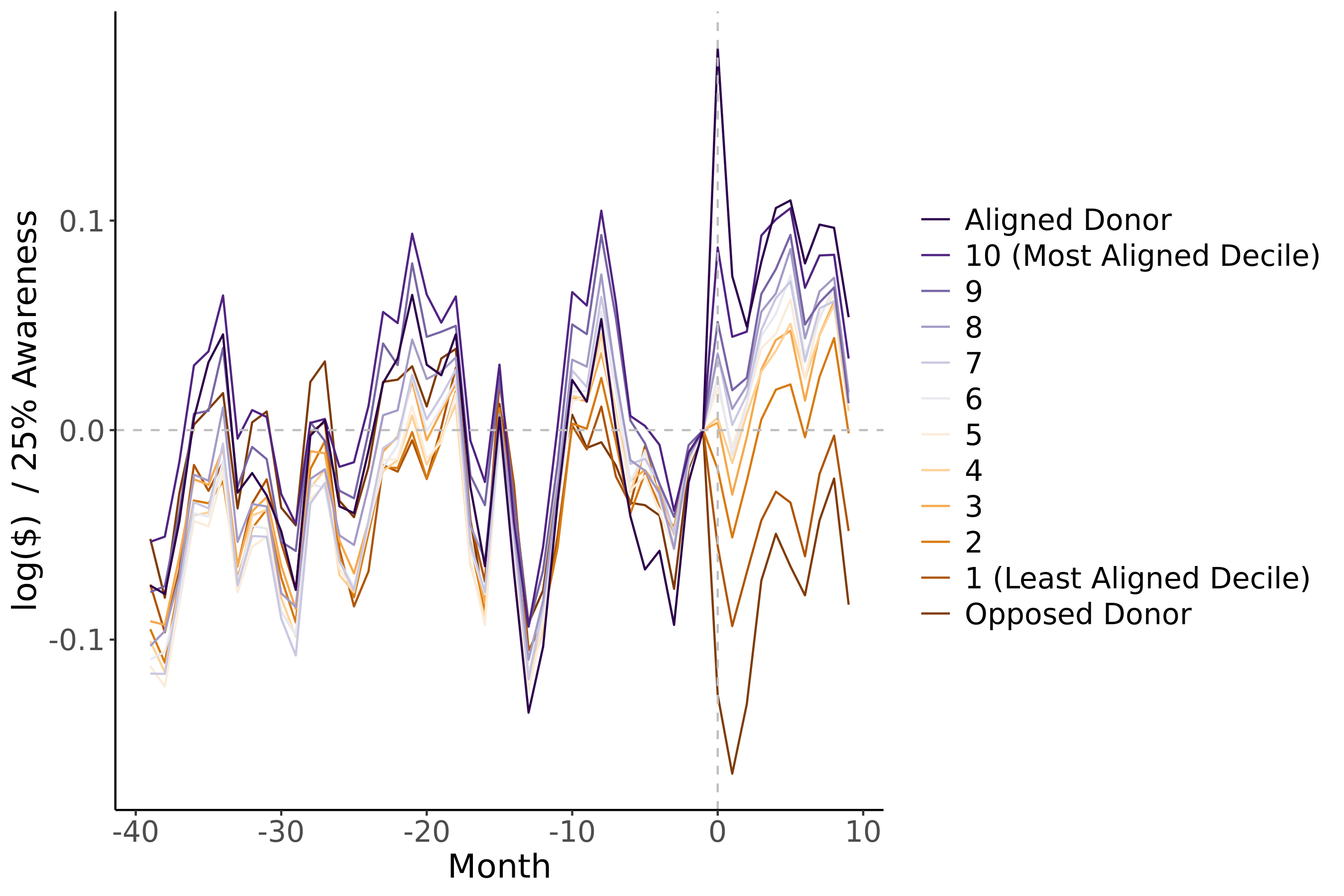}}
\par\end{centering}
\begin{centering}
\subfloat[Panel B: Response Effects by Group (Shifting Levels to Match Estimated
Overall Impact)]{\begin{centering}
\par\end{centering}
\centering{}\includegraphics[width=0.9\textwidth]{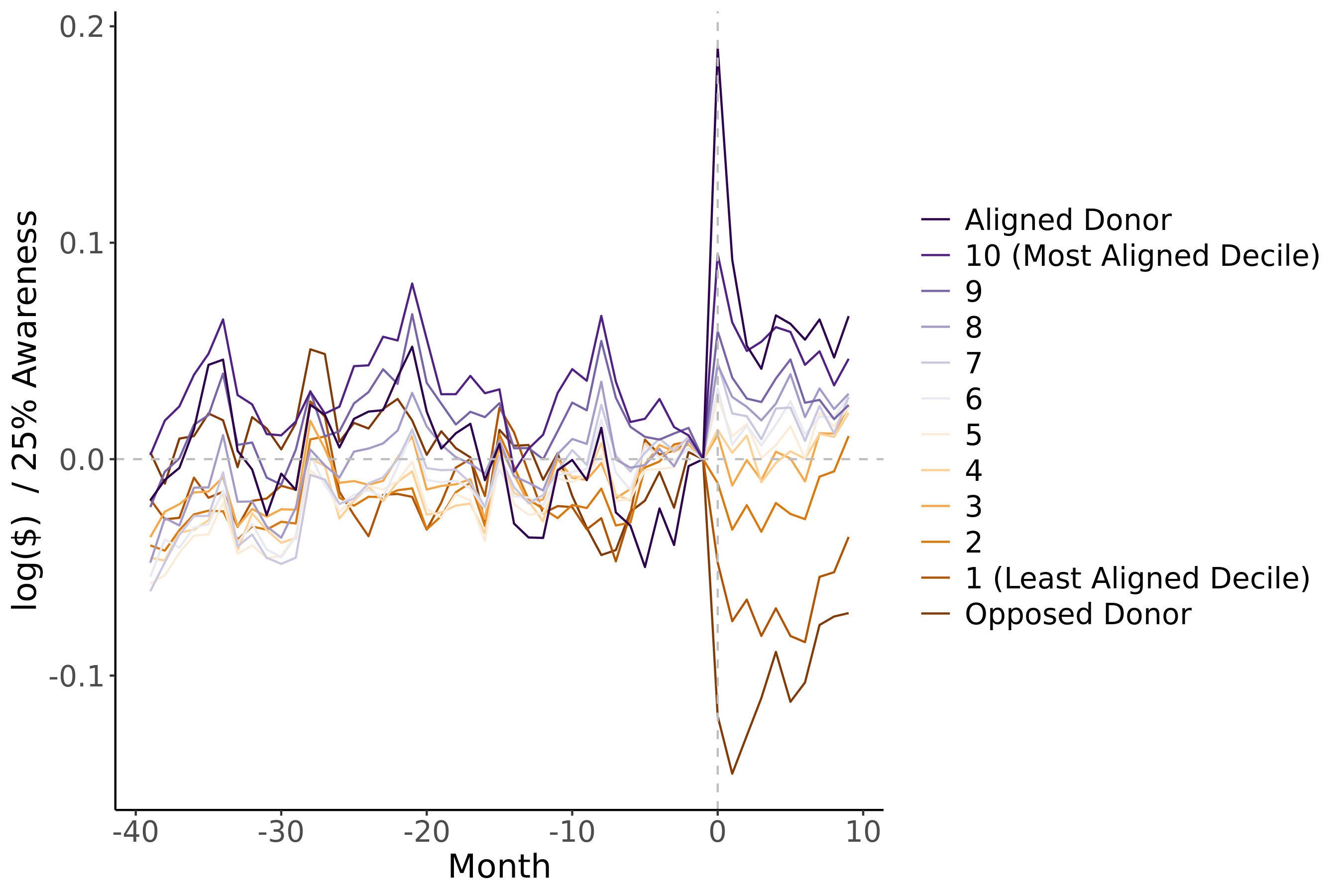}}
\par\end{centering}
{\footnotesize Note: Figure shows changes in log consumption at firms
in the months surrounding their social stances by consumer social
alignment groups. Panels A and B have been modified to show 3 years
of pre-event data (i.e., 39 pre-event 4-week ``months''), with all
other specifications following Figure \ref{fig:consumption-levels-bygroup}
and Figure \ref{fig:consumption-responses-bygroup} Panel B, respectively.}{\footnotesize\par}
\end{figure}
\pagebreak{}
\par\end{center}

\begin{center}
\begin{figure}[H]
\begin{centering}
\caption{Response Levels by Group (2-Year Post-Period) \label{fig:consumption-responses-bygroup-longpost}}
\subfloat[Panel A: Response Levels by Group (vs.\ Group's Consumption at All
Other Firms)]{\begin{centering}
\par\end{centering}
\centering{}\includegraphics[width=0.82\linewidth]{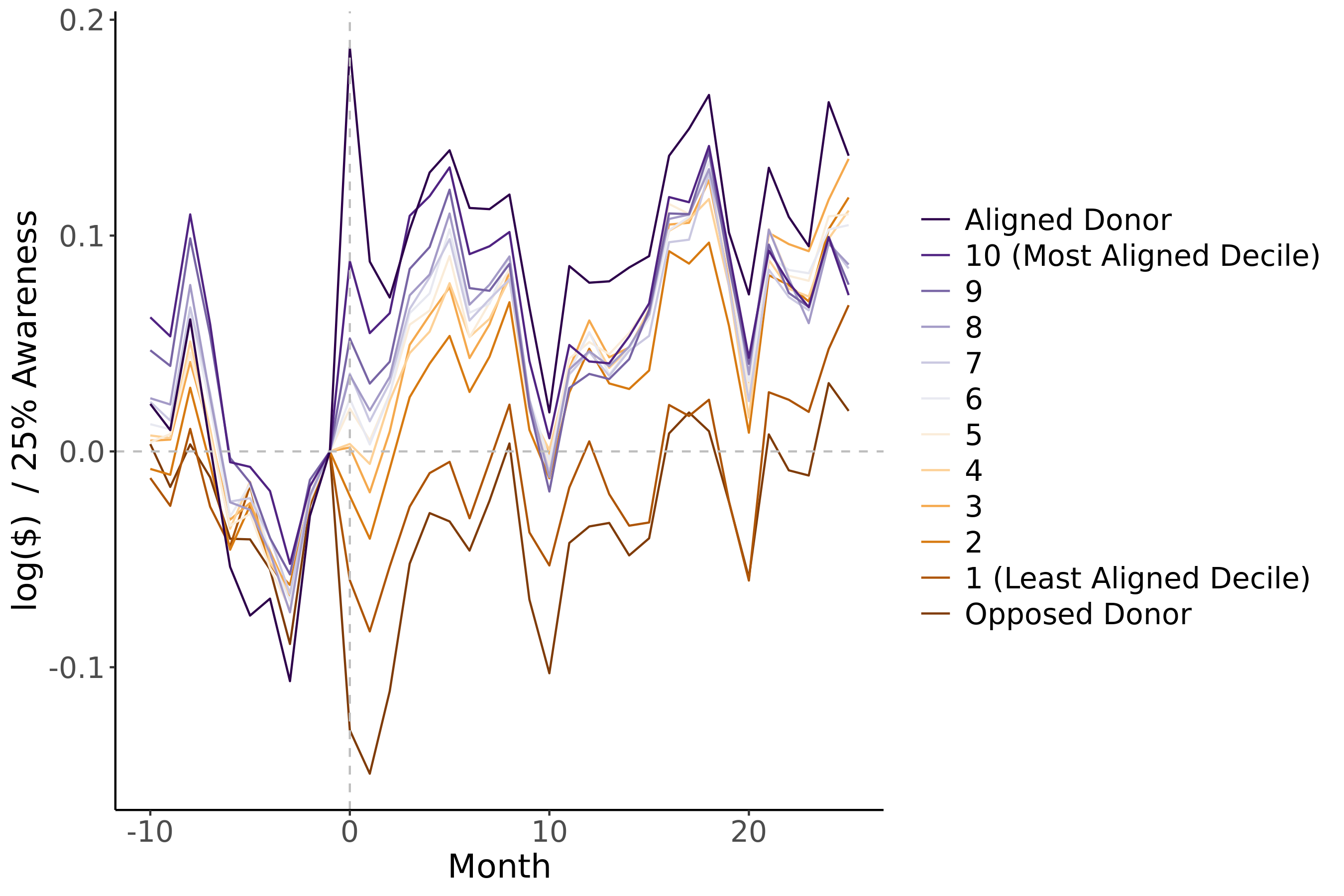}}
\par\end{centering}
\begin{centering}
\subfloat[Panel B: Response Effects by Group (Shifting Levels to Normalize 5th
and 6th Decile to Zero)]{\begin{centering}
\par\end{centering}
\centering{}\includegraphics[width=0.82\textwidth]{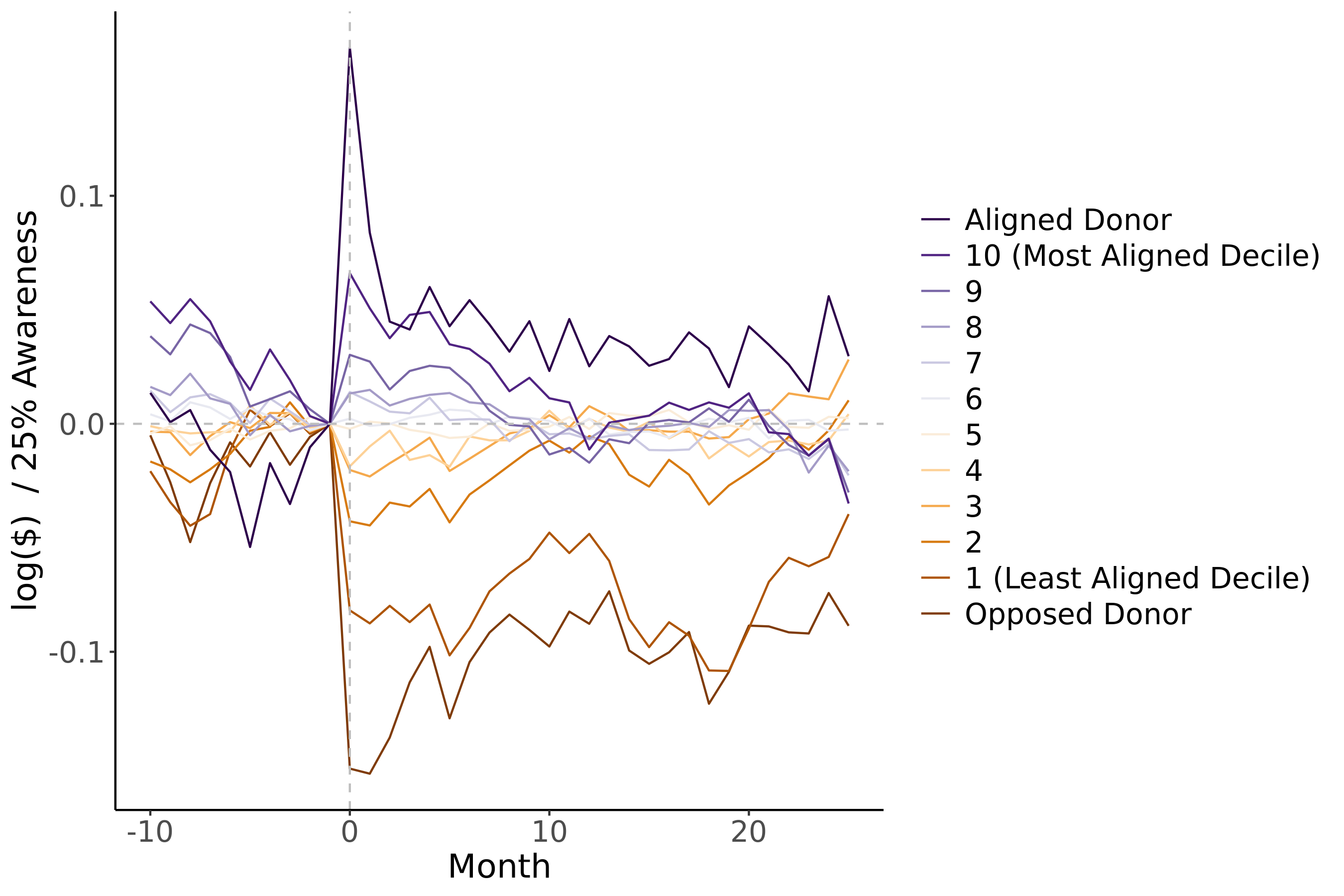}}
\par\end{centering}
{\footnotesize Note: Figure shows changes in log consumption at firms
in the months surrounding their social stances by consumer social
alignment groups. Panel A modifies Figure \ref{fig:consumption-levels-bygroup}
to show two years of post-event data (i.e., 26 post-event 4-week ``months''),
with all other specifications following this appendix figure. Panel
B then shifts all levels within a period such that the resulting consumption
responses of the middle two deciles of non-donors sum to zero.}{\footnotesize\par}
\end{figure}
\pagebreak{}
\begin{figure}[H]
\begin{centering}
\caption{Social Stance Consumption Impacts, by Cluster Alignment \label{fig:consumption-responses-by-stancetype}}
\subfloat[Panel A: Overall Average Consumption Impacts]{\begin{centering}
\par\end{centering}
\centering{}\includegraphics[width=0.8\textwidth]{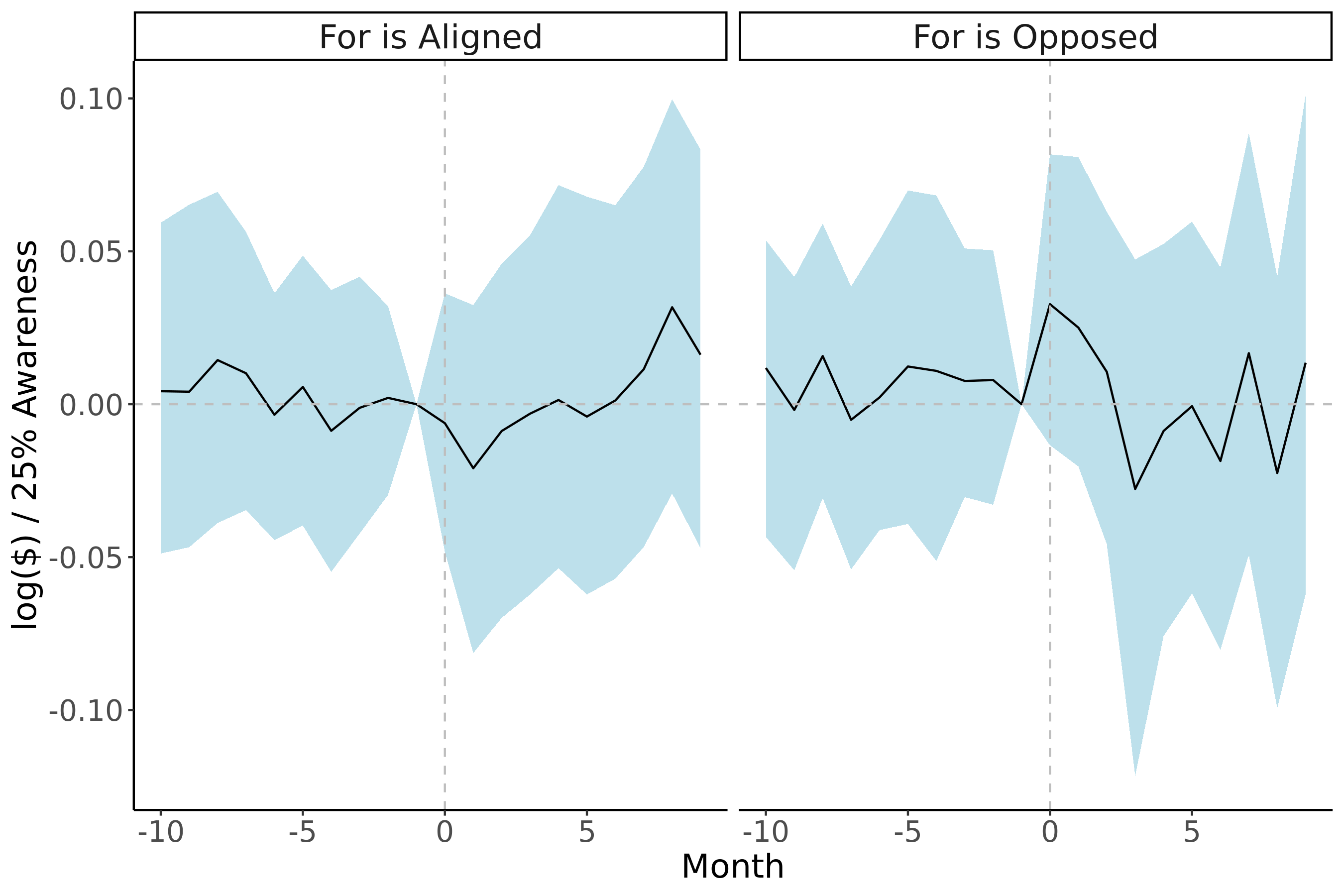}}
\par\end{centering}
\begin{centering}
\subfloat[Panel B: Consumption Responses, by Group]{\begin{centering}
\par\end{centering}
\centering{}\includegraphics[width=1\linewidth]{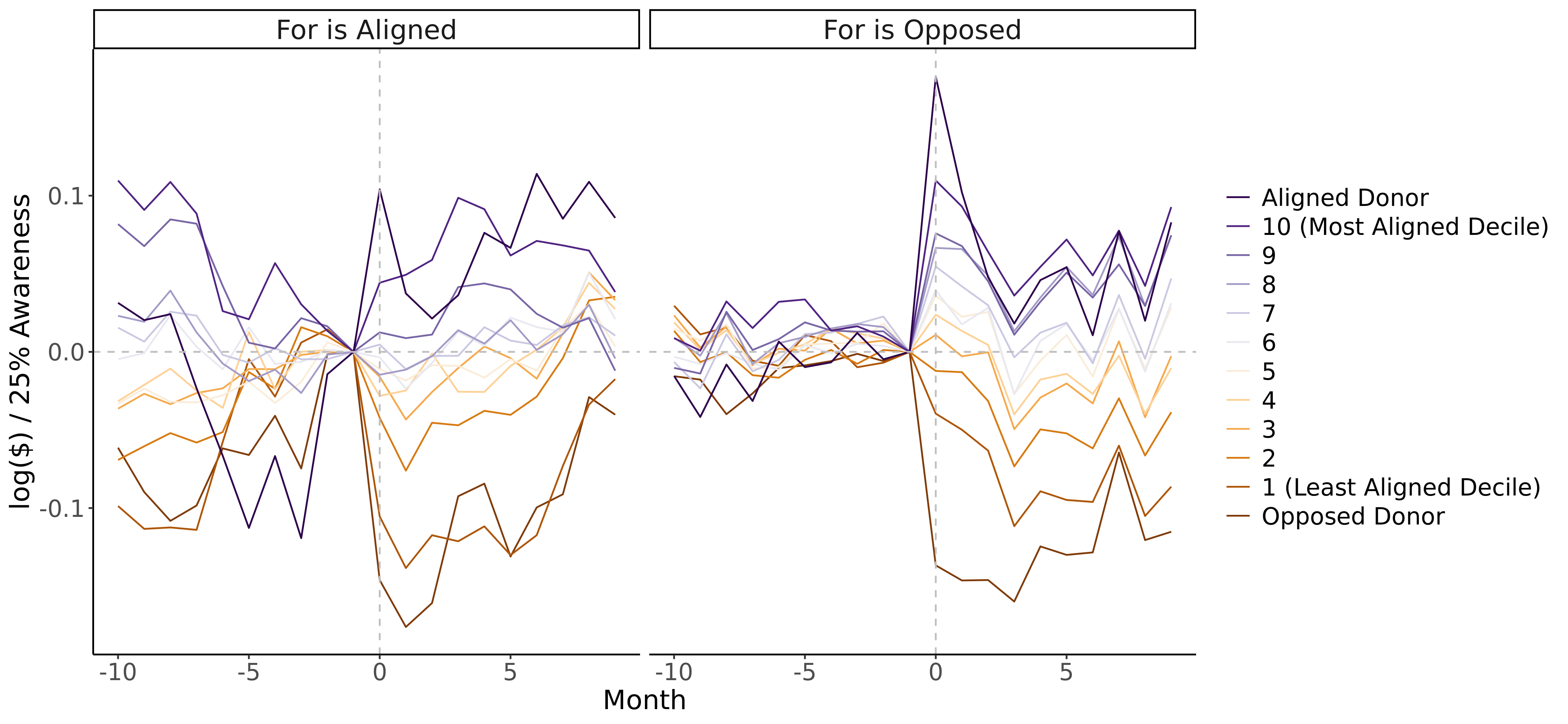}}
\par\end{centering}
{\footnotesize Note: Figure shows estimated overall and disaggregated
consumption impacts separately among events where donors in the For
cluster are likely aligned with vs.\ opposed to the firm's stance.
Panel A extends Figure \ref{fig:consumption-responses-bygroup} Panel
A to separately show estimated overall consumption impacts. 95\% confidence
intervals are constructed using a wild cluster bootstrap approach
that accounts for uncertainty in our synthetic difference-in-differences
counterfactuals. Panel B extends Figure \ref{fig:consumption-responses-bygroup}
Panel B to separately show estimated consumption responsiveness by
social alignment group. Groups in Panel B are colored according to
their likely alignment with the firm's stance.}{\footnotesize\par}
\end{figure}
\pagebreak{}
\begin{figure}[H]
\begin{centering}
\caption{Interpretation of News by Alignment, in BrandIndex\label{fig:brandindex_favorability}}
\subfloat[Panel A: Net Favorability of News]{\begin{centering}
\par\end{centering}
\centering{}\includegraphics[width=0.76\textwidth]{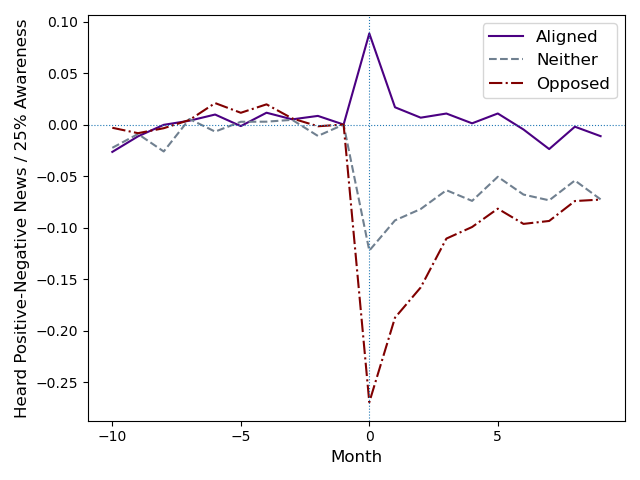}}
\par\end{centering}
\begin{centering}
\subfloat[Panel B: Net Favorability toward Firm]{\begin{centering}
\par\end{centering}
\centering{}\includegraphics[width=0.76\textwidth]{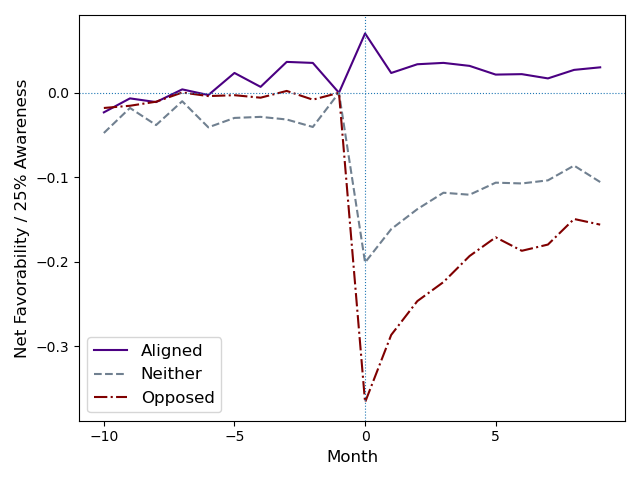}}
\par\end{centering}
{\footnotesize Note: Figure shows changes in favorability around firms'
social stance events by social alignment, based on BrandIndex responses.
Panel A shows favorability regarding news about the firm, coded as
$1$ if the respondent reported having heard ``Positive'' news about
the firm in the last two weeks, $-1$ if reported having heard ``Negative''
news, and $0$ otherwise. Panel B shows favorability toward the firm
more generally, again coded as $+1$ (Positive), $-1$ (Negative),
or $0$ (Neutral). Responses are scaled relative to consumer awareness,
averaged across firms using a $\tau^{2}_{j}$-weighted average, and
normalized relative to the month before a firm's event ($t=-1$).}{\footnotesize\par}
\end{figure}
\pagebreak{}
\begin{figure}[H]
\begin{centering}
\caption{Favorability Levels by Alignment, in BrandIndex\label{fig:brandindex_favorability_levels}}
\subfloat[Panel A: Net Favorability of News]{\begin{centering}
\par\end{centering}
\centering{}\includegraphics[width=0.76\textwidth]{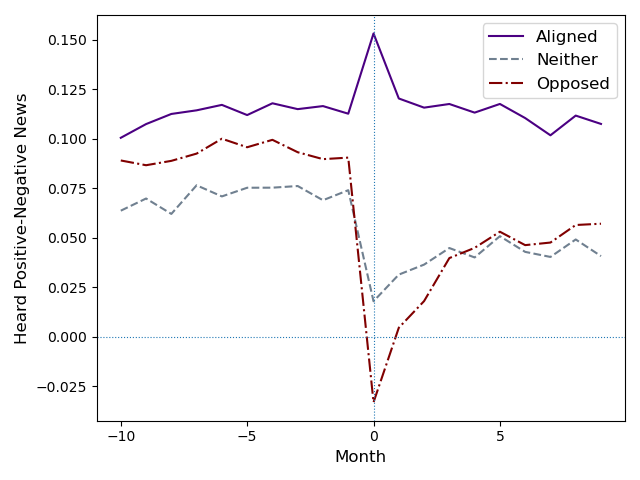}}
\par\end{centering}
\begin{centering}
\subfloat[Panel B: Net Favorability toward Firm]{\begin{centering}
\par\end{centering}
\centering{}\includegraphics[width=0.76\textwidth]{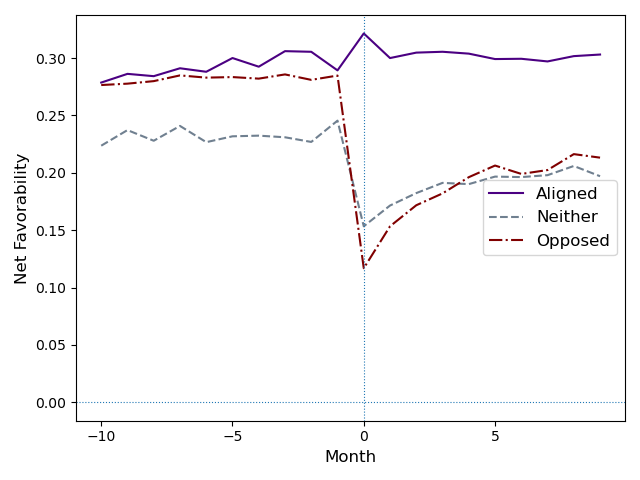}}
\par\end{centering}
{\footnotesize Note: Figure shows favorability levels around firms'
social stance events by social alignment, based on BrandIndex responses.
Panel A shows favorability regarding news about the firm, coded as
$1$ if the respondent reported having heard ``Positive'' news about
the firm in the last two weeks, $-1$ if reported having heard ``Negative''
news, and $0$ otherwise. Panel B shows favorability toward the firm
more generally, again coded as $+1$ (Positive), $-1$ (Negative),
or $0$ (Neutral). Figure differs from Figure \ref{fig:brandindex_favorability}
in that responses are not scaled relative to consumer awareness nor
normalized relative to month $t=-1$. Favorability levels are averaged
across firms using a $\tau_{j}$-weighted average.}{\footnotesize\par}
\end{figure}
\pagebreak{}
\begin{figure}[H]
\begin{centering}
\caption{Impact on Self-Reported Purchase Behavior by Alignment, in BrandIndex\label{fig:brandindex_purchasebehavior}}
\subfloat[Panel A: Would Consider Purchase at Firm]{\begin{centering}
\par\end{centering}
\centering{}\includegraphics[width=0.76\textwidth]{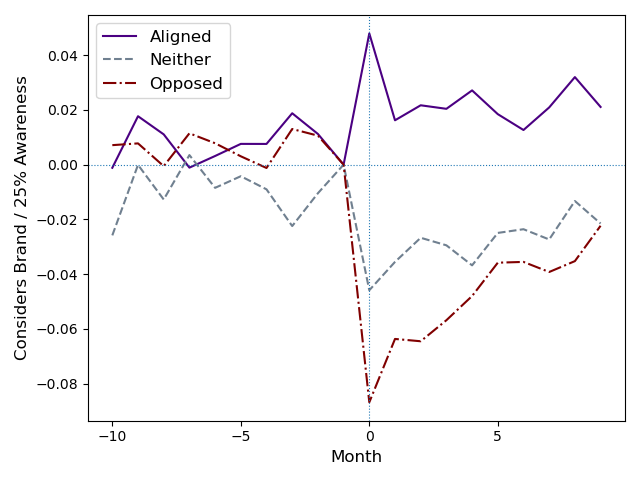}}
\par\end{centering}
\begin{centering}
\subfloat[Panel B: Intend to Purchase from Firm]{\begin{centering}
\par\end{centering}
\centering{}\includegraphics[width=0.76\textwidth]{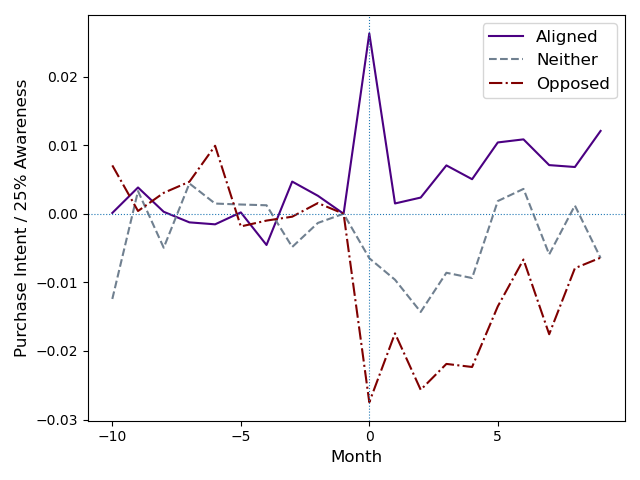}}
\par\end{centering}
{\footnotesize Note: Figure shows changes in self-reported purchase
behavior at firms in the months surrounding their social stances,
based on BrandIndex responses. Panel A shows an indicator for whether
a respondent would consider purchasing from the firm when next shopping
in that firm's market (1 if yes; 0 if no), and Panel B shows an indicator
for whether a respondent would be most likely to purchase from that
firm. Responses are scaled relative to consumer awareness, averaged
across firms using a $\tau^{2}_{j}$-weighted average, and normalized
relative to the month before a firm's event ($t=-1$).}{\footnotesize\par}
\end{figure}
\pagebreak{}
\begin{figure}[H]
\begin{centering}
\caption{Channels for Learning About Firm Stance by Alignment, in BrandIndex\label{fig:brandindex_learningchannels}}
\subfloat[Panel A: Word-of-Mouth Exposure]{\begin{centering}
\par\end{centering}
\centering{}\includegraphics[width=0.76\textwidth]{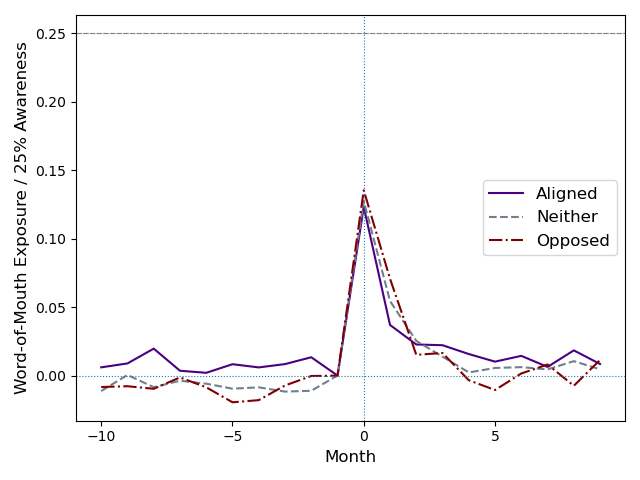}}
\par\end{centering}
\begin{centering}
\subfloat[Panel B: Ad Awareness]{\begin{centering}
\par\end{centering}
\centering{}\includegraphics[width=0.76\textwidth]{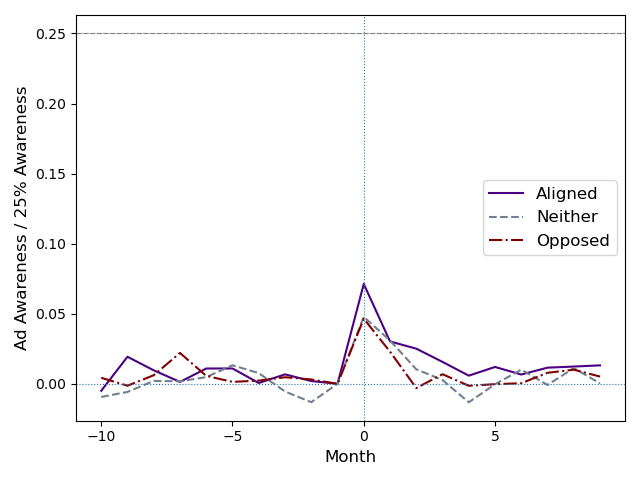}}
\par\end{centering}
{\footnotesize Note: Figure shows changes in exposure to information
about firms, based on BrandIndex responses. Panel A shows an indicator
for whether a respondent recently talked with someone about the brand
(in-person, online, or through social media), and Panel B shows an
indicator for whether a respondent recently saw an advertisement from
that firm. Responses are scaled relative to consumer awareness, averaged
across firms using a $\tau^{2}_{j}$-weighted average, and normalized
relative to the month before a firm's event ($t=-1$). A dashed gray
horizontal line at 0.25 in each plot shows our 25 percent consumer
awareness benchmark.}{\footnotesize\par}
\end{figure}
\pagebreak{}
\par\end{center}

\begin{center}
\begin{figure}[H]
\begin{centering}
\caption{Advertising Expenditures of Event-Study Firms\label{fig:adspend}}
\par\end{centering}
\begin{centering}
\includegraphics[width=1\textwidth]{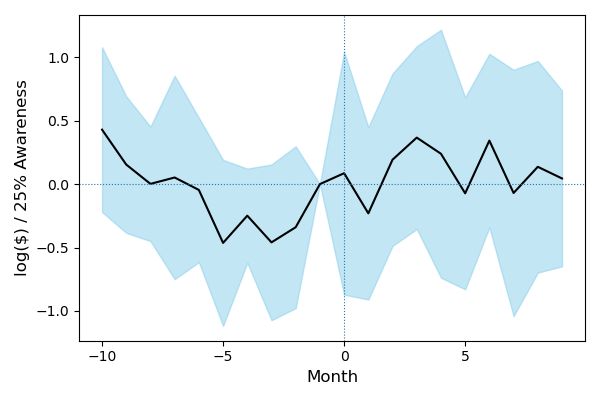}
\par\end{centering}
\raggedright{}{\footnotesize Note: Figure shows advertising expenditures
for firms in the months surrounding their social stances. Advertising
expenditures are calculated in thousands of dollars, summing advertising
expenditures within an event-month across channels in Nielsen's Ad
Intel dataset. To avoid dropping zero expenditures, we add one to
this value for all firm-month totals before taking logs. Outcomes
are scaled relative to consumer awareness, averaged across firms using
a $\tau^{2}_{j}$-weighted average, and normalized relative to the
month before a firm's event ($t=-1$). }{\footnotesize\par}
\end{figure}
\pagebreak{}
\par\end{center}

\begin{center}
\begin{figure}[H]
\begin{centering}
\caption{Price Indices of Event-Study Firms\label{fig:priceindex}}
\par\end{centering}
\begin{centering}
\subfloat[Panel A: Laspeyres Price Index, Brand+Banner]{\begin{centering}
\par\end{centering}
\centering{}\includegraphics[width=0.49\textwidth]{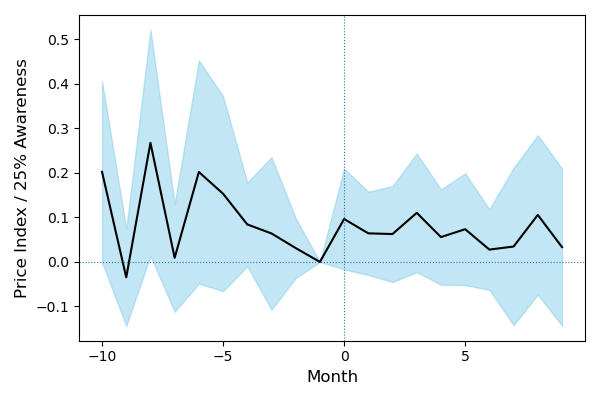}}\subfloat[Panel B: Paasche Price Index, Brand+Banner]{\begin{centering}
\par\end{centering}
\centering{}\includegraphics[width=0.49\textwidth]{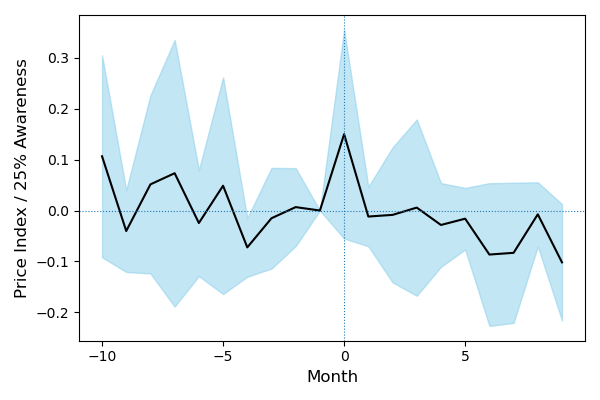}}
\par\end{centering}
\begin{centering}
\subfloat[Panel C: Laspeyres Price Index, Brand-Only]{\begin{centering}
\par\end{centering}
\centering{}\includegraphics[width=0.49\textwidth]{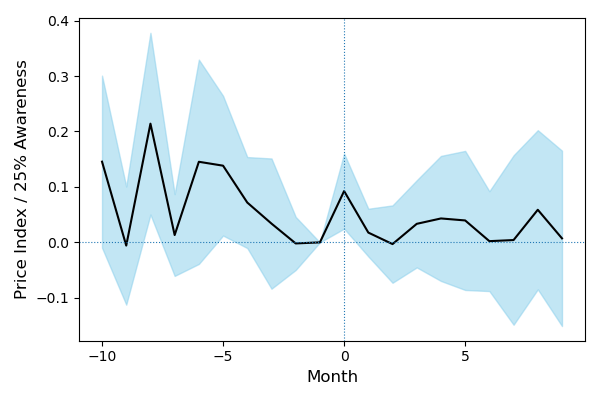}}\subfloat[Panel D: Paasche Price Index, Brand-Only]{\begin{centering}
\par\end{centering}
\centering{}\includegraphics[width=0.49\textwidth]{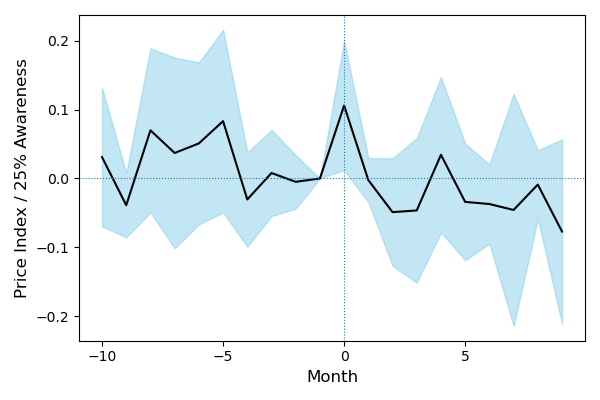}}
\par\end{centering}
\raggedright{}{\footnotesize Note: Figure shows price indices for firms
in the months surrounding their social stances. Panels A-B are based
on the products that belong to a firm's brand or that are sold under
its banner (i.e., in its stores or website, regardless of branding),
and show Laspeyres and Paasche price indices, respectively. Panels
C-D show analogous Laspeyres and Paasche price indices when restricting
only to products with the firm's brand (excluding other products sold
within its stores but without its brand). Outcomes are scaled relative
to consumer awareness, averaged across firms using a $\tau^{2}_{j}$-weighted
average, and normalized relative to the month before a firm's event
($t=-1$). Data consist of receipt-captured information from Numerator's
omni-channel consumer panel, and all plots use event-month $t=-1$
as the base period.}{\footnotesize\par}
\end{figure}
\pagebreak{}
\begin{figure}[H]
\begin{centering}
\caption{Cumulative Stock Price Returns of Event-Study Firms\label{fig:stock-prices}}
\vspace*{-0.3cm}\subfloat[Panel A: Cumulative Return (No Adjustment)]{\begin{centering}
\par\end{centering}
\centering{}\includegraphics[width=0.49\textwidth]{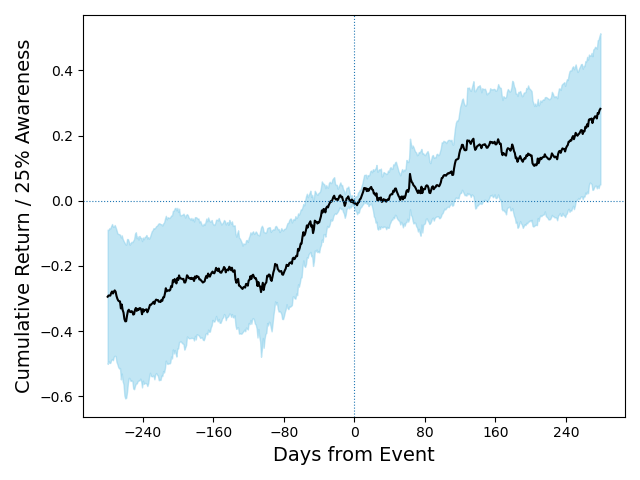}}
\par\end{centering}
\begin{centering}
\vspace*{-0.1cm}\subfloat[Panel B: Cumulative Abnormal Return (Market-Adjusted)]{\begin{centering}
\par\end{centering}
\centering{}\includegraphics[width=0.49\textwidth]{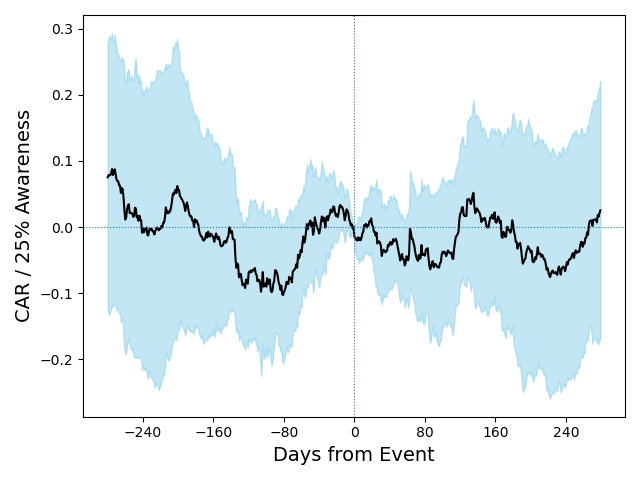}}
\par\end{centering}
\begin{centering}
\vspace*{-0.1cm}\subfloat[{Panel C: CAR (Market-Adjusted, {[}$-10$,$27${]}-Day Window)}]{\begin{centering}
\par\end{centering}
\centering{}\includegraphics[width=0.49\textwidth]{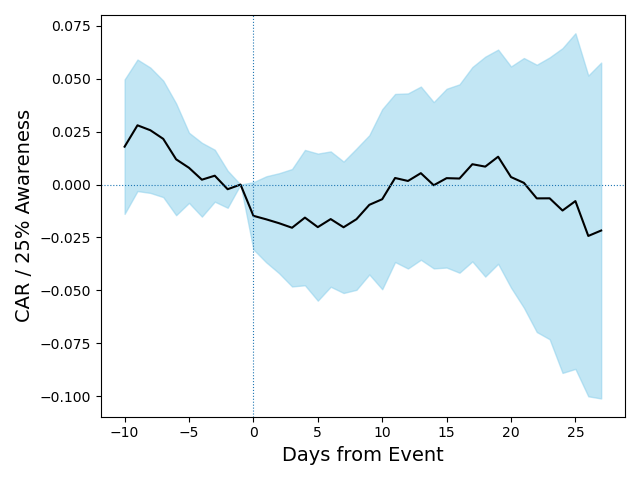}}
\par\end{centering}
\raggedright{}{\footnotesize Note: Figure shows cumulative stock returns
(Panel A) and cumulative abnormal stock returns (Panels B-C) for firms
around their social stances. Abnormal returns are defined in excess
of the CRSP value-weighted market return. Outcomes are scaled relative
to consumer awareness, averaged across firms using a $\tau^{2}_{j}$-weighted
average, and normalized relative to the day before a firm's event
($t=-1$). }{\footnotesize\par}
\end{figure}
\pagebreak{}
\par\end{center}

\begin{center}
\begin{figure}[H]
\begin{centering}
\caption{Cumulative Abnormal Stock Returns (Alternative Benchmarks)\label{fig:stock-prices-altrisk}}
\vspace*{-0.33cm}\subfloat[Panel A: vs.\ Market Model]{\begin{centering}
\par\end{centering}
\centering{}\includegraphics[width=0.49\textwidth]{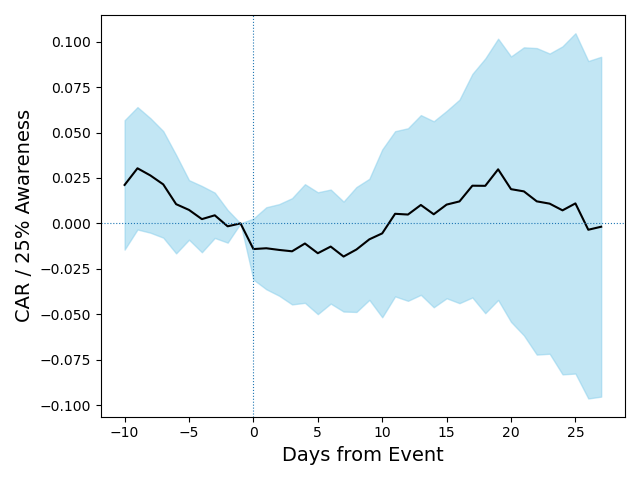}}
\par\end{centering}
\begin{centering}
\vspace*{-0.15cm}\subfloat[Panel B: vs.\ Fama-French 3-Factor Model]{\begin{centering}
\par\end{centering}
\centering{}\includegraphics[width=0.49\textwidth]{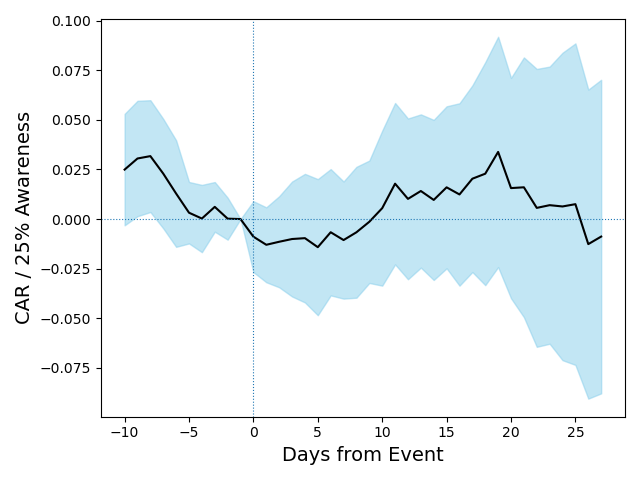}}
\par\end{centering}
\begin{centering}
\vspace*{-0.15cm}\subfloat[Panel C: vs.\ 3-Factor Model with Momentum]{\begin{centering}
\par\end{centering}
\centering{}\includegraphics[width=0.49\textwidth]{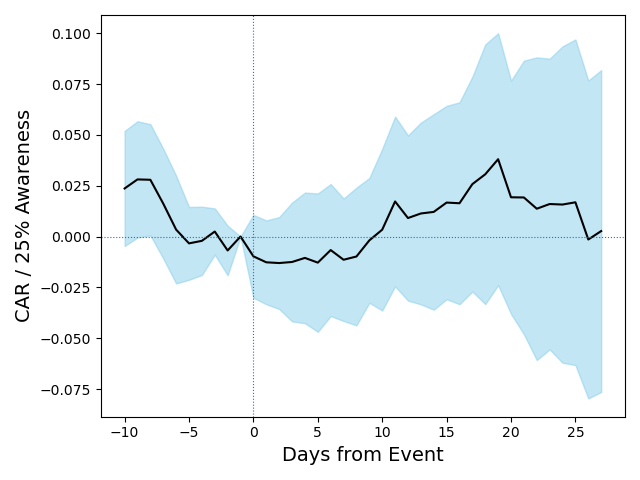}}
\par\end{centering}
\raggedright{}{\footnotesize Note: Figure shows cumulative abnormal
stock returns for firms in the days surrounding their social stances.
Abnormal returns are defined according to CAPM (Panel A), the Fama-French
3-Factor Model (Panel B), or the Fama-French Plus Momentum model (Panel
C). Outcomes are scaled relative to consumer awareness, averaged across
firms using a $\tau^{2}_{j}$-weighted average, and normalized relative
to the day before a firm's event ($t=-1$).}{\footnotesize\par}
\end{figure}
\pagebreak{}
\par\end{center}

\begin{center}
\begin{figure}[H]
\begin{centering}
\caption{Job Posting Behavior of Event-Study Firms\label{fig:jobpostings}}
\subfloat[Panel A: New Job Posting Flow]{\begin{centering}
\par\end{centering}
\centering{}\includegraphics[width=0.49\textwidth]{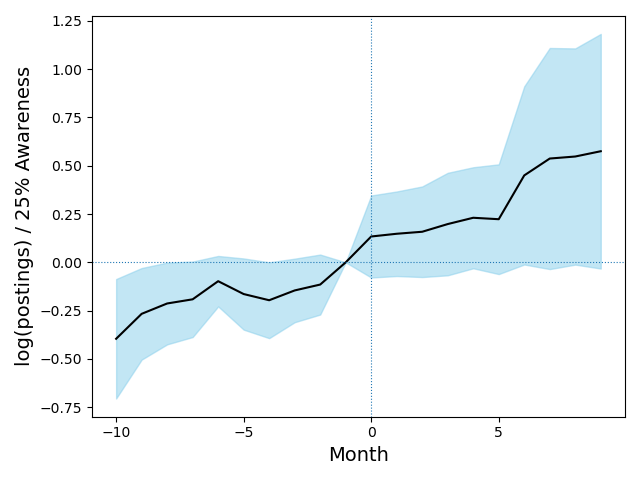}}\subfloat[Panel B: New Job Posting Flow (vs.\ Other U.S.\ Firms)]{\begin{centering}
\par\end{centering}
\centering{}\includegraphics[width=0.49\textwidth]{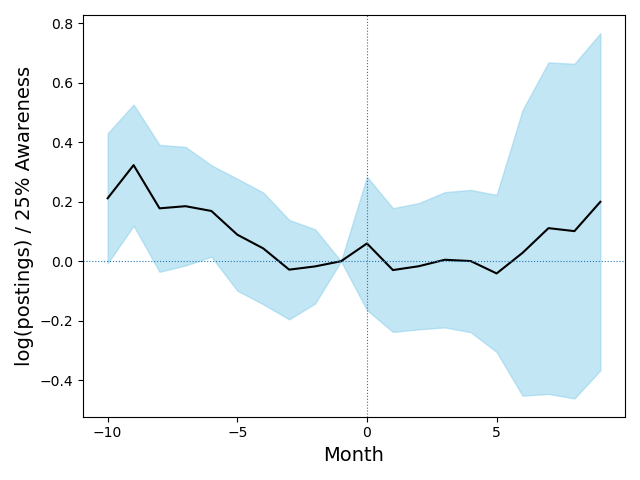}}
\par\end{centering}
\begin{centering}
\subfloat[Panel C: Average Posted Salary]{\begin{centering}
\par\end{centering}
\centering{}\includegraphics[width=0.49\textwidth]{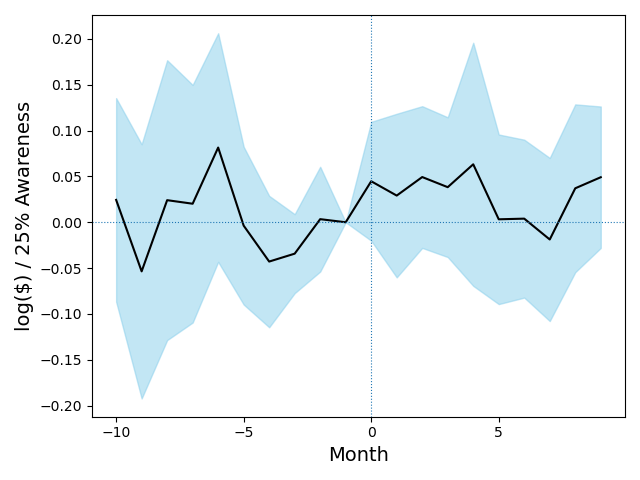}}\subfloat[Panel D: Average Posted Salary (vs.\ Other U.S.\ Firms)]{\begin{centering}
\par\end{centering}
\centering{}\includegraphics[width=0.49\textwidth]{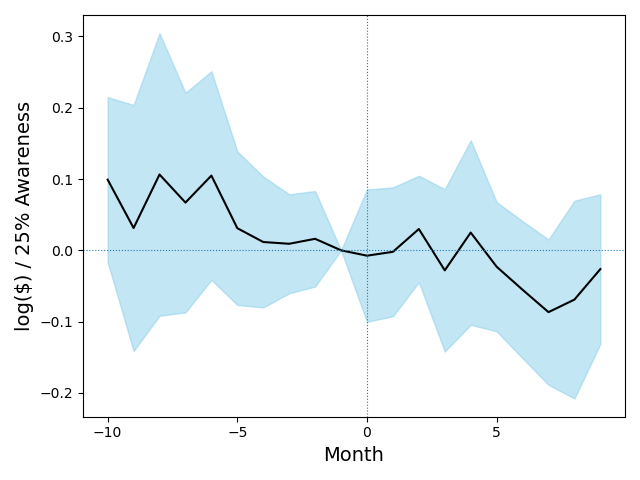}}
\par\end{centering}
\raggedright{}{\footnotesize Note: Figure shows new job postings (Panels
A-B) and average posted salaries (Panels C-D) for firms in the months
surrounding their social stances. Panel A counts the number of new
job postings by a firm in a given event-month across three job posting
datasets from Revelio Labs; these three datasets are sourced from
LinkUp, LinkedIn, and job aggregator websites, respectively. To avoid
dropping zeros when taking logs, we add one to all firm-month job
posting totals before taking logs. Panel B normalizes relative to
U.S.\ trends by calculating an analogous log total across all other
U.S.\ job postings in these data, and then subtracting this value
from our event-firm series. Panel C plots the log of the average posted
salary across a firm's job postings from a given event-month. Panel
D similarly subtracts the log average salary across all other U.S.\ job
postings in these data. Outcomes are scaled relative to consumer awareness,
averaged across firms using a $\tau^{2}_{j}$-weighted average, and
normalized relative to the month before a firm's event ($t=-1$).}{\footnotesize\par}
\end{figure}
\pagebreak{}
\begin{figure}[H]
\begin{centering}
\caption{Worker Flows of Event-Study Firms\label{fig:workerflows}}
\subfloat[Panel A: Inflows]{\begin{centering}
\par\end{centering}
\centering{}\includegraphics[width=0.75\textwidth]{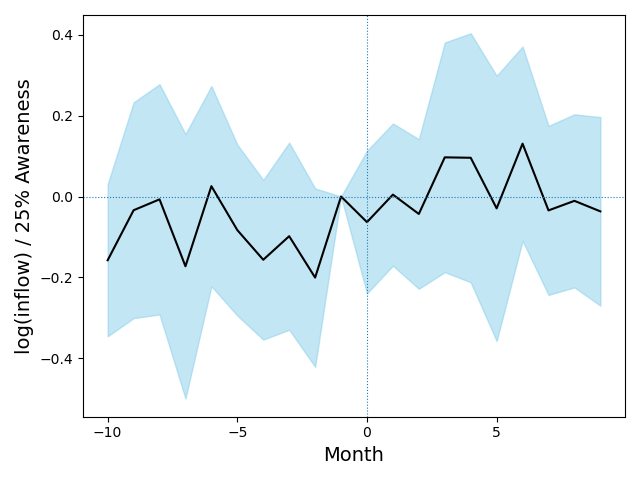}}
\par\end{centering}
\begin{centering}
\subfloat[Panel B: Outflows]{\begin{centering}
\par\end{centering}
\centering{}\includegraphics[width=0.75\textwidth]{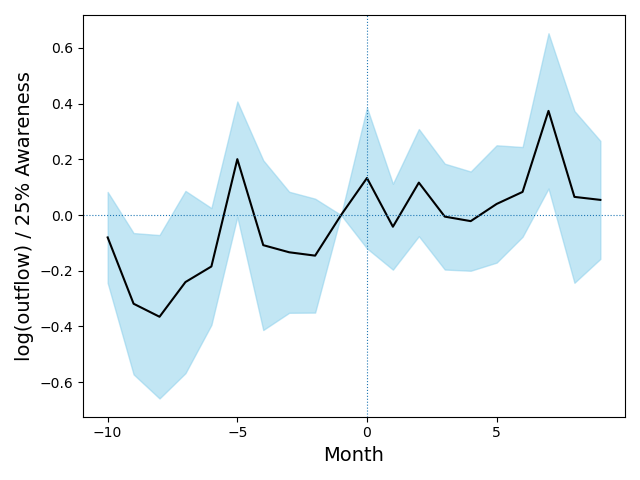}}
\par\end{centering}
\raggedright{}{\footnotesize Note: Figure shows worker flows for firms
in the months surrounding their social stances, based on LinkedIn
employment histories accessed via Revelio Labs. Panel A plots inflows,
calculated as the log number of employees working for an event-study
firm in event-month $t$ that weren't working for the firm in event-month
$t-1$. Panel B plots outflows, calculated as the log number of employees
who were working for an event-study firm in event-month $t-1$ but
not in $t$. To avoid dropping zeros, we add one to all totals before
taking logs. Outcomes are scaled relative to consumer awareness, averaged
across firms using a $\tau^{2}_{j}$-weighted average, and normalized
relative to the month before a firm's event ($t=-1$).}{\footnotesize\par}
\end{figure}
\pagebreak{}
\begin{figure}[H]
\begin{centering}
\caption{Glassdoor Employee Reviews of Event-Study Firms\label{fig:glassdoorreviews}}
\subfloat[Panel A: Overall Rating]{\begin{centering}
\par\end{centering}
\centering{}\includegraphics[width=0.49\textwidth]{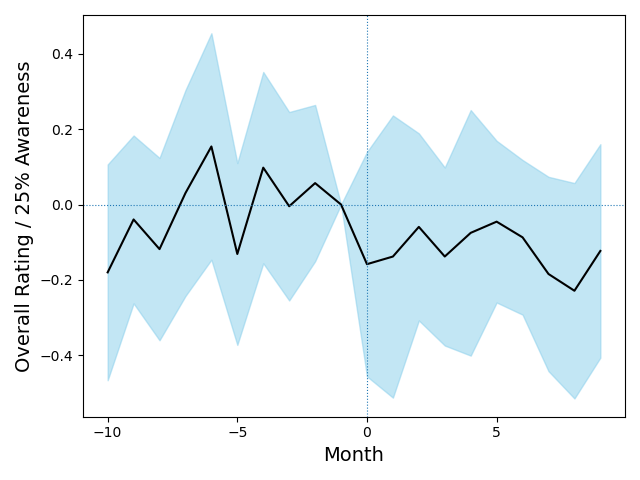}}\subfloat[Panel B: Culture and Values Rating]{\begin{centering}
\par\end{centering}
\centering{}\includegraphics[width=0.49\textwidth]{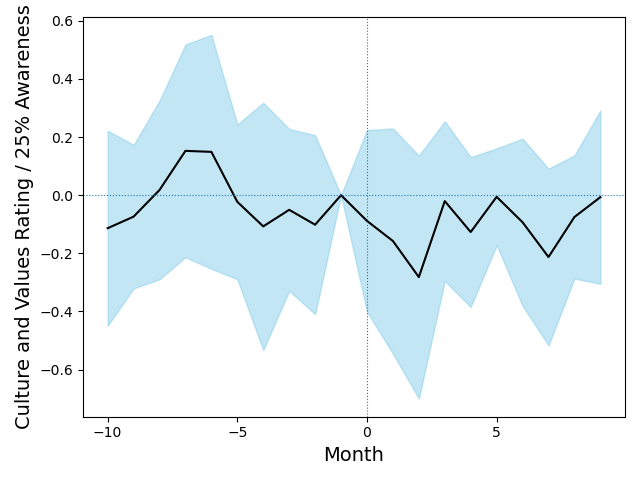}}
\par\end{centering}
\begin{centering}
\subfloat[Panel C: Overall Rating (by County Alignment Tercile)]{\begin{centering}
\par\end{centering}
\centering{}\includegraphics[width=0.49\textwidth]{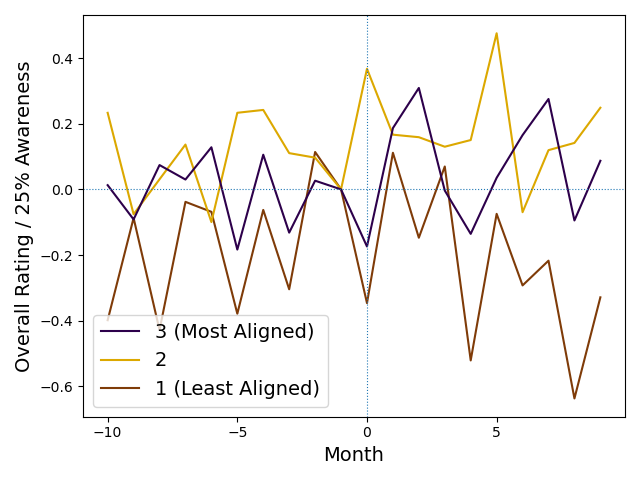}}\subfloat[Panel D: Culture and Values Rating (by County Alignment)]{\begin{centering}
\par\end{centering}
\centering{}\includegraphics[width=0.49\textwidth]{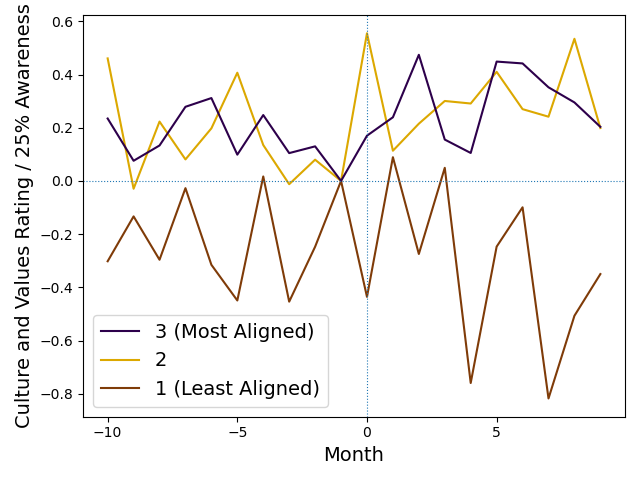}}
\par\end{centering}
\raggedright{}{\footnotesize Note: Figure shows employee reviews of
firms in the months surrounding their social stances, based on Glassdoor
data accessed via Revelio Labs. Reviews are anonymous and are provided
on a 1-5 star scale. Panel A shows the average overall rating for
the firm, while Panel B shows the average rating for the firm on the
``Culture and Values'' dimension specifically. Panels C-D respectively
plot average ratings on these two dimensions now split by alignment
tercile, inferring alignments from the vote share of the review's
county. Outcomes are scaled relative to consumer awareness, averaged
across firms using a $\tau^{2}_{j}$-weighted average, and normalized
relative to the month before a firm's event ($t=-1$).}{\footnotesize\par}
\end{figure}
\pagebreak{}
\begin{figure}[H]
\begin{centering}
\caption{Cumulative Net Sales Impact, by Alignment Direction \label{fig:cumulativeimpacts-by-stancetype}}
\includegraphics[width=1\textwidth]{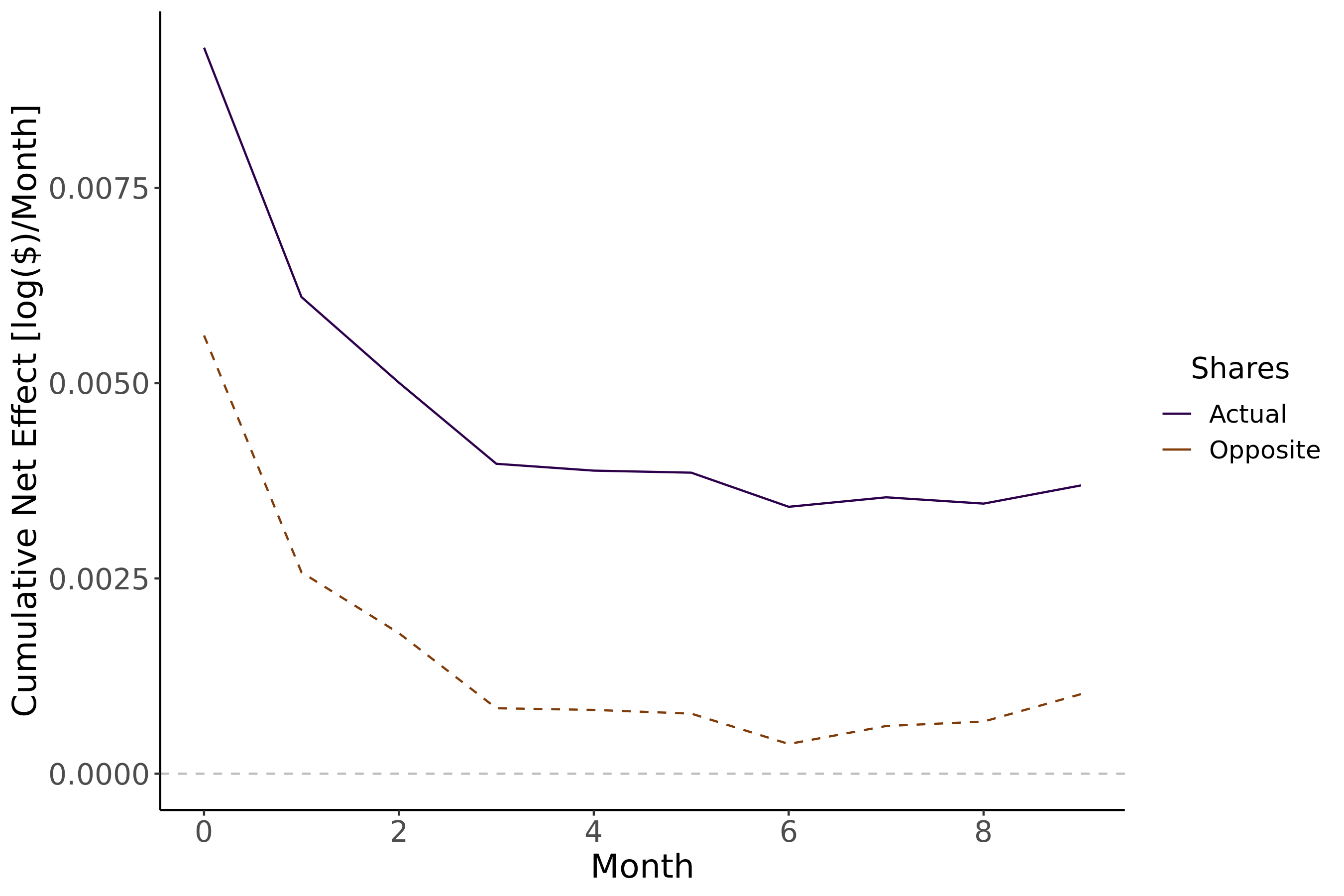}
\par\end{centering}
{\footnotesize Note: Figure shows the cumulative monthly impact of
aggregated group-specific consumption responses after a given number
of months. These cumulative monthly impacts are estimated using the
average stance consumption response estimates from Figure \ref{fig:consumption-responses-bygroup}
Panel B. These effects are aggregated across groups using either the
$\tau_{j}$-weighted average baseline consumption shares shown in
Figure \ref{fig:baseline_shares} of firms' actual consumer base,
or alternatively by the (reversed) baseline shares they would have
faced had they taken counterfactual stances in the opposite For/Against
direction on the same issue topic. See Section \ref{sec:supply-side}
for detail.}{\footnotesize\par}
\end{figure}
\pagebreak{}
\begin{figure}[H]
\begin{centering}
\caption{Variation in Baseline Group Consumption Shares, by Market\label{fig:baseline-share-examples}}
\subfloat[Panel A: By Geography]{\begin{centering}
\par\end{centering}
\centering{}\includegraphics[width=0.83\textwidth]{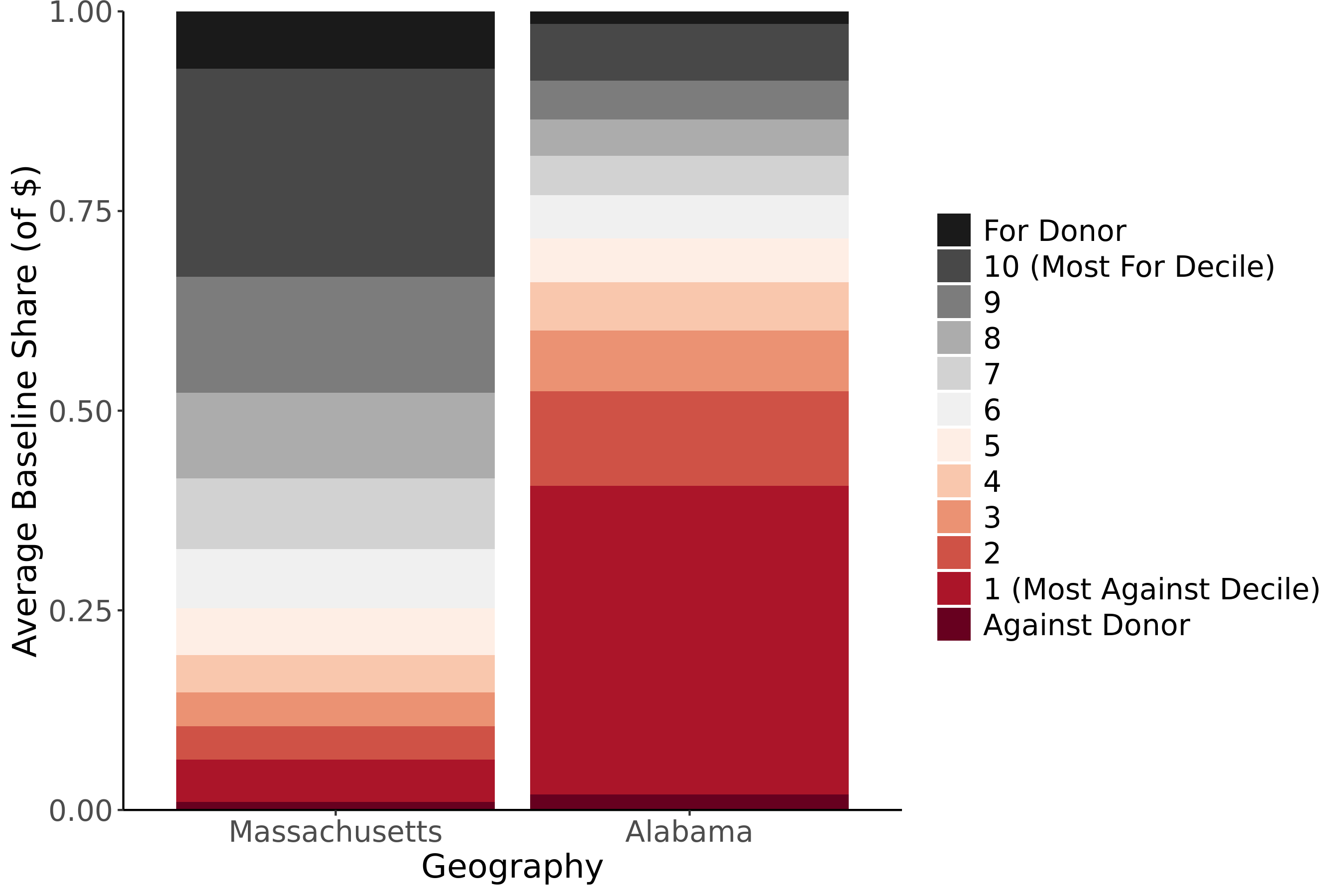}}
\par\end{centering}
\begin{centering}
\subfloat[Panel B: By Industry]{\begin{centering}
\par\end{centering}
\centering{}\includegraphics[width=0.83\textwidth]{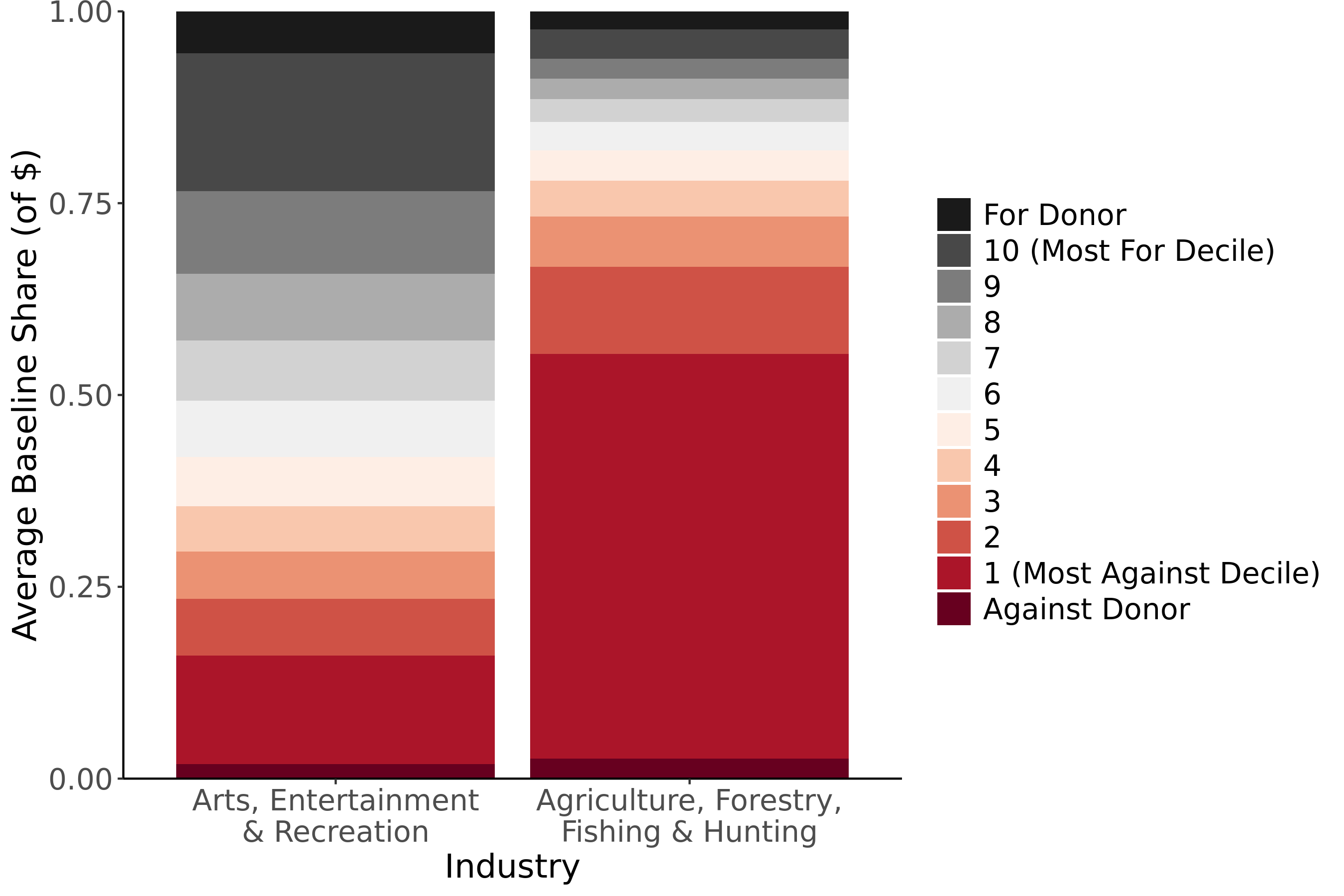}}
\par\end{centering}
{\footnotesize Note: Figure shows the share of consumption (in \$)
by group, which we refer to as baseline shares, in different markets.
Consumer groups are defined as described in Section \ref{sec:social-preferences-baseline-shares},
ordering consumers based on their predicted social alignment with
positions in the For donation cluster. Panel A shows each group's
share of consumption (in \$) by state for Massachusetts and Alabama
separately, aggregating across all transactions throughout the period
studied (2008-2023Q1) made by consumers based in the specified state.
Panel B shows each group's share of consumption (in \$) separately
for two example industries (``Arts, Entertainment \& Recreation''
and ``Agriculture, Forestry, Fishing \& Hunting''), aggregating
all transactions made at firms in the specified industry throughout
the period studied.}{\footnotesize\par}
\end{figure}
\pagebreak{}
\begin{figure}[H]
\begin{centering}
\caption{Cumulative Net Sales Impacts by State Baseline Shares and Hypothetical
Stance Direction\label{fig:impacts-bystate}}
\subfloat[Panel A: Hypothetical Stance, Direction Aligned with For Cluster]{\begin{centering}
\par\end{centering}
\centering{}\includegraphics[width=0.9\textwidth]{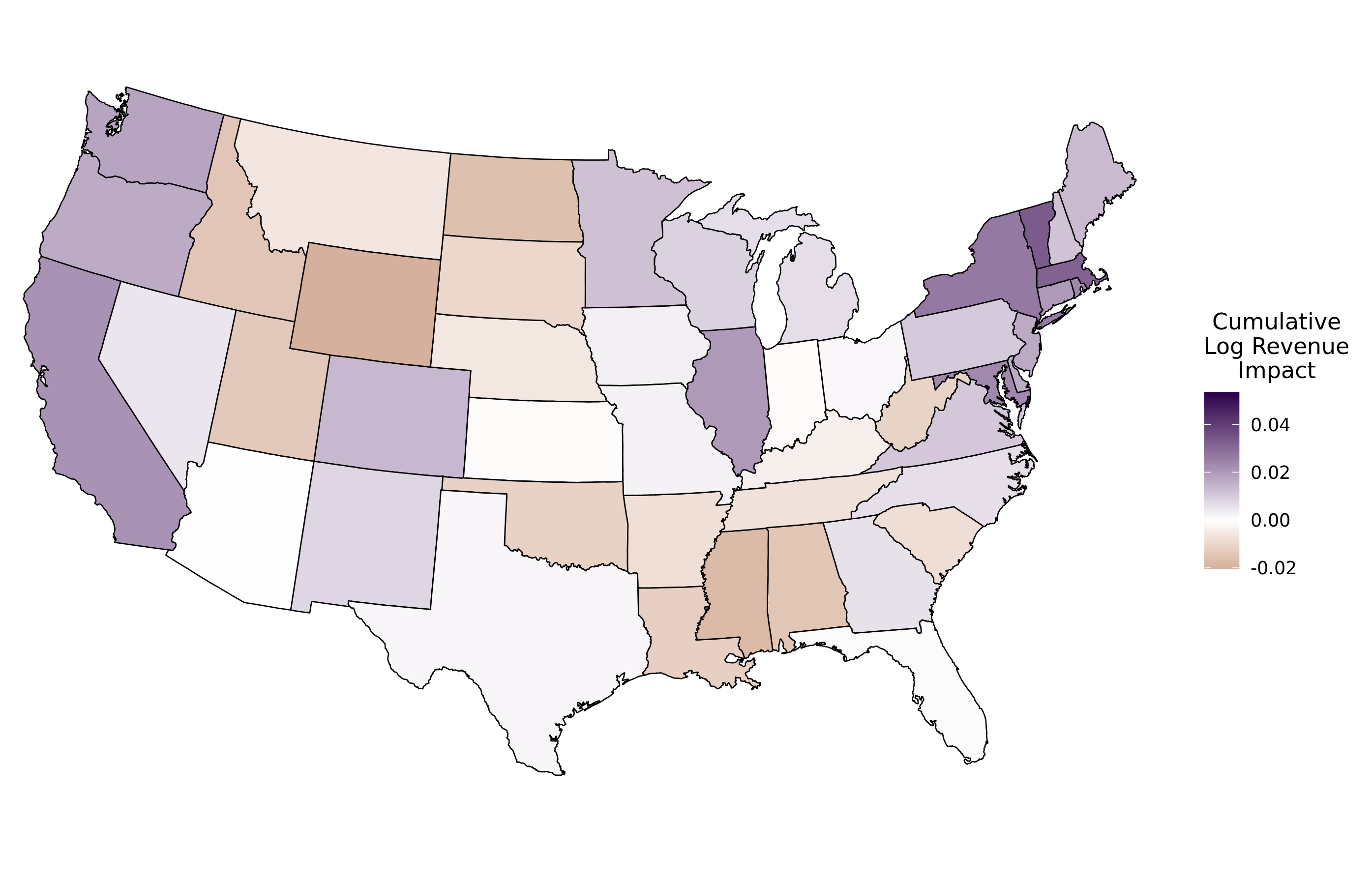}}
\par\end{centering}
\begin{centering}
\vspace*{-0.35cm}\subfloat[Panel B: Hypothetical Stance, Direction Aligned with Against Cluster]{\begin{centering}
\par\end{centering}
\centering{}\includegraphics[width=0.9\textwidth]{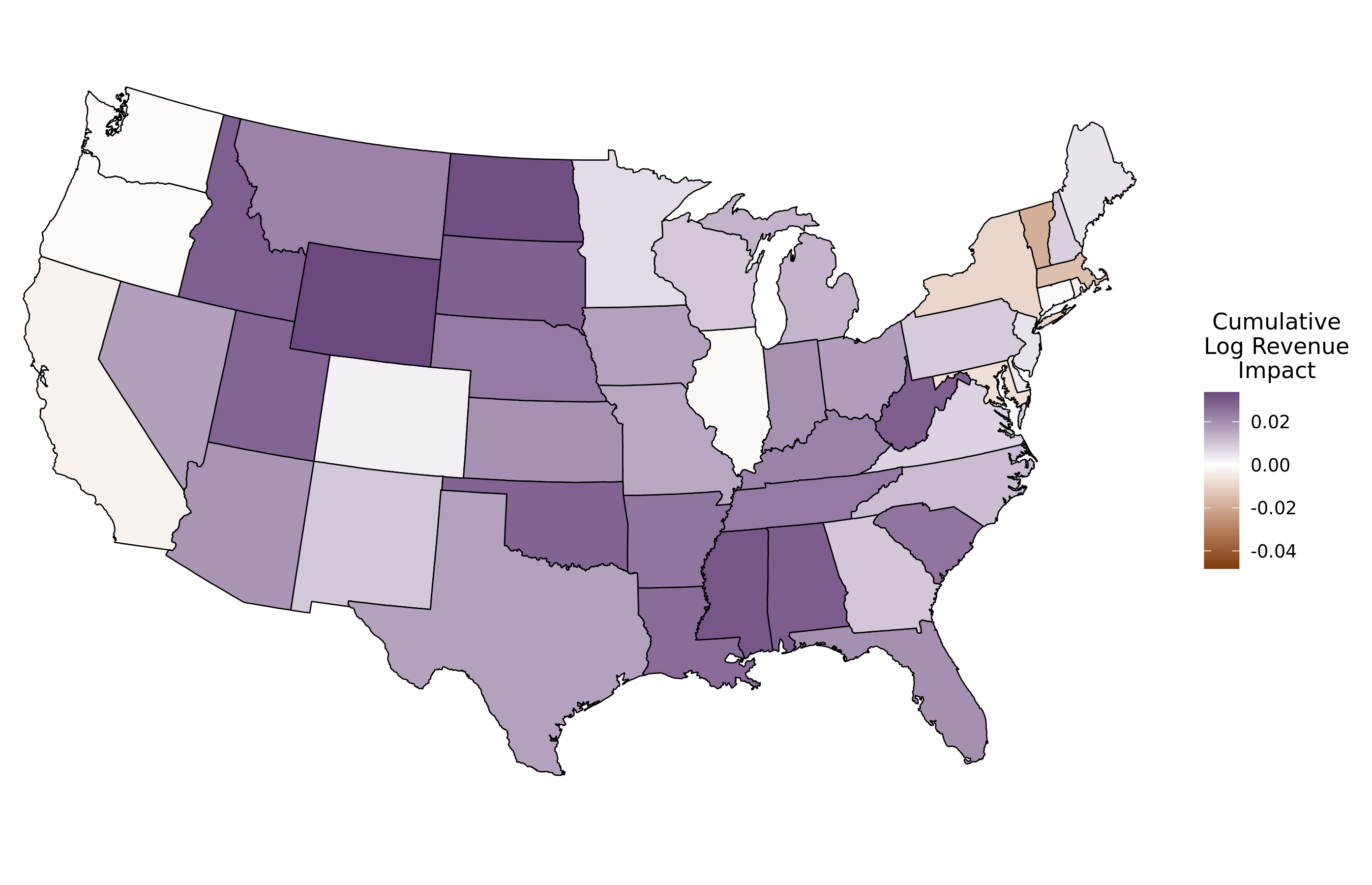}}
\par\end{centering}
{\footnotesize\vspace*{-0.35cm}Note: Figure shows the average monthly
log sales impact of a firm's social stance after five months, using
average consumption response estimates identified in Figure \ref{fig:consumption-responses-bygroup}
Panel B combined with state-specific baseline shares. These baseline
shares are constructed using all consumption within a state, as shown
in Figure \ref{fig:baseline-share-examples} Panel A for Massachusetts
and Alabama. Panels A and B, respectively, show the average monthly
log sales impact (per 25 percent consumer awareness) induced by an
average stance in the same direction vs.\ in the opposite direction
as positions in the For donation cluster.}{\footnotesize\par}
\end{figure}
\pagebreak{}
\begin{figure}[H]
\begin{centering}
\caption{BrandIndex-Based Consumer Awareness, by Event\label{fig:awareness-lines}}
\subfloat[Panel A: Share Hearing News about Firm]{\begin{centering}
\par\end{centering}
\centering{}\includegraphics[width=0.92\textwidth]{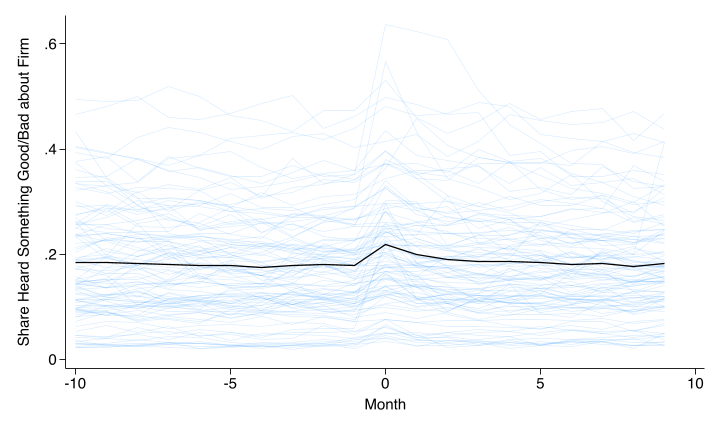}}
\par\end{centering}
\begin{centering}
\subfloat[Panel B: Consumer Awareness of Firm Social Stances]{\begin{centering}
\par\end{centering}
\centering{}\includegraphics[width=0.92\textwidth]{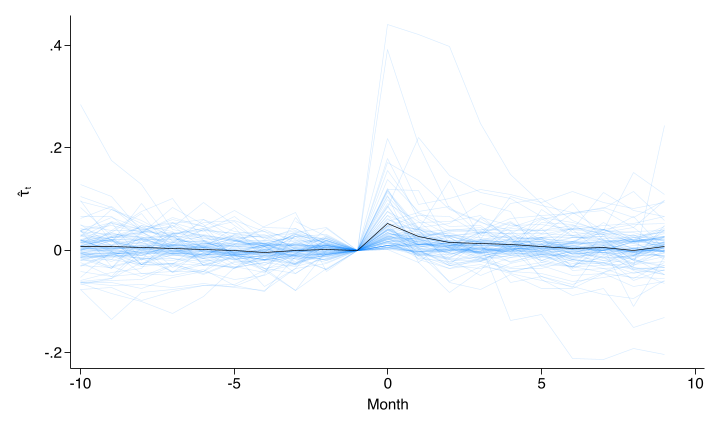}}
\par\end{centering}
{\footnotesize Note: Figure shows time trends related to our BrandIndex-based
consumer awareness measure, plotting separate blue lines for each
event covered by BrandIndex as well as a black line for the mean across
events. Panel A shows monthly trends in the share of consumers who
report having good or bad news about the firm in the last two weeks,
as described in Appendix Section \ref{subsec:event-size-detail}.
Denoting this share as $a_{jt}$ for event-firm $j$ in month $t$,
Panel B then shows trends in $\hat{\tau}_{jt}:=\frac{a_{jt}-a_{j,-1}}{1-a_{j,-1}}$.
The value for a given line in month $t=0$ gives our estimate of the
share of consumers who were aware of that firm's social stance event
(see Appendix Section \ref{subsec:event-size-detail} for detail).}{\footnotesize\par}
\end{figure}
\pagebreak{}
\begin{figure}[H]
\begin{centering}
\caption{News Coverage of Event-Study Firms\label{fig:social_stance_news}}
\subfloat[Panel A: Mean News Coverage of Social Stance Event-Study Firms Over
Time]{\begin{centering}
\par\end{centering}
\centering{}\includegraphics[width=0.77\textwidth]{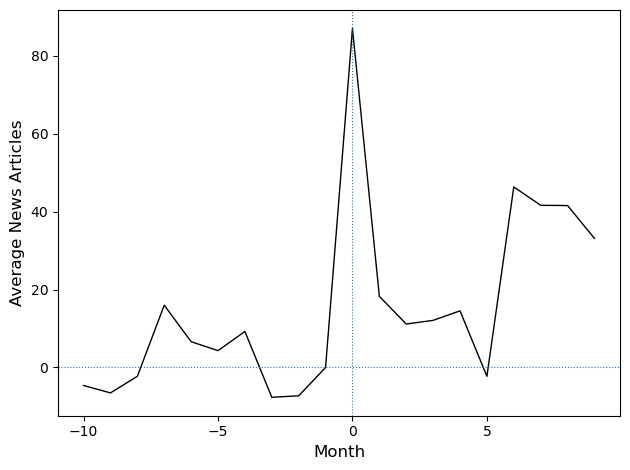}}
\par\end{centering}
\begin{centering}
\subfloat[Panel B: News-Based Event Size Distribution (Histogram)]{\begin{centering}
\par\end{centering}
\centering{}\includegraphics[width=0.77\textwidth]{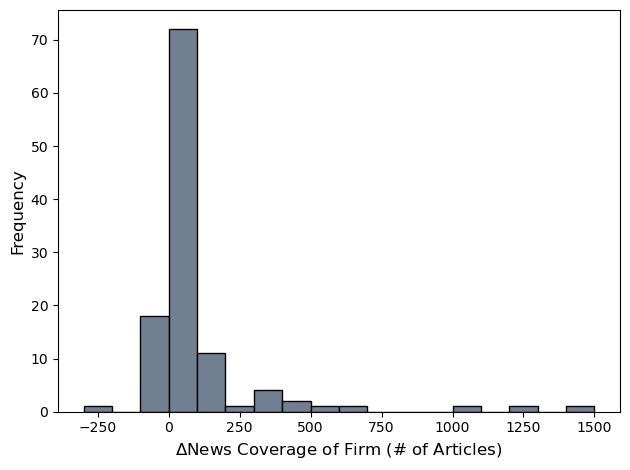}}
\par\end{centering}
{\footnotesize Note: Figure shows changes in news coverage of firms
in the months surrounding their social stances. Panel A shows the
number of TDM ProQuest U.S.\ Newsstream articles about the firm,
averaged by month across event-study firms. Social stance firms are
the subject of $\articlepremonthavg$ articles in month $t=-1$, and
this value is normalized to zero in Panel A. Panel B shows a histogram
summarizing across events the change in news coverage of the event-study
firm between months $t=-1$ and $t=0$, which is an alternative proxy
for an event's size or salience as defined in Section \ref{sec:event-selection-size}.}{\footnotesize\par}
\end{figure}
\pagebreak{}
\begin{figure}[H]
\begin{centering}
\caption{News Coverage of Stance (LLM Classification)\label{fig:social_stance_news_llm}}
\subfloat[Panel A: Mean News Coverage of Social Stance Over Time]{\begin{centering}
\par\end{centering}
\centering{}\includegraphics[width=0.77\textwidth]{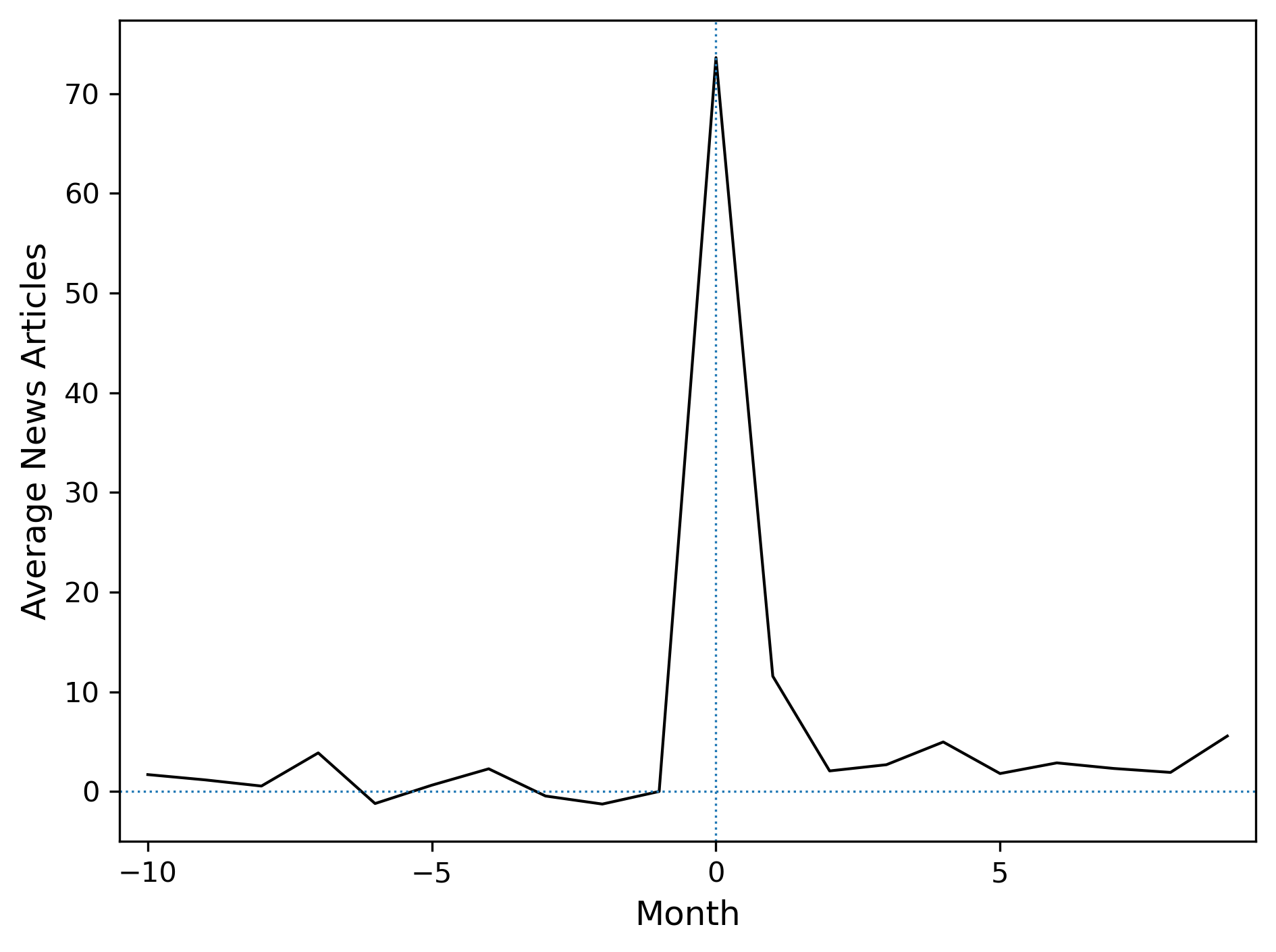}}
\par\end{centering}
\begin{centering}
\subfloat[Panel B: Stance-News-Based Event Size Distribution (Histogram)]{\begin{centering}
\par\end{centering}
\centering{}\includegraphics[width=0.77\textwidth]{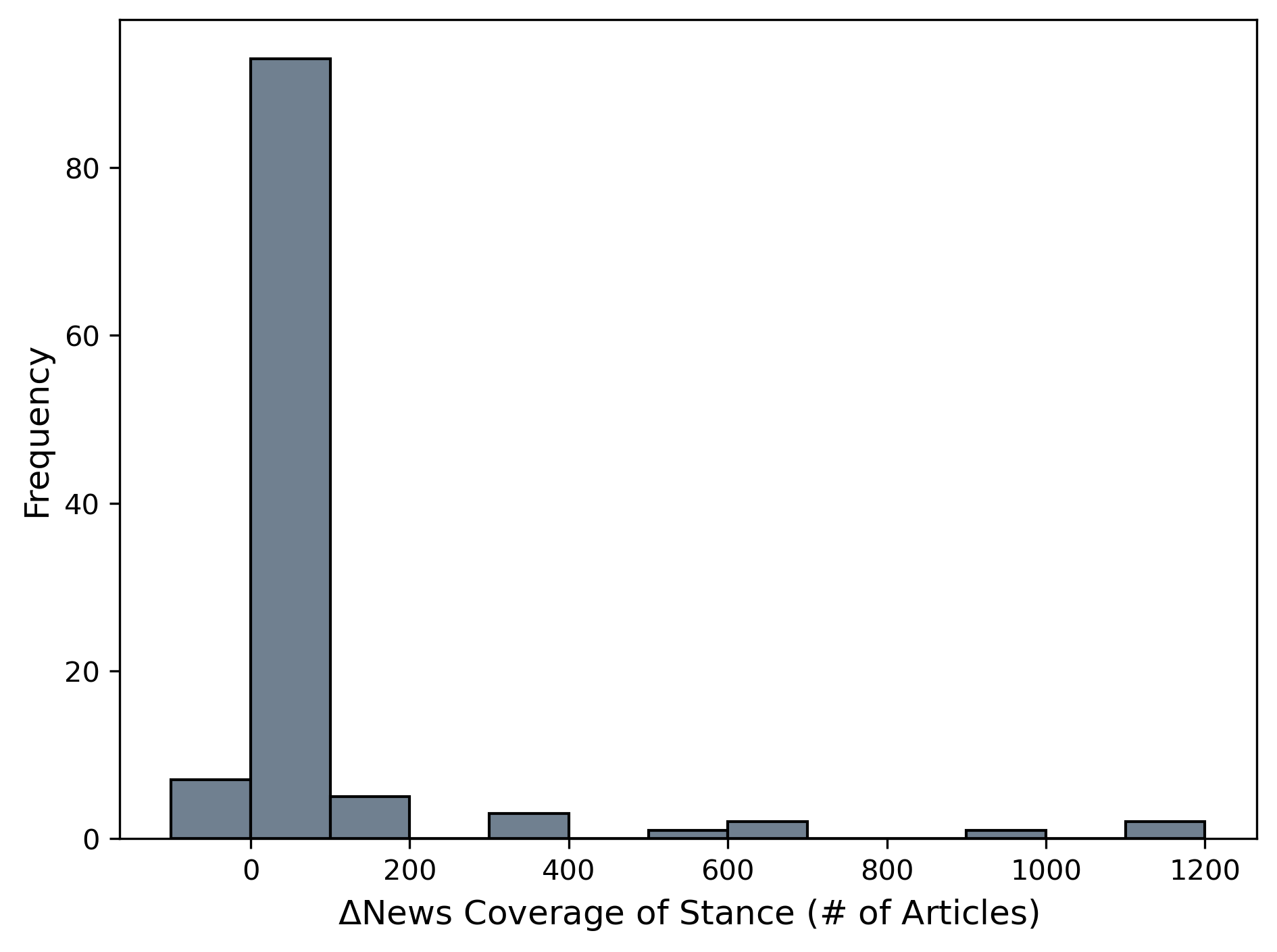}}
\par\end{centering}
{\footnotesize Note: Figure follows Figure \ref{fig:social_stance_news}
in showing changes in news coverage of firms in the months surrounding
their social stances. It modifies this previous figure by using an
LLM (gpt-4o-mini) to identify and filter to articles that specifically
discuss a controversial social stance taken by the event-study firm.
Panel A shows the number of TDM ProQuest U.S.\ Newsstream articles
about the firm's stance, averaged by month across event-study firms.
Panel B shows a histogram summarizing across events the change in
news coverage of the firm's stance between months $t=-1$ and $t=0$.}{\footnotesize\par}
\end{figure}
\pagebreak{}
\begin{figure}[H]
\begin{centering}
\caption{Google Trends Searches for Event-Study Firms\label{fig:google-trends}}
\includegraphics[width=1\textwidth]{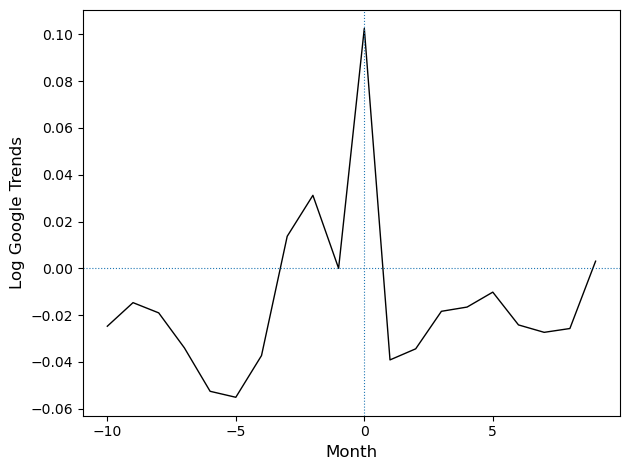}
\par\end{centering}
{\footnotesize Note: Figure shows changes in log Google Trends searches
for firms in the months surrounding their social stances. Changes
are normalized relative to the month before a firm's event ($t=-1$).
Months are defined as 4-week periods relative to the firm's event.}{\footnotesize\par}
\end{figure}
\par\end{center}
\pagebreak{}
\par\end{flushleft}

{\footnotesize\vspace*{-0.7cm}{\footnotesize \setstretch{0.79}\begin{longtable}{ccccl}
\caption{Generic List of All Social Stance Events} \label{tab:event-list-All} \\
\toprule
$\hat{\tau}_j$ & Description & Imputed $\hat{\tau}_j$ & Year & Direction \\
\midrule
\endfirsthead
\caption[]{Generic List of All Social Stance Events} \\
\toprule
$\hat{\tau}_j$ & Description & Imputed $\hat{\tau}_j$ & Year & Direction \\
\midrule
\endhead
\midrule
\multicolumn{5}{r}{Continued on next page} \\
\midrule
\endfoot
\bottomrule
\endlastfoot
0.440 & Comments against same-sex marriage & No & 2012 & Against \\
0.392 & Endorsement of controversial racial justice protester & No & 2018 & For \\
0.272 & Contraceptive-related corporate policy & Yes & 2014 & Against \\
0.217 & Gun control stance and policies & No & 2018 & For \\
0.179 & Removed brand of controversial figure & No & 2017 & For \\
0.171 & Transgender bathroom policy & No & 2016 & For \\
0.165 & Pro-immigration stance and ad campaign & Yes & 2017 & For \\
0.156 & Controversial diversity training & No & 2020 & For \\
0.137 & Criticism of protesters & No & 2017 & Against \\
0.121 & Severed ties with pro-gun group & No & 2018 & For \\
0.119 & Severed ties with pro-gun group & No & 2018 & For \\
0.117 & Prioritized privacy over law enforcement cooperation & No & 2016 & For \\
0.116 & Stance on voting legislation & No & 2021 & For \\
0.113 & Controversial diversity and inclusion campaign & No & 2015 & For \\
0.106 & Pulled ads from controversial program & No & 2011 & Against \\
0.104 & Banned account of controversial figure & No & 2021 & For \\
0.098 & Supported controversial public policy & No & 2021 & For \\
0.091 & Opposition to controversial figure & No & 2018 & For \\
0.086 & Removed popular figure for anti-LGBTQ remarks & No & 2013 & For \\
0.086 & Removed controversial video and app & No & 2021 & For \\
0.084 & Opposition to anti-LGBTQ legislation & No & 2022 & For \\
0.075 & Donated to controversial cause & No & 2018 & Against \\
0.074 & Banned controversial group & No & 2020 & For \\
0.068 & Opposed immigration restrictions & No & 2017 & For \\
0.067 & Supported same-sex marriage & No & 2012 & For \\
0.062 & Removed brand of controversial figure & No & 2015 & For \\
0.062 & Supported same-sex marriage & Yes & 2015 & For \\
0.060 & Supported controversial figure & No & 2017 & Against \\
0.055 & Donated to controversial causes & No & 2021 & Against \\
0.055 & Stance on voting legislation & No & 2021 & For \\
0.055 & Employee fired over anti-diversity memo & No & 2017 & For \\
0.052 & Supported controversial figure & No & 2016 & Against \\
0.051 & Continued ties with pro-gun group & Yes & 2018 & Against \\
0.049 & Supported Black Lives Matter & Yes & 2016 & For \\
0.049 & Severed ties with pro-gun group & No & 2018 & For \\
0.048 & Pro-LGBTQ+ campaign & No & 2014 & For \\
0.048 & Supported abortion access & No & 2021 & For \\
0.047 & Produced program perceived as anti-transgender & No & 2021 & Against \\
0.045 & Response to immigration restrictions & No & 2017 & Against \\
0.044 & Suspended sale of some types of firearms & No & 2012 & For \\
0.042 & Opposition to controversial figure & No & 2016 & For \\
0.041 & Opposition to anti-LGBTQ legislation & No & 2016 & For \\
0.040 & Pro-LGBTQ+ stance & No & 2018 & For \\
0.040 & Opposition to anti-LGBTQ legislation & No & 2016 & For \\
0.040 & Asked customers not to bring their firearms in stores & No & 2019 & For \\
0.038 & Supported abortion access & No & 2021 & For \\
0.038 & Donations to causes perceived as anti-abortion & No & 2021 & Against \\
0.037 & Pro-LGBTQ+ advertisement & No & 2015 & For \\
0.035 & Criticized comment by controversial figure & No & 2017 & For \\
0.034 & Asked customers not to bring their firearms in stores & No & 2014 & For \\
0.034 & Controversial donations by key stakeholder & Yes & 2017 & Against \\
0.034 & Stopped promoting brand of controversial figure & No & 2017 & For \\
0.032 & Pulled ads from controversial program & No & 2017 & For \\
0.032 & Donations to causes perceived as anti-abortion & No & 2019 & Against \\
0.031 & Severed ties with controversial figure & No & 2017 & For \\
0.030 & Controversial donations by key stakeholder & No & 2021 & Against \\
0.029 & Opposed anti-LGBTQ legislation & Yes & 2016 & For \\
0.029 & Donations to causes perceived as pro-gun & No & 2018 & Against \\
0.029 & Severed ties with pro-gun group & No & 2018 & For \\
0.027 & Supported controversial figure & No & 2019 & Against \\
0.027 & Perceived opposition to controversial figure & No & 2016 & For \\
0.026 & Opposed anti-LGBTQ legislation & Yes & 2015 & For \\
0.026 & Opposed immigration restrictions & No & 2017 & For \\
0.026 & Supported Black Lives Matter & No & 2020 & For \\
0.025 & Continued selling brand of controversial figure & No & 2017 & Against \\
0.025 & Opposition to controversial figure & No & 2021 & Against \\
0.024 & Supported gun control & No & 2018 & For \\
0.023 & Pulled ads from controversial program & No & 2018 & For \\
0.023 & Supported pro-choice abortion activist & Yes & 2016 & For \\
0.023 & Imposed health mandates & No & 2020 & For \\
0.023 & Supported Black Lives Matter & No & 2016 & For \\
0.022 & Pulled ads from controversial program & Yes & 2018 & For \\
0.022 & Supported abortion access & No & 2021 & For \\
0.022 & Severed ties with pro-gun group & No & 2018 & For \\
0.021 & Confirmed severed ties with pro-gun group & No & 2018 & For \\
0.021 & Removed brand of controversial figure & No & 2017 & For \\
0.020 & Supported Black Lives Matter and racial justice initiatives & No & 2020 & For \\
0.020 & Removed book from controversial figure & No & 2014 & For \\
0.019 & Imposed health mandates & Yes & 2022 & For \\
0.017 & Criticism of healthcare reform & No & 2012 & Against \\
0.017 & Removed brand of controversial figure & No & 2021 & For \\
0.017 & Asked customers not to openly carry firearms in stores & No & 2019 & For \\
0.016 & Supported controversial figure & Yes & 2016 & Against \\
0.016 & Opposed immigration restrictions & No & 2017 & For \\
0.016 & Donated to controversial cause & Yes & 2020 & Against \\
0.016 & Transgender locker room policy & Yes & 2015 & For \\
0.016 & Severed ties with pro-gun group & No & 2018 & For \\
0.016 & LGBTQ+ support/inclusion & No & 2017 & For \\
0.015 & Opposed immigration restrictions & No & 2017 & For \\
0.015 & Ad subverting gender stereotypes & No & 2011 & For \\
0.013 & Pulled ads from controversial program & No & 2018 & For \\
0.013 & Asked customers not to bring their firearms in stores & No & 2016 & For \\
0.010 & Opposed controversial figure and changes to national monuments & No & 2017 & For \\
0.010 & Severed ties with military-style weapons makers & No & 2018 & For \\
0.010 & Pulled ads from controversial program & No & 2018 & For \\
0.010 & Halted orders from supplier over gun sales & No & 2018 & For \\
0.009 & Donations to causes perceived as anti-abortion & No & 2021 & Against \\
0.009 & Removed brand of controversial figure & No & 2017 & For \\
0.009 & Stopped promoting brand of controversial figure & No & 2017 & For \\
0.009 & Severed ties with pro-gun group & No & 2018 & For \\
0.009 & Opposed anti-abortion legislation & No & 2019 & For \\
0.008 & Stated intent to continue flying confederate-era flag & No & 2017 & Against \\
0.008 & Severed ties with controversial figure & No & 2017 & For \\
0.006 & Severed ties with pro-gun group & No & 2018 & For \\
0.006 & Ad criticizing policy of controversial figure & No & 2018 & For \\
0.005 & Continued ties with controversial figure & No & 2021 & Against \\
0.004 & Severed ties with controversial figure & No & 2014 & For \\
0.004 & Pulled ads from controversial program & No & 2018 & For \\
0.004 & Stopped selling controversial family's brand & No & 2017 & For \\
0.004 & Severed ties with pro-gun group & No & 2018 & For \\
0.004 & Opposed controversial figure and his supporters & Yes & 2017 & For \\
0.003 & Affirmative action hiring initiative & No & 2020 & For \\
0.002 & Pulled ads from controversial program & No & 2017 & For \\
0.002 & Key stakeholder supported controversial figure & Yes & 2021 & Against \\
0.001 & Pulled ads from controversial platform & No & 2016 & For \\
0.000 & Opposed immigration restrictions & No & 2017 & For \\
\end{longtable}
}\vspace*{-0.6cm}}{\footnotesize\par}

\begin{spacing}{0.8}
\noindent\begin{flushleft}
{\footnotesize Note: Table shows the complete list of social stance
events we analyze. For each event, we provide the year, whether the
stance was aligned with positions in the For vs.\ Against donation
cluster, our estimate of the share of consumers who were aware of
this event ($\hat{\tau}_{j}$), an indication of whether $\hat{\tau}_{j}$
was imputed from Google Trends data and news reports because this
event was not covered by BrandIndex data, and a brief description
of each event. These descriptions are generic, as we are unable to
identify the firms included in our analysis under the terms of the
agreement with our data provider.}
\par\end{flushleft}
\end{spacing}

\end{document}